\documentclass[iop,onecolumn]{emulateapj}

\usepackage{amsmath}

\newcommand{\Bpar}{B_\parallel}
\newcommand{\Bperp}{B_\perp}
\newcommand{\didl}{\frac{\partial I_i}{\partial \lambda}}
\newcommand{\didlnoi}{\frac{\partial I(\lambda)}{\partial \lambda}}
\newcommand{\didltwo}{\frac{\partial^2 I_i}{\partial \lambda^2}}
\newcommand{\didltwonoi}{\frac{\partial^2 I(\lambda)}{\partial \lambda^2}}
\newcommand{\coschi}{\cos 2 \chi}
\newcommand{\sinchi}{\sin 2 \chi}

\begin{document}

\title{Bayesian inference of solar and stellar magnetic fields \\ in the weak-field approximation}
\author{A. Asensio Ramos}
\affil{Instituto de Astrof\'{\i}sica de Canarias, 38205, La Laguna, Tenerife, Spain \\
Departamento de Astrof\'{\i}sica, Universidad de La Laguna, E-38205 La Laguna, Tenerife, Spain \email{aasensio@iac.es}}

\begin{abstract}
The weak-field approximation is one of the simplest models that allows us to relate
the observed polarization induced by the Zeeman effect with the magnetic field vector 
present on the plasma of interest. It is usually applied for diagnosing
magnetic fields in the solar and stellar atmospheres. A fully Bayesian approach to the
inference of magnetic properties in unresolved structures is presented. The analytical
expression for the marginal posterior distribution is obtained, from which 
we can obtain statistically relevant information about the model parameters. The role 
of a-priori information is discussed and a hierarchical procedure is presented that gives robust results
that are almost insensitive to the precise election of the prior. The strength of the formalism
is demonstrated through an application to IMaX data. Bayesian methods can optimally
exploit data from filter-polarimeters given the scarcity of spectral information as compared
with spectro-polarimeters. The effect of noise and how it degrades our ability to extract information from the Stokes profiles
is analyzed in detail.
\end{abstract}


\section{Introduction}
Extracting information about the magnetic field vector from spectro-polarimetric observations
is not devoid of difficulties. The main one is that, practically always, one has to go through a modeling
phase. This modeling typically consists of setting an atmospheric model that depends on some parameters
which one wants to infer. Such a procedure, usually known as spectro-polarimetric inversion, has allowed to extract
extremely valuable information about the behavior of the magnetic field in the
solar photosphere and chromosphere \citep[see e.g.,][and references therein]{bellotrubio_spw4_06}.

It is hard to summarize in a few lines the history of spectro-polarimetric inversions from the
first steps back in the 1970s. The initially proposed models were of low complexity because the quality of the observations
and the computing power did not allow the use of more elaborate 
models \citep[e.g.,][]{auer_heasly_house77,skumanich_lites87,lites_skumanich90,keller90}.
Although the assumptions on which these models are based may not be exactly fulfilled 
in the solar atmosphere, their simplicity allowed to put the cornerstone for quantitative
spectropolarimetry. In fact, these models are still in use for interpreting high-quality
observations from the most advanced instruments \citep{Lagg04,orozco_hinode07,borrero_vfisv10}.
Later on, inversion codes based on the concept of response functions \citep{landi_response77} have 
facilitated the inversion of high-quality Stokes profiles making it possible
to infer vertical stratifications of the magnetic properties of the atmosphere \citep{sir92,socas_trujillo_ruiz00,frutiger00}.

After the enormous success of standard inversion methods based on least-squares optimization,
it is time to study in depth the inversion process itself and introduce more powerful
techniques. This is what has been done recently by \cite{asensio_martinez_rubino07}, who
treated the inversion process as a Bayesian probabilistic inference problem.
These techniques, which allow to fully exploit the information encoded
in the Stokes profiles once a model has been proposed to explain them, 
have been used recently by \cite{asensio_hinode09,asensio_spw6_10} to
infer that fields in the quietest regions of the solar internetwork appear to
be quasi-isotropically distributed, reinforcing the previous results of
\cite{marian_clv08}. We consider that the Bayesian approach is the best choice, specially in those
cases in which the spectro-polarimetric signal is at the noise level or the
wavelength covering is very sparse (like in filter-polarimeters). Additionally,
it gives the opportunity to quantitatively compare different models and use
different models as a committee to gain insight on a common physical parameter 
\citep{asensio_spw6_10}.

The approach followed by \cite{asensio_martinez_rubino07}
based on a Markov Chain Montecarlo sampler is completely general so that it can
cope with very complex radiative transfer forward problems. The main drawback is
that it can become costly in terms of computational time, although not prohibitive,
as already shown by \cite{asensio_hinode09}. Obviously, Bayesian inference cannot 
be compared to standard inversion methods just by the computing cost, because
the amount of information obtained is much richer. In order to motivate the
application of Bayesian inference, we consider in this paper 
the assumption of weak field \citep{landi_landolfi04} for the inference
of magnetic fields from the observation of Stokes profiles. This is one of the most
straightforward approximations one can consider to relate the Stokes profiles and the
magnetic field vector. In spite of
its simplicity, its range of applicability is very broad and it is
systematically applied in different fields, from solar to stellar
physics, as shown in the next section. As we show, the simplicity of the model allows us to obtain analytical
expression for some of the posterior distributions of the model parameters. We consider
that the weak-field approximation is of practical application while simultaneously being
simple enough to demonstrate the power of Bayesian inference and show their fundamental 
points.

Section 2 describes
the model. Section 3 shows the simple non-hierarchical approach and also sets
the notation used throughout the paper. Section 4 presents the robust hierarchical
model for the case in which the noise standard deviation is known and when it is
inferred from the data. Finally, the method is demonstrated with some examples in section 5.


\section{The Zeeman weak-field approximation}
When the splitting produced in a given spectral line via the Zeeman effect by the
presence of a deterministic magnetic field ($\bar g \Delta \lambda_B$, with $\bar g$ the 
effective Land\'e factor) is smaller than the line broadening
($\Delta \lambda_D$), the line
is said to be well modeled in the weak-field regime \citep[][]{landi73}.
Writing the expressions for $\Delta \lambda_B$ and $\Delta \lambda_D$
\citep[e.g.,][]{landi_landolfi04}, a line is in the weak-field approximation
when the magnetic field strength fulfills:
\begin{equation}
B < \frac{4 \pi m c}{\bar g \lambda_0 e_o} \sqrt{\frac{2kT}{M}+v_\mathrm{mic}},
\end{equation}
where $m$ and $e_0$ are the electron mass and charge, respectively, $c$ is the speed of light,
$k$ is the Boltzmann constant, $M$ is the mass of the species, $\lambda_0$ is
the central wavelength of the spectral line under consideration and $v_\mathrm{mic}$ is the microturbulent
velocity. For an iron line at $\lambda_0=5000$ \AA, using $v_\mathrm{mic}=1$ km s$^{-1}$
and $T=5800$ K, we end up with:
\begin{equation}
\bar g B < 2400 \mathrm{G}.
\end{equation}
In principle, this is more than enough to deal with photospheric magnetic
fields outside from active regions in the solar atmosphere observed
with magnetically sensitive lines in the optical (with $\bar g>1$) or even active
regions observed in lines with a weak magnetic sensitivity. Since 
the thermal width is enhanced in the chromosphere due to the increased 
temperature and the field strength is known to be smaller, the weak-field
approximation is especially interesting for inferring magnetic fields
at chromospheric heights.

In this approximation, a straightforward relation between the magnetic
properties of the plasma and the emergent Stokes parameters exist.
In present day observations of the weakly magnetized regions of the solar
atmosphere we cannot state that we are resolving all magnetic structures. In 
case that our resolution element is not filled with an unidirectional magnetic
field vector, we can mimic the loss of signal by assuming that the observed signal in the
pixel is obtained as the average of a magnetic component with a relative
weight $f$ and a non-magnetic component with the weight $1-f$. The following 
equations hold for Stokes $V$ at first order in the field strength
and for Stokes $Q$, $U$ at second order in the field strength \citep[e.g.,][]{landi_landolfi04}:
\begin{eqnarray}
V(\lambda) &=& \alpha f \Bpar \didlnoi \nonumber \\
Q(\lambda) &=& \beta f \Bperp^2 \coschi \didltwonoi \nonumber \\
U(\lambda) &=& \beta f \Bperp^2 \sinchi \didltwonoi,
\label{eq:circ_lin_pol}
\end{eqnarray}
as functions of $\Bpar$, the projection of the magnetic field vector along the line-of-sight (LOS), 
$\Bperp$, the component of the vector perpendicular to the LOS, $\chi$, the field azimuth and $I(\lambda)$, the wavelength
variation of the intensity across the spectral line. The
proportionality constants $\alpha$ and $\beta$ have the values:
\begin{equation}
\alpha = -4.67 \times 10^{-13} \bar{g} \lambda^2, \qquad \beta = -5.45 \times 10^{-26} \bar{G} \lambda^4,
\end{equation}
with the wavelength $\lambda$ in \AA\ and the components of the magnetic field measured in G.
The factor $\bar{g}$ is the effective Land\'e factor and $\bar{G}$ is the
equivalent for linear polarization \citep[e.g.,][]{landi_landolfi04}. Both factors measure the sensitivity
of the spectral line to the presence of a magnetic field. The previous expressions are only valid
when $\Bpar$, $\Bperp$, the field azimuth $\chi$, the line-of-sight velocity, the Doppler width, and any
broadening mechanism are constant with height in the line formation region. Additionally, the expression
for Stokes $Q$ and $U$ can only be applied for non-saturated lines. Since the
field azimuth is constant with height in the atmosphere, it is possible to define the total
linear polarization $L = (Q^2+U^2)^{1/2}$ which can be written as:
\begin{equation}
L =  f \Bperp^2 \left| \beta \didltwonoi \right|,
\label{eq:lin_pol}
\end{equation}
where the absolute value is a consequence of the definition of $L$. From this point,
we assume that $Q(\lambda)$, $U(\lambda)$ and $V(\lambda)$, the observed wavelength variation of
the linear and circular polarization profiles, respectively, can be correctly modeled
with the aid of Eqs. (\ref{eq:circ_lin_pol}).

In spite of and thanks to its simplicity, the weak-field approximation is broadly
applied for the inference of solar and stellar magnetic fields from the observation
of Stokes profiles. A limited selection includes the detection and diagnostic of
magnetic fields in: central stars of planetary nebulae \citep{jordan05}, white dwarfs \citep{aznar_cuadrado04}
pulsating stars \citep{silvester09}, hot subdwarfs \citep{otoole05}, Ap and Bp stars \citep{wade00}
and chemically peculiar stars \citep{bagnulo02}. The least-squares deconvolution (LSD) technique that
is widely used for the detection of magnetic fields in solar-type stars \citep{donati97} is
also fundamentally based on the weak-field approximation. Many synoptic magnetographs like those
of Big Bear \citep{spirock01,varsik95} are calibrated using this formalism or equivalent. The
circular polarization observed in chromospheric lines like the 10830 \AA\ multiplet of \ion{He}{1}
is very well modeled under the weak-field approximation \citep{merenda06,asensio_trujillo_hazel08}. It is
even used to produce modern vector magnetograms like those obtained with the IMaX instrument 
\citep{imax11} onboard the Sunrise balloon \citep{sunrise10}.

\section{Non-hierarchical Bayesian approach}
The weak-field approximation is a model for the interpretation of observed Stokes profiles
that depends on the four-vector of parameters $(f,\Bpar,\Bperp,\chi)$. Under
the Bayesian approach \citep[see][and references therein]{asensio_martinez_rubino07,asensio_hinode09}, 
all knowledge gained about the parameters when some
dataset $D$ is presented to the model is encoded on the posterior probability
distribution function $p(f,\Bpar,\Bperp,\chi|D)$. Using the Bayes theorem, this
posterior distribution can be written as:
\begin{equation}
p(f,\Bpar,\Bperp,\chi|D) = \frac{p(D|f,\Bpar,\Bperp,\chi) p(f,\Bpar,\Bperp,\chi)}{p(D)},
\label{eq:bayes_theorem}
\end{equation}
where $p(D|f,\Bpar,\Bperp,\chi)$ is the likelihood distribution that takes
into account the influence of data on our knowledge of the model
parameters and $p(f,\Bpar,\Bperp,\chi)$ is the prior
distribution that accounts for all information about the parameters known
in advance. The term $p(D)$ is the evidence or marginal posterior (the area below
the multidimensional posterior) that, for parameter inference, is just an unimportant 
multiplicative constant that we neglect in the following. We analyze now the analytical form of all 
quantities present in Eq. (\ref{eq:bayes_theorem}).

\subsection{Likelihood}
When Stokes profiles are observed with a spectro-polarimeter or a
filter-polarimeter we measure the circular and linear
polarization at $N$ discrete wavelength points. Consequently,
the observables are $\{(V_i,Q_i,U_i)$, $i=1,\ldots,N\}$,
where $V_i$, $Q_i$ and $U_i$ represent the value of the circular and
linear polarization at wavelength $\lambda_i$, respectively.
Assuming that the observations are corrupted with uncorrelated Gaussian noise 
with zero mean and variance $\sigma_n^2$, the likelihood function is given by
the following expression:
\begin{eqnarray}
p(D|f,\Bpar,\Bperp,\chi) &=& (2\pi)^{-3N/2} \sigma_n^{-3N} 
\exp \left\{ -\frac{1}{2\sigma_n^2} \left[ \sum_{i=1}^N \left( V_i - \alpha f \Bpar \didl \right)^2 \right. \right. \nonumber \\
&+& \left. \left. \sum_{i=1}^N \left( Q_i - \beta f \Bperp^2 \coschi \didltwo \right)^2 + 
\sum_{i=1}^N \left( U_i - \beta f \Bperp^2 \sinchi \didltwo \right)^2 \right] \right\}
\end{eqnarray}
where we have assumed that the total likelihood is given by the product of the likelihood
for each wavelength point and Stokes parameter. Note that we have assumed the same noise
variance for Stokes $Q$, $U$ and $V$. If this is not the case, the likelihood can be modified
accordingly and the following calculations are modified accordingly. 
Note also that we have assumed that the uncertainty in the first and second
intensity derivatives are negligible. If this is not the case, its effect can be
introduced by substituting $\sigma_n^2$ by the variance of the terms inside
the parenthesis, which might also include correlations between $(V_i,Q_i,U_i)$
and $\didl$ or $\didltwo$.
It is important to point out that the likelihood is only a Gaussian distribution if we deal
with the observables $Q(\lambda)$ and $U(\lambda)$, and not for $L(\lambda)$. In this case,
it would follow a Rayleigh distribution.
Due to the simplicity of the weak-field model, the exponent of the likelihood can be easily factorized
so that quantities related to observables are isolated from the model parameters. After some
algebra, we end up with:
\begin{equation}
p(D|f,\Bpar,\Bperp,\chi) = (2\pi)^{-3N/2} \sigma_n^{-3N} \exp \left[ -(A_1 + A_2 f^2 \Bpar^2 + A_3 f^2 \Bperp^4 - 2 A_4 f \Bpar 
- 2 (A_5 \coschi + A_6 \sinchi) f \Bperp^2) \right],
\label{eq:likelihood}
\end{equation}
where the quantities $A_i$ are the only ones that depend on the observations and
are given by:
\begin{eqnarray}
A_1 &=& (2\sigma_n^2)^{-1} \sum_i (V_i^2 + Q_i^2 + U_i^2), \qquad A_2 = (2\sigma_n^2)^{-1} \alpha^2 \sum_i \left(\didl \right)^2, \qquad
A_3 = (2\sigma_n^2)^{-1} \beta^2 \sum_i \left( \didltwo \right)^2, \nonumber \\
A_4 &=& (2\sigma_n^2)^{-1} \alpha \sum_i V_i \didl, \qquad A_5 = (2\sigma_n^2)^{-1} \beta \sum_i Q_i \didltwo,
\qquad A_6 = (2\sigma_n^2)^{-1} \beta \sum_i U_i \didltwo.
\label{eq:A_values}
\end{eqnarray}
More intuition can be gained if we complete the squares in the likelihood function, so that:
\begin{eqnarray}
p(D|f,\Bpar,\Bperp,\chi) &=& (2\pi)^{-3N/2} \sigma_n^{-3N} \exp \left\{ -f^2\left[A_2 \left(\Bpar -\frac{A_4}{A_2f} \right)^2
+ A_3 \left( \Bperp^2 - \frac{A_5 \coschi + A_6 \sinchi}{A_3f} \right)^2 \right] \right\} \nonumber \\
&\times& \exp \left\{ -\left[A_1 - \frac{A_4^2}{A_2} - \frac{(A_5 \coschi + A_6 \sinchi)^2}{A_3}\right] \right\},
\label{eq:likelihood_completesquare}
\end{eqnarray}

\subsection{Priors}
\label{sec:priors}
One of the advantages of the Bayesian approach is that all a-priori information about model
parameters is made explicit in the formalism. This is made through the probability distribution 
$p(f,\Bpar,\Bperp,\chi)$. For simplicity, we assume that the prior distribution factorizes, so that:
\begin{equation}
p(f,\Bpar,\Bperp,\chi) = p(f) p(\Bpar) p(\Bperp) p(\chi).
\end{equation}
Correlations between model parameters will then be obtained from the observed data through the likelihood.
We do not have any preference for any specific value or the filling factor and the field azimuth, so 
we set uniform flat priors in the interval $[0,1]$ and $[0,2\pi]$, respectively. As a 
consequence, $p(f)=\Pi(f-1/2)$ and $p(\chi)=(2\pi)^{-1}\Pi((2\pi)^{-1}(\chi-\pi))$, where
$\Pi(x)$ is the standard rectangular function. If one wants
to carry out the inference assuming $f=1$, it is enough to set $p(f)=\delta(f-1)$ and
the following formulae are still valid. Concerning the components of the
magnetic field, it is interesting to use priors that give low preference
to very large values. We know that values above a few thousand G are not realistic. Instead of
using a truncated flat prior, for computational purposes and inspired on physical
considerations, we use different functional
forms trying to be as non-informative as possible while maintaining 
physical constraints. For $\Bpar$ we propose a Gaussian distribution
with a large variance $\sigma_\parallel^2$, while for $\Bperp$ we propose
a Rayleigh distribution with large variance $\sigma_\perp^2$, a consequence of 
the assumption of Gaussianity for the two components of the magnetic field perpendicular to the line-of-sight (LOS).
If $\sigma_\parallel=\sigma_\perp$, we end up with an isotropic distribution
for the magnetic field vector.
The two parameters $\sigma_\parallel$ and $\sigma_\perp$ are known as hyperparameters because they parametrize 
the priors. The complete prior distribution is given by:
\begin{equation}
p(f,\Bpar,\Bperp,\chi) =  \Pi(f-1/2) \frac{1}{2\pi} \Pi\left(\frac{\chi-\pi}{2\pi} \right) \left[\frac{1}{\sqrt{2\pi}\sigma_\parallel}
\exp \left(-\frac{\Bpar^2}{2\sigma_\parallel^2}\right) \right]
\left[ \sigma_\perp^{-2} \Bperp \exp \left(-\frac{\Bperp^2}{2\sigma_\perp^2}\right) \right].
\label{eq:prior}
\end{equation}
We note that any other functional form for the prior can be used
if they are based on sensible physical intuition,
although some of the following integrals may require more numerical work 
for their evaluation. As a rule, if data contains sufficient information about the
model parameters, the results should be almost insensitive to the election of the prior as long as
the prior gives non-negligible probability to the regions of high likelihood. For this
reason, it is fundamental to compare the posterior with the chosen prior to know if data
has added information to the inference problem. If they are very similar, the results
depend on the specific election of the prior and should, in principle, be discarded.

\subsection{Posterior}
The full posterior distribution results from the application of Eq. (\ref{eq:bayes_theorem}):
\begin{eqnarray}
p(f,\Bpar,\Bperp,\chi|D) &=& \Pi(f-1/2) \Pi\left(\frac{\chi-\pi}{2\pi} \right) (2\pi)^{-(3N+3)/2}
\sigma_n^{-3N} \frac{1}{\sigma_\parallel \sigma_\perp^2}  \Bperp 
\exp \left\{ -\left[A_1 + \left(A_2 f^2+\frac{1}{2 \sigma_\parallel^2} \right) \Bpar^2 + \right. \right. \nonumber \\
&+& \left. \left. A_3 f^2 \Bperp^4 - 2 A_4 f \Bpar - \left[2 f (A_5 \coschi + A_6 \sinchi)- \frac{1}{2\sigma_\perp^2} \right] \Bperp^2 \right] \right\}.
\label{eq:posterior}
\end{eqnarray}
Since in the non-hierarchical model that we are dealing with in this section the normalizing
constants are just a scale, they can be dropped without effect. They will be however important in
the following section when marginalizing out the prior parameters.
The following rewriting is also convenient for
evaluating some of the marginal posterior for the components of the magnetic field
vector:
\begin{equation}
p(f,\Bpar,\Bperp,\chi|D) = \Pi(f-1/2) \Pi\left(\frac{\chi-\pi}{2\pi} \right) 
(2\pi)^{-(3N+3)/2} \sigma_n^{-3N} \frac{1}{\sigma_\parallel \sigma_\perp^2} \Bperp \exp \left[ -(B_1 -2 B_2 f + B_3 f^2) \right],
\end{equation}
where
\begin{equation}
B_1 = A_1 + \frac{\Bpar^2}{2 \sigma_\parallel^2}  + \frac{\Bperp^2}{2\sigma_\perp^2} , \qquad
B_2 = A_4 \Bpar + (A_5 \coschi + A_6 \sinchi) \Bperp^2, \qquad
B_3 = A_2 \Bpar^2 + A_3 \Bperp^4.
\end{equation}
The joint posterior distribution contains all information about the parameters of the weak-field 
approximation for a set of observed Stokes profiles. Different versions of the posterior can
be built if different priors are assigned.

\subsection{Maximum-a-posteriori and maximum-likelihood solutions}
The maximum a-posteriori (MAP) solution to the problem
(the one producing the largest posterior inside the prior volume) can be obtained
by finding the value of $(f,\Bpar,\Bperp,\chi)_\mathrm{MAP}$ that maximizes Eq. (\ref{eq:posterior}),
which is equivalent to minimizing $-\ln p(f,\Bpar,\Bperp,\chi|D)$. The MAP solution is, therefore, the
one minimizing:
\begin{eqnarray}
-\ln p(f,\Bpar,\Bperp,\chi|D) &=& -\ln \Bperp + A_1 + \left(A_2 f^2+\frac{1}{2 \sigma_\parallel^2} \right) \Bpar^2 + A_3 f^2 \Bperp^4 - 2 A_4 f \Bpar \nonumber \\
&-&\left[2 f (A_5 \coschi + A_6 \sinchi) - \frac{1}{2\sigma_\perp^2} \right] \Bperp^2 + \mathrm{cte}.
\end{eqnarray}
The solution can be found by solving the following non-linear system of equations:
\begin{eqnarray}
-A_4 \Bpar - A_5 \Bperp^2 + A_2 \Bpar^2 f + A_3 \Bperp^4 f &=& 0 \nonumber \\
-A_4 f + \Bpar \left( \frac{1}{2 \sigma_\parallel^2} + A_2 f^2 \right) &=& 0 \nonumber \\
-\frac{1}{\Bperp} + 4A_3 \Bperp^3 f^2 - 2\Bperp \left[2(A_5 \coschi + A_6 \sinchi) f - \frac{1}{2\sigma_\perp^2} \right] &=& 0. \nonumber \\
4f\Bperp^2 A_5 \sinchi - 4f\Bperp^2 A_6 \coschi&=&0.
\label{eq:log_posterior}
\end{eqnarray}
The last equation is the standard estimation of the field azimuth which, provided that $\Bperp \neq 0$ and
$f \neq 0$, results in:
\begin{equation}
\tan 2 \chi = \frac{A_6}{A_5},
\end{equation}
which can then be understood as the maximum-a-posteriori estimation of the field azimuth.
Additionally, particularizing to the case of a longitudinal magnetograph (i.e., if we do not measure
linear polarization), we find the obvious solution $f \Bpar = A_4/A_2$, that has been
already found by \cite{marian_luis09}.
Note also that the standard maximum-likelihood solution (also known as least-squares solution) is obtained following the same
scheme after setting $\sigma_\parallel \to \infty$ and $\sigma_\perp \to \infty$ and dropping the term $\ln \Bperp$ from Eq. (\ref{eq:log_posterior}).

\subsection{Marginal posteriors}
To fully take into account the presence of degeneracies among
the model parameters in the Bayesian approach, we have to compute the
marginal posterior distributions for every parameters individually. The 
marginal posterior is obtained by integrating out all parameters but the
one of interest. For instance, the posterior for $\Bpar$ is
computed as:
\begin{equation}
p(\Bpar|D) = \int_0^{2\pi} \mathrm{d}\chi \int_0^1 \mathrm{d}f \int_0^\infty \mathrm{d}\Bperp \, p(f,\Bpar,\Bperp,\chi|D).
\end{equation}
This integration will introduce into the probability
distribution of a given parameter all possible values of the rest of
model parameters weigthed by their associated probabilities. The
advantage of the weak-field model is that some of these
integrals can be obtained analytically in closed form. 
Due to the 
complexity of the integrals, we have been unable to find closed analytical
expressions for the one-dimensional marginal distributions. However, we present
in App. \ref{sec:appendix_posteriors} the expressions for the 2-dimensional marginal posteriori
$p(f,\chi|D)$ and for the 3-dimensional marginal posteriors $p(f,\Bpar,\Bperp|D)$, 
$p(\Bpar,\Bperp,\chi|D)$, $p(f,\Bperp,\chi|D)$ and $p(f,\Bpar,\chi|D)$.

The shape of the marginal posterior is fundamental to decide whether a parameter
is constrained or not by the observations. Although the solution to the
problem would be to give the full marginal posteriors, it is sometimes
interesting for presentation purposes to give a summary of the distribution.
If there is a clear peak with
tails falling to zero (like a Gaussian, but can often be asymmetric), the value
of the parameter at the peak (the most probable) is the so-called marginal 
maximum-a-posteriori (MMAP) solution or mode. A confidence interval can be
put by just integrating the posterior until a given fraction of the total
area is obtained. Another possibility to summarize the distribution is to
give the median value (where the cumulative probability distribution equals
$1/2$), together with a confidence region. Finally, it is possible to
compute moments of the quantity $x$ using the standard definition:
\begin{equation}
\langle x^n \rangle = \frac{\int x^n p(x|D) \, \mathrm{d}x }{\int p(x|D) \, \mathrm{d}x },
\end{equation}
and the confidence interval is obtained like in the MMAP.

\subsection{Marginal posterior of derived quantities}
Since all the information about the model parameters is contained in the posterior
probability distribution function, information about any derived quantity can be 
computed from it using the machinery presented in App. \ref{sec:appendix}.

\subsubsection{Magnetic flux density}
One of the quantities of interest is the magnetic flux density,
defined as $F_\parallel = f \Bpar$. Following the change of variables formulae, we can
write the probability distribution function of the magnetic flux density as:
\begin{eqnarray}
p(F_\parallel \geq 0|D) &=& \int_{F_\parallel}^\infty p \left( f=\frac{F_\parallel}{\Bpar}, \Bpar \Bigg| D \right) \frac{1}{|\Bpar|} \mathrm{d}\Bpar \nonumber \\
p(F_\parallel < 0|D) &=& \int_{-\infty}^{-F_\parallel} p \left( f=\frac{F_\parallel}{\Bpar}, \Bpar \Bigg| D \right) \frac{1}{|\Bpar|} \mathrm{d}\Bpar
\label{eq:posterior_fparallel}
\end{eqnarray}
where the joint distribution $p(f,\Bpar|D)$ is obtained by marginalizing $\Bperp$ and $\chi$ from
the full posterior. Since 
$0 \leq f \leq 1$, $|\Bpar| \geq |F_\parallel|$ has to be fulfilled, which produces
the separation of the integral in two different cases depending on the sign of $F_\parallel$.
The integration on $\Bpar$ has to be carried out numerically since it is not possible to obtain
a closed expression. 
The previous approach is unnecessarily complex for such a simple quantity like the magnetic flux 
density. In this special case, it can be computed following a different path, noting that the posterior for $F_\parallel$ 
can also be obtained directly because:
\begin{equation}
V(\lambda) = \alpha F_\parallel \didlnoi,
\end{equation}
and neither $Q(\lambda)$ nor $U(\lambda)$ do depend on $F_\parallel$. In such a case, it is
easy to write a posterior for this variable:
\begin{equation}
p(F_\parallel|D) = p(F_\parallel) (2\pi)^{-N/2} \sigma_n^{-N} \exp \left\{ -\frac{1}{2\sigma_n^2} \left[
\sum_i V_i^2 + F_\parallel^2 \sum_i \alpha^2 \left(\didl\right)^2 -2 F_\parallel \sum_i \alpha V_i \didl \right] \right\}.
\label{eq:posterior_fparallel_nonhierarchical}
\end{equation}
The only remaining ingredient is the prior distribution $p(F_\parallel)$.
Since $F_\parallel=f \Bpar$ and the priors for $f$ and $\Bpar$ have been discussed
in Sec. \ref{sec:priors}, the prior for $F_\parallel$ can be obtained following App. \ref{sec:appendix},
thus resulting in:
\begin{equation}
p(F_\parallel) = \frac{1}{2\sqrt{2\pi}\sigma_\parallel} E_1 \left(\frac{F_\parallel^2}{2\sigma_\parallel^2} \right),
\end{equation}
where $E_1(x)$ is the first exponential integral \citep[e.g.,][]{abramowitz72}. When data is sufficiently informative so that
the prior distribution becomes unimportant, the posterior for the magnetic flux density is
Gaussian with mean and variance given by \citep{marian_luis09}:
\begin{eqnarray}
\mu(F_\parallel) = \frac{\sum_i \alpha V_i \didl}{\sum_i \alpha^2 \left( \didl \right)^2} \nonumber \\
\sigma^2(F_\parallel) = \frac{\sigma^2}{\sum_i \alpha^2 \left( \didl \right)^2}.
\label{eq:mean_variance_fparallel}
\end{eqnarray}

\subsubsection{Magnetic field inclination}
A similar approach can be followed to compute the marginal posterior for the field inclination
from the joint distribution of $\Bpar$ and $\Bperp$.
To this end, we apply again the rules of App. \ref{sec:appendix} to the change of
variables $\theta = \arctan(\Bperp / \Bpar)$ and obtain:
\begin{eqnarray}
p(\theta \geq 0|D) &=& |1+\tan^2 \theta| \int_{0}^{\infty} \mathrm{d}\Bpar p(\Bpar,\Bperp=\Bpar \tan \theta|D) |\Bpar| \nonumber \\
p(\theta < 0|D) &=& |1+\tan^2 \theta| \int_{-\infty}^{0} \mathrm{d}\Bpar p(\Bpar,\Bperp=\Bpar \tan \theta|D) |\Bpar|
\label{eq:posterior_inclination}
\end{eqnarray}
where the joint distribution is computed by marginalizing $f$ and $\chi$ from the posterior
and the limits of integration have been adapted to fulfill $\Bperp = \Bpar \tan(\theta) \geq 0$. 

Using the same procedure, the prior distribution for the inclination can be obtained by plugging 
Eq. (\ref{eq:prior}) into Eq. (\ref{eq:posterior_inclination}). The
results is, after some algebra, that $p(\theta)=|\sin \theta|$, an obvious result because the prior we
are using assumes that the field is isotropically distributed.

\subsubsection{Magnetic field strength and energy}
Exactly the same procedure can be applied to obtain the posterior for the magnetic field
strength, $B=\sqrt{\Bpar^2+\Bperp^2}$, and to the magnetic field energy, 
$E_\mathrm{mag}=B^2/8\pi$, obtaining:
\begin{eqnarray}
p(B|D) &=& \int_{-B}^{B} \mathrm{d}\Bpar \left| \frac{B}{\sqrt{B^2-\Bpar^2}} \right| p\left(\Bpar,\sqrt{B^2-\Bpar^2} \Big|D \right) \nonumber \\
p(E_\mathrm{mag}|D) &=& \int_{-\sqrt{8\pi E_\mathrm{mag}}}^{\sqrt{8\pi E_\mathrm{mag}}} \mathrm{d}\Bpar \left| \frac{1}{\sqrt{8\pi E_\mathrm{mag}-\Bpar^2}} \right|
p\left(\Bpar,\sqrt{8\pi E_\mathrm{mag}-\Bpar^2} \Big|D\right),
\end{eqnarray}
where the joint probability distribution is computed by marginalizing $f$ and $\chi$ from the posterior.

\section{Hierarchical Bayesian approach}
In principle, the values of $\sigma_\parallel$ and $\sigma_\perp$ in the previous formalism
are understood to be fixed and given a-priori. One of the disadvantages of this approach is that, if there is
not sufficient information about the model parameters encoded in the observations, the
results will surely be sensitive to the election of these numbers (see \S\ref{sec:results} for an example). 
However, we can take advantage of the fact that, in the Bayesian formalism, every unknown can be 
considered a random variable. We can put an appropriate prior on them and
extend the formalism in a hierarchical way and let data determine their
values. Since the priors are put over hyperparameters (parameters of the prior distributions), they are known
as hyperpriors. In principle, nothing avoids us to make hyperpriors depend on additional 
hyperparameters over which appropriate priors are assigned and continue the
hierarchical structure \citep[see, e.g.,][for more details on hierarchical models]{gregory05,gelman_bayesian03}.
The immediate effect of the hierarchical approach is that results are much less dependent on the
specific choice of hyperparameters and we make the problem essentially free of parameters. In some 
sense, this happens because we allow priors adapt to data.
It can be demonstrated that some standard regularization schemes applied to least-squares techniques
are indeed particular cases of hierarchical models. 

\subsection{Known noise variance}
Using trivial probability calculus, the full posterior can be expressed by marginalizing out 
the hyperparameters, thus:
\begin{equation}
p(f,\Bpar,\Bperp,\chi|D) = \int \mathrm{d}\sigma_\parallel \mathrm{d}\sigma_\perp p(D|f,\Bpar,\Bperp,\chi) 
p(f,\Bpar,\Bperp,\chi|\sigma_\parallel,\sigma_\perp) p(\sigma_\parallel,\sigma_\perp),
\label{eq:prior_marginal}
\end{equation}
where we have used the fact that the likelihood does not depend on the parameters $\sigma_\parallel$
and $\sigma_\perp$. Note that this marginalization takes into account \emph{all} possible
values of the hyperparameters weighted by their probability. The 
term $p(f,\Bpar,\Bperp|\sigma_\parallel,\sigma_\perp)$ is the prior defined
in Eq. (\ref{eq:prior}) and the new distribution $p(\sigma_\parallel,\sigma_\perp)$ gives
prior information about the hyperparameters. Since they behave as scale parameters (they can
take values spanning several decades), a good option is to use a uniform 
prior on logarithmic scale, so that:
\begin{equation}
p(\sigma_\parallel,\sigma_\perp) = \frac{1}{\sigma_\parallel} \frac{1}{\sigma_\perp}.
\end{equation}
This prior, also known as Jeffreys' prior \citep{jeffreys61,mackay03}, has
the remarkable property of being the only non-informative prior for scale-invariant quantities.
This complicates the analytical solution of the problem requiring more
purely numerical integrations but also introduces additional stabilization. 
Although the Jeffreys' prior is improper (its integral is not finite), the
integral of the priors over $\sigma_\parallel$ and $\sigma_\perp$ are
finite but also improper. The integral of Eq. (\ref{eq:prior_marginal}) can be finally 
carried out for all possible values of $\sigma_\parallel$ and $\sigma_\perp$ to give:
\begin{equation}
p(f,\Bpar,\Bperp,\chi|D) = p(D|f,\Bpar,\Bperp) \Pi(f-1/2) \frac{1}{4\pi} \Pi\left(\frac{\chi-\pi}{2\pi} \right) \frac{1}{|\Bpar| \Bperp},
\label{eq:posterior_hierarchical1}
\end{equation}
where the likelihood $p(D|f,\Bpar,\Bperp)$ is given by Eq. (\ref{eq:likelihood}). The
marginalization process demonstrates that the Jeffreys' prior is a natural election
for $\Bpar$ and $\Bperp$. In other words, we could have started our calculation
by selecting a Jeffreys' prior in Eq. (\ref{eq:prior})
without considering the hierarchical approach.

In case no information is present in the data (flat likelihood), the prior for $F_\parallel$ can be
obtained by substitution of the previous expression in Eq. (\ref{eq:posterior_fparallel}):
\begin{equation}
p(F_\parallel) = \frac{1}{|F_\parallel|},
\label{eq:prior_fparallel_hierarchical1}
\end{equation}
so the posterior is given by Eq. (\ref{eq:posterior_fparallel_nonhierarchical}).

It is also possible to introduce
new free hyperparameters $\sigma_\mathrm{min}$ and $\sigma_\mathrm{max}$ that
define a lower and upper estimation of the width of the prior distributions, respectively. 
Although they constitute a new set of free parameters,
the hierarchical character offers the advantage that results
are much less sensitive to their exact values. Because we do not 
expect magnetic fields in the solar atmosphere larger than $\sim 4000$ G or we do
not expect to detect fields below $\sim 0.1$ G, it is sensible 
to choose $\sigma_\mathrm{min}=0.1$ G and $\sigma_\mathrm{max}=4000$ G. These values
can be adapted for specific cases (inverting Stokes profiles observed in
active regions or the quiet Sun) but the results are really insensitive to 
the specific values, as shown in \S\ref{sec:results}. The following prior is the
resulting one after marginalizing out $\sigma_\parallel$ and $\sigma_\perp$:
\begin{eqnarray}
p(f,\Bpar,\Bperp,\chi) &=& \Pi(f-1/2) \frac{1}{4\pi} \Pi\left(\frac{\chi-\pi}{2\pi} \right) 
\frac{1}{\Bpar \Bperp} \left[ 
\mathrm{erf} \left( \frac{\Bpar}{\sqrt{2}\sigma_\mathrm{min}} \right) -
\mathrm{erf} \left( \frac{\Bpar}{\sqrt{2}\sigma_\mathrm{max}} \right)\right] 
\nonumber \\
&\times& \left[ 
\exp \left( -\frac{\Bperp^2}{2\sigma_\mathrm{max}^2} \right) -
\exp \left( -\frac{\Bperp^2}{2\sigma_\mathrm{min}^2} \right)\right],
\label{eq:prior_hierarchical_known_noise}
\end{eqnarray}
which converges to Eq. (\ref{eq:posterior_hierarchical1}) when $\sigma_\mathrm{min}=0$ and $\sigma_\mathrm{max} \to \infty$.
The function $\mathrm{erf}(x)$ is the error function \citep[e.g.,][]{abramowitz72}.
This yields the following general expression for the posterior that we use when the noise variance is known:
\begin{eqnarray}
p(f,\Bpar,\Bperp,\chi|D) &=& \frac{1}{2} \Pi(f-1/2) \Pi\left(\frac{\chi-\pi}{2\pi} \right) (2\pi)^{-(3N+1)/2} \sigma_n^{-3N} \nonumber \\
&\times& \exp\left[-\left(A_1 + A_2 f^2 \Bpar^2 + A_3 f^2 \Bperp^4 - 2 A_4 f \Bpar - 2 (A_5 \coschi + A_6 \sinchi) f \Bperp^2\right)\right] \nonumber \\
&\times& \frac{1}{\Bpar \Bperp} \left[ 
\mathrm{erf} \left( \frac{\Bpar}{\sqrt{2}\sigma_\mathrm{min}} \right) -
\mathrm{erf} \left( \frac{\Bpar}{\sqrt{2}\sigma_\mathrm{max}} \right)\right] 
\left[ 
\exp \left( -\frac{\Bperp^2}{2\sigma_\mathrm{max}^2} \right) -
\exp \left( -\frac{\Bperp^2}{2\sigma_\mathrm{min}^2} \right)\right].
\label{eq:posterior_hierarchical_known_noise}
\end{eqnarray}
Again, we have been unable to obtain the 1-dimensional marginal distributions, but
App. \ref{sec:appendix_posteriors} presents the 3-dimensional marginal posteriors $p(f,\Bpar,\Bperp|D)$
and $p(\Bpar,\Bperp,\chi|D)$.
In this generalized case, under non-informative data, the prior for $F_\parallel$ results in:
\begin{equation}
p(F_\parallel) = \frac{1}{F_\parallel} \left[ 
\mathrm{erf} \left( \frac{F_\parallel}{\sqrt{2}\sigma_\mathrm{min}} \right)-
\mathrm{erf} \left( \frac{F_\parallel}{\sqrt{2}\sigma_\mathrm{max}} \right) \right]
+\frac{1}{\sqrt{2 \pi} \sigma_\mathrm{min}} E_1 \left(\frac{F_\parallel^2}{2\sigma_\mathrm{min}^2}\right)-
\frac{1}{\sqrt{2 \pi} \sigma_\mathrm{max}} E_1 \left(\frac{F_\parallel^2}{2\sigma_\mathrm{max}^2} \right)
\label{eq:prior_fparallel_knownnoise}
\end{equation}
which reduces to Eq. (\ref{eq:prior_fparallel_hierarchical1}) when $\sigma_\mathrm{min}=0$ and
$\sigma_\mathrm{max} \to \infty$. Substitution in
Eq. (\ref{eq:posterior_fparallel_nonhierarchical}) gives the posterior for $F_\parallel$.

\subsection{Unknown noise variance}
Another step forward in the hierarchical model can be given if the noise standard deviation 
is not known with precision. Sometimes the peculiarities of the observation
induce that estimating the observational uncertainty is not an easy task. It is
typically estimated assuming ergodicity and calculating the variance of the signal
on a continuum window, where the polarimetric signal is assumed to be zero. 
This might be flawed if, for instance, the polarimetric signal is very broad
and the assumption of zero signal in the continuum window is not correct.
Likewise, when there is a large velocity field producing a large Doppler
shift, neighboring spectral lines can enter into the continuum window
and give wrong estimation of the observational uncertainty. This is especially
relevant when when observing with filter-polarimeters, where the information
of the continuum is usually contained in one or two spectral samples.
In such a case, it is possible to consider $\sigma_n$ as a random variable which
is marginalized at the end. Although we assume that the variance of the 
observational uncertainty is unknown, we postulate that the noise follows a
Gaussian distribution. This distribution, in the absence of detailed information about the
correct distribution apart from the finite variance, is the one suggested by the
principle of maximum entropy. Therefore:
\begin{equation}
p(f,\Bpar,\Bperp,\chi|D) = \int \mathrm{d}\sigma_\parallel \mathrm{d}\sigma_\perp \mathrm{d}\sigma_n \, p(D|f,\Bpar,\Bperp,\chi,\sigma_n)
p(f,\Bpar,\Bperp,\chi|\sigma_\parallel,\sigma_\perp) p(\sigma_\parallel,\sigma_\perp) p(\sigma_n).
\label{eq:prior_marginal_noise}
\end{equation}
Using a Jeffreys' prior over $\sigma_n$, so that $p(\sigma_n)=\sigma_n^{-1}$, integrating over the full
domain in $\sigma_n$, $\sigma_\parallel$ and $\sigma_\perp$, we obtain
the following posterior:
\begin{equation}
p(f,\Bpar,\Bperp,\chi|D) = \frac{1}{2} \Pi(f-1/2) \Pi\left(\frac{\chi-\pi}{2\pi} \right) \frac{1}{|\Bpar| \Bperp} (2\pi)^{-(3N+1)/2}
C^{-3N/2},
\label{eq:posterior_hierarchical_unknown_noise}
\end{equation}
where 
\begin{equation}
C= A_1' + A_2' f^2 \Bpar^2 + A_3' f^2 \Bperp^4 - 2 A_4' f \Bpar - 2 (A_5' \coschi + A_6' \sinchi) f \Bperp^2,
\end{equation}
the $A'_i$ are obtained from the $A_i$ defined in Eq. (\ref{eq:A_values}) but dropping the terms $(2\sigma_n^2)^{-1}$
and $N$ is the number of wavelength points in the profiles.
If we limit the integration to the range $[\sigma_{n_\mathrm{min}},\sigma_{n_\mathrm{max}}]$ because
we know that noise cannot be too small or too large, but keep the integration on the full space
for $\sigma_\parallel$ and $\sigma_\perp$ we end up with the following posterior:
\begin{eqnarray}
p(f,\Bpar,\Bperp,\chi|D) &=& \frac{1}{2} \Pi(f-1/2) \Pi\left(\frac{\chi-\pi}{2\pi} \right) \frac{1}{|\Bpar| \Bperp} (2\pi)^{-(3N+1)/2} \nonumber \\
&\times& C^{-3N/2} \left[ \Gamma \left(\frac{3N}{2},\frac{C}{2\sigma_{n_\mathrm{max}}^2} \right)
-\Gamma \left(\frac{3N}{2},\frac{C}{2\sigma_{n_\mathrm{min}}^2} \right) \right],
\label{eq:posterior_hierarchical_unknown_noise2}
\end{eqnarray}
which reduces to Eq. (\ref{eq:posterior_hierarchical_unknown_noise}) when $\sigma_{n_\mathrm{min}} \to 0$ and $\sigma_{n_\mathrm{max}} \to \infty$
due to the properties of the incomplete Gamma function $\Gamma(a,x)$ \citep[e.g.,][]{abramowitz72}.
It is also possible to limit the integration volume in $\sigma_n$, $\sigma_\parallel$ and $\sigma_\perp$,
arriving at the most general hierarchical model we consider that can be applied when the noise variance is unknown:
\begin{eqnarray}
p(f,\Bpar,\Bperp,\chi|D) &=& \frac{1}{2} \Pi(f-1/2) \Pi\left(\frac{\chi-\pi}{2\pi} \right) (2\pi)^{-(3N+1)/2} C^{-3N/2} 
\left[ \Gamma \left(\frac{3N}{2},\frac{C}{2\sigma_{n_\mathrm{max}}^2} \right)
-\Gamma \left(\frac{3N}{2},\frac{C}{2\sigma_{n_\mathrm{min}}^2} \right) \right] \nonumber \\
&\times&\frac{1}{\Bpar \Bperp} \left[ 
\mathrm{erf} \left( \frac{\Bpar}{\sqrt{2}\sigma_\mathrm{min}} \right) -
\mathrm{erf} \left( \frac{\Bpar}{\sqrt{2}\sigma_\mathrm{max}} \right)\right] 
\left[ 
\exp \left( -\frac{\Bperp^2}{2\sigma_\mathrm{max}^2} \right) -
\exp \left( -\frac{\Bperp^2}{2\sigma_\mathrm{min}^2} \right)\right].
\label{eq:posterior_hierarchical_unknown_noise_final}
\end{eqnarray}

Concerning the posterior for the magnetic flux density, marginalizing Eq. (\ref{eq:posterior_fparallel_nonhierarchical}) 
over $\sigma_n$ with the Jeffreys' prior, we find:
\begin{equation}
p(F_\parallel|D) = p(F_\parallel) (2\pi)^{-3N/2} D^{-N/2} \left[ \Gamma \left(\frac{N}{2},\frac{D}{2\sigma_{n_\mathrm{max}}^2} \right)
-\Gamma \left(\frac{N}{2},\frac{D}{2\sigma_{n_\mathrm{min}}^2} \right) \right],
\end{equation}
where
\begin{equation}
D = \sum_i V_i^2 + F_\parallel^2 \sum_i \alpha^2 \left(\didl\right)^2 -2 F_\parallel \sum_i \alpha V_i \didl.
\end{equation}
The prior distribution $p(F_\parallel)$ can be that of Eq. (\ref{eq:prior_fparallel_hierarchical1}) or
the most general one of Eq. (\ref{eq:prior_fparallel_knownnoise}).

\section{Illustrative examples}
\label{sec:results}
\subsection{Observed profile}
\begin{figure*}
\centering
\includegraphics[width=0.8\textwidth]{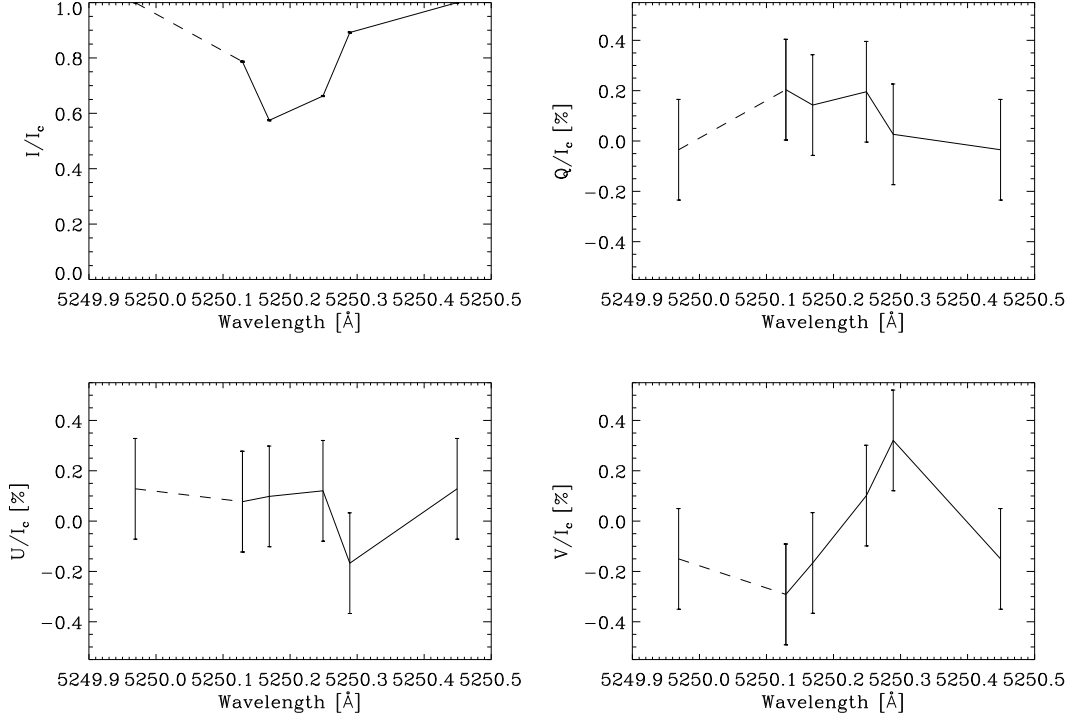}
\caption{Stokes profiles observed with IMaX and used in the examples. The profiles consist
of 5 wavelength points that have been obtained with a Fabry-Perot filter-polarimeter. For
a better visualization, we have repeated the continuum point at 5250.45 \AA\ symmetrically
on the red wing, and connected it with dashed lines. We have also marked the uncertainty
(estimated standard deviation of the noise) in each observed point with error bars.
\label{fig:imax_profiles}}
\end{figure*}

In order to illustrate the behavior of the non-hierarchical and hierarchical models
that we have developed, let us consider extracting information under the weak-field
approximation from the wavelength variation of a selected Stokes profile observed with IMaX 
\citep{imax04,imax11} onboard Sunrise \citep{sunrise10} in a quiet region of the solar atmosphere. 
The observed profile is shown
in Fig. \ref{fig:imax_profiles}. The reason for choosing this instrument is to
show the power of Bayesian methods to deal with cases in which there is reduced
information in the observables. The number of sampled points is just 5 which makes 
the typical shape of the polarization profiles hardly indistinguishable. In order
to apply the previous formulation we need to have an estimation of the first
and second derivative of the Stokes $I$ profile. The poor spectral sampling makes
this operation delicate. However, we have decided to apply a numerical derivative 
based on a standard 3-point Lagrange interpolation of the profile. This could have
been improved by, for instance, fitting the intensity profile to a Gaussian and
using this fitted profile to carry out the derivative. We have verified that
differences are negligible with both approaches.

\begin{figure*}
\centering
\includegraphics*[viewport=12 10 547 372,width=0.31\textwidth]{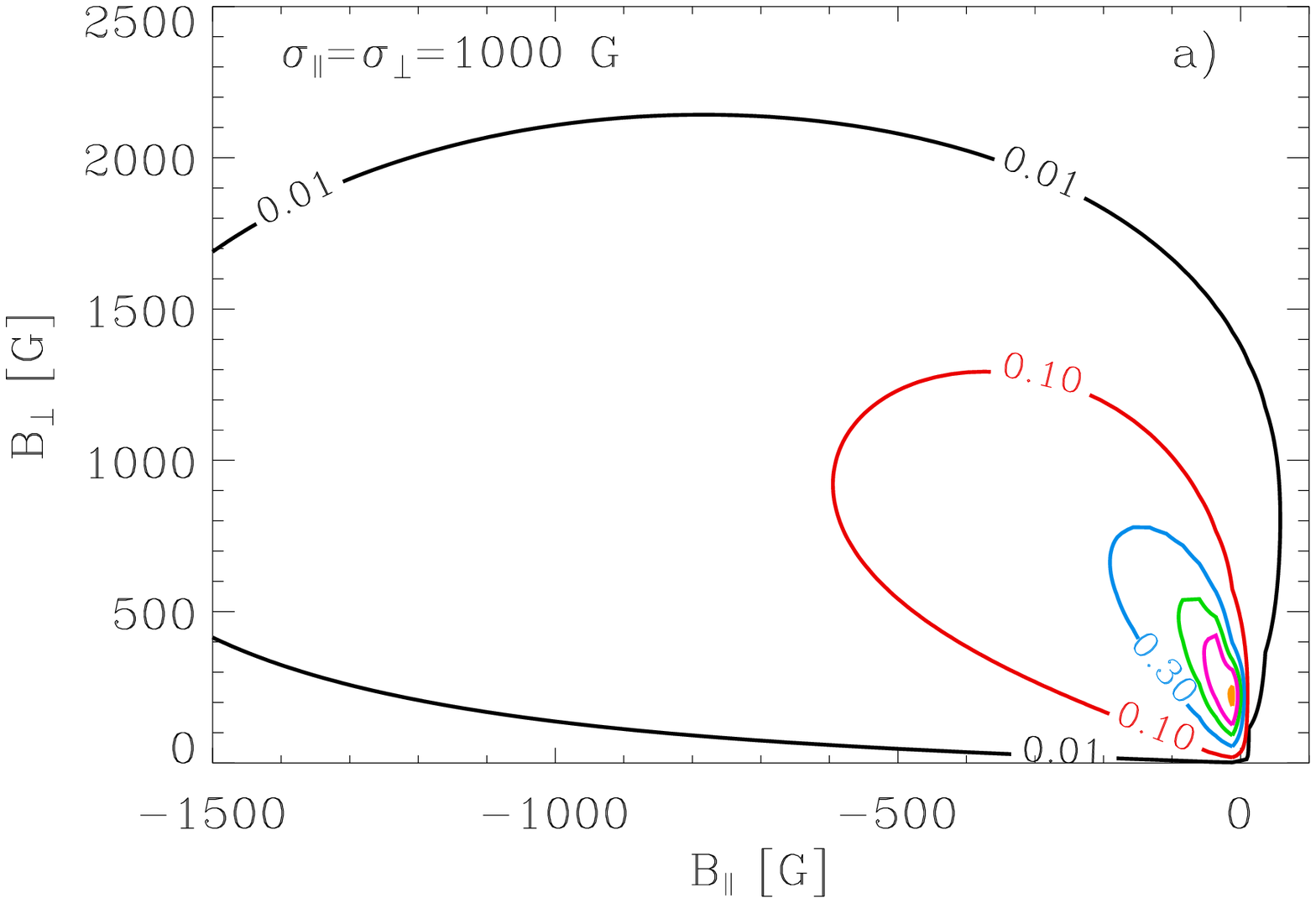}
\includegraphics*[viewport=12 10 547 372,width=0.31\textwidth]{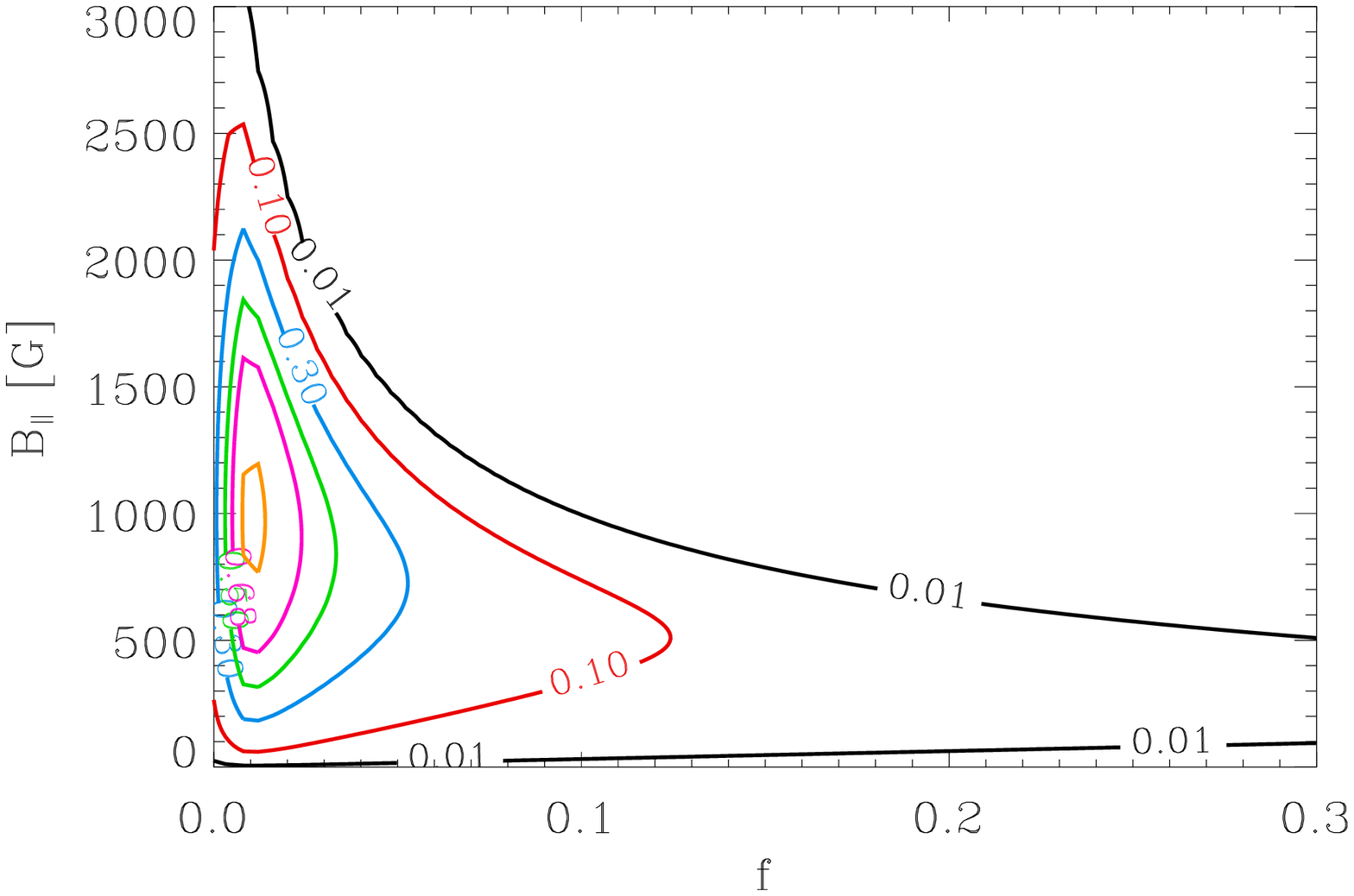}
\includegraphics*[viewport=12 10 547 372,width=0.31\textwidth]{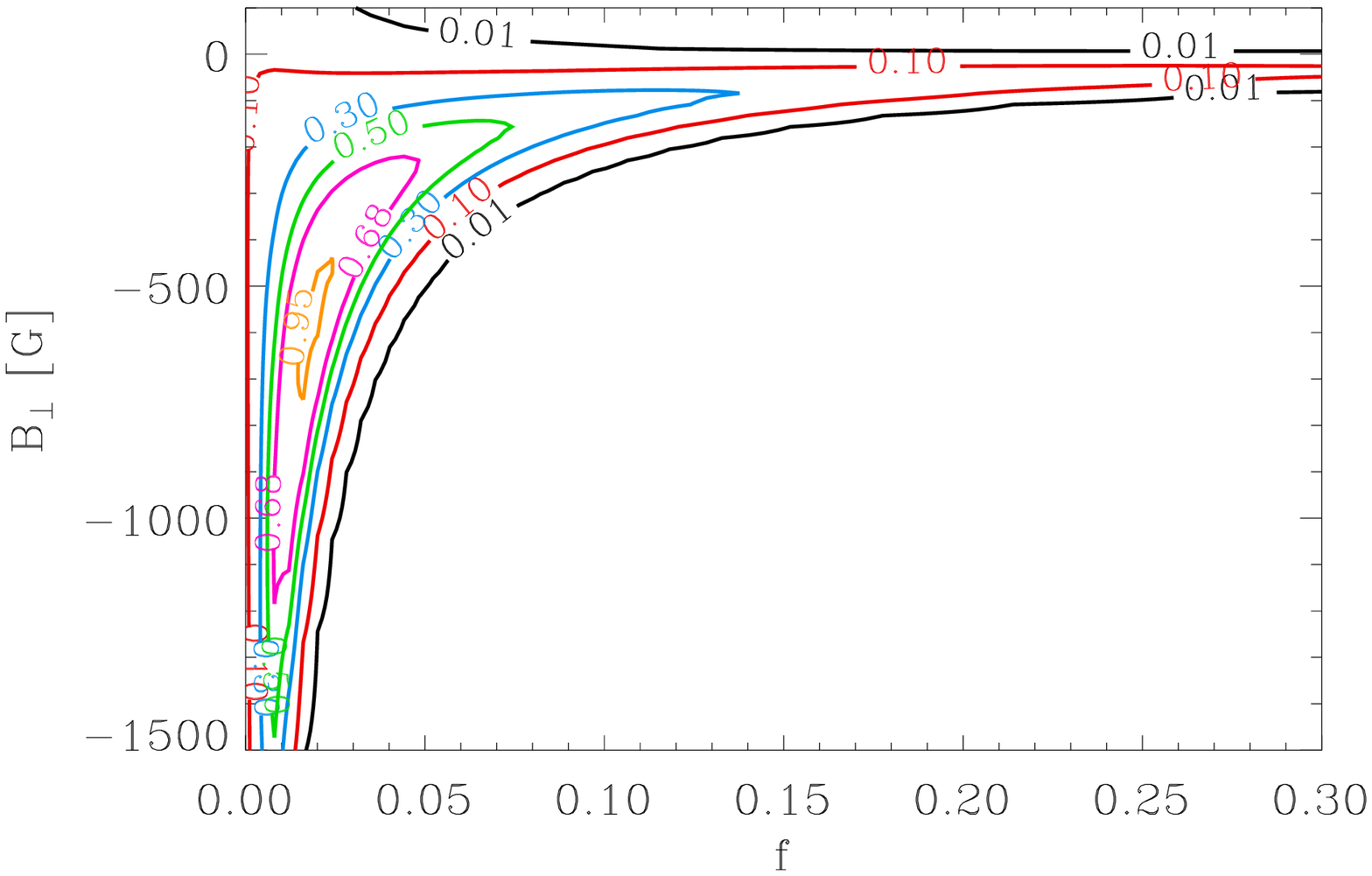}
\includegraphics*[viewport=12 10 547 372,width=0.31\textwidth]{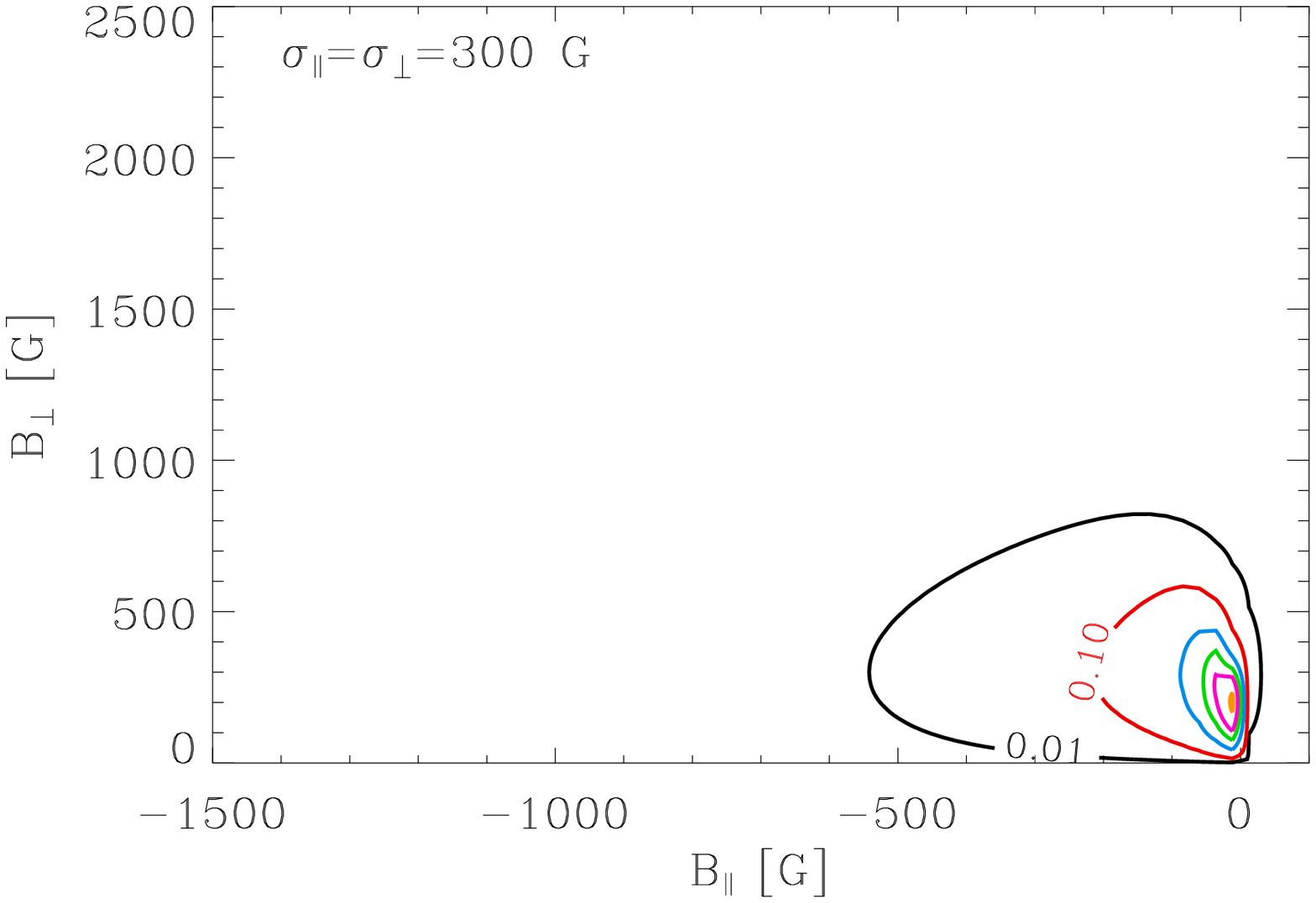}
\includegraphics*[viewport=12 10 547 372,width=0.31\textwidth]{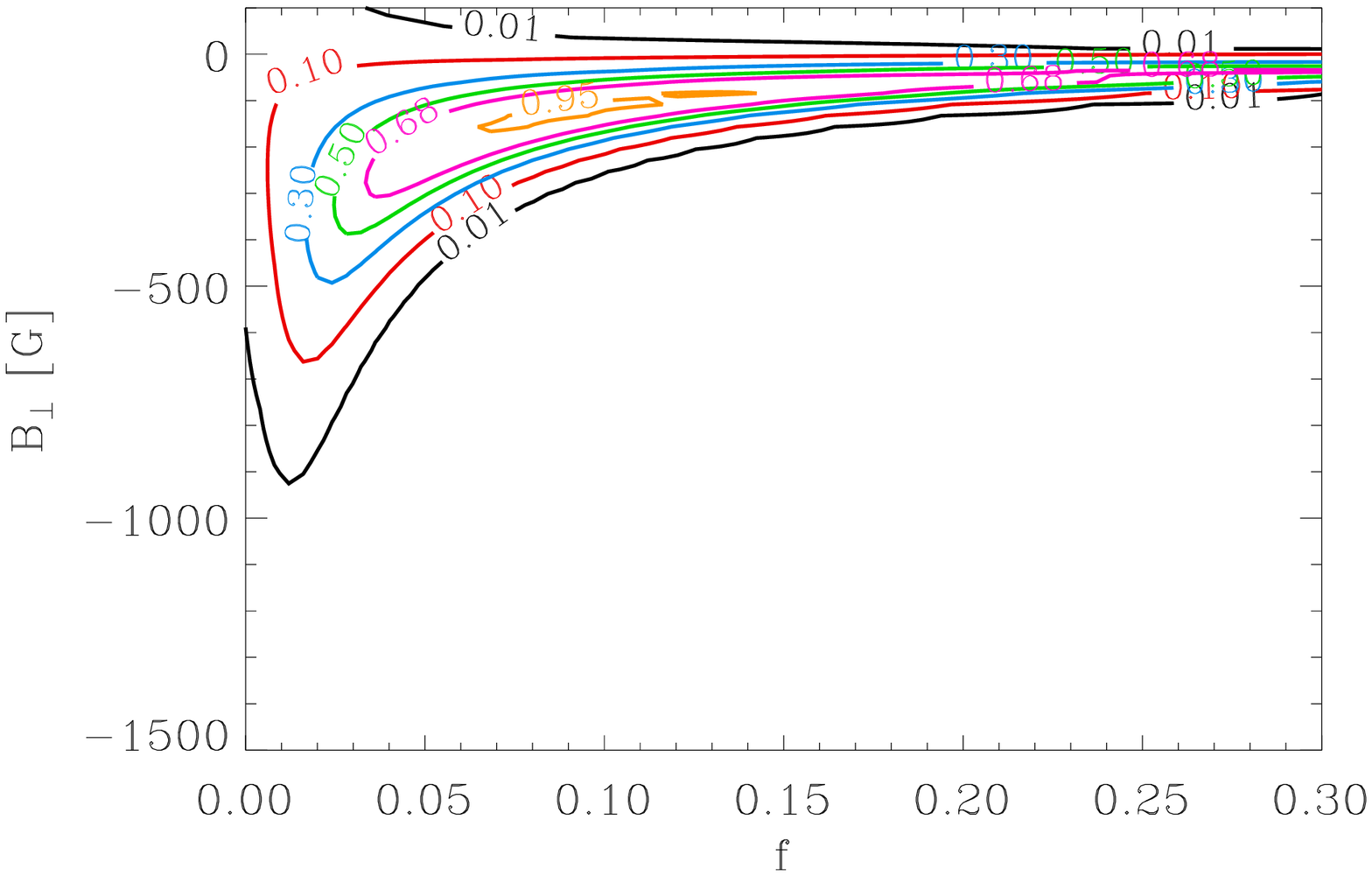}
\includegraphics*[viewport=12 10 547 372,width=0.31\textwidth]{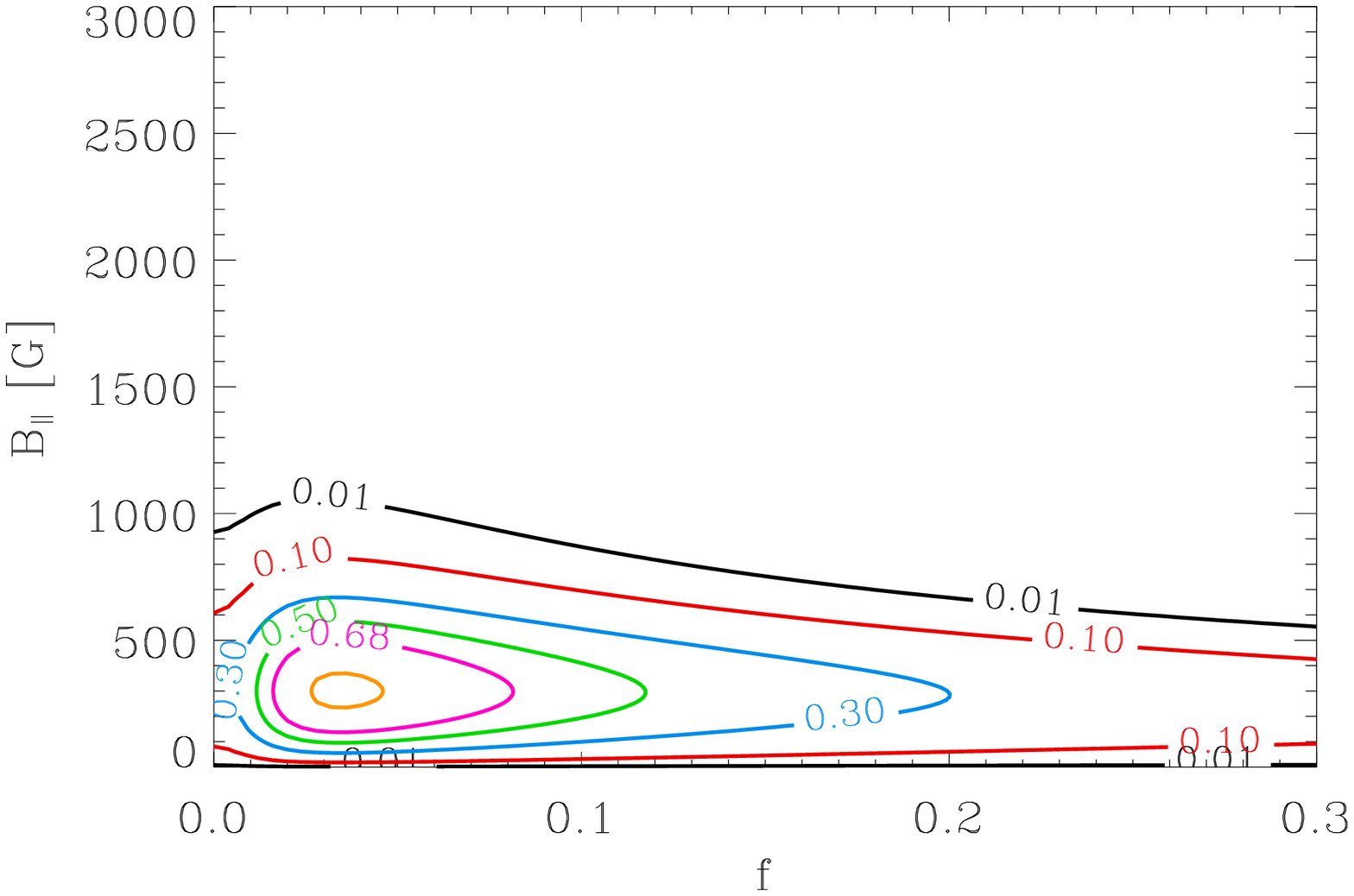}
\includegraphics*[viewport=12 10 547 372,width=0.31\textwidth]{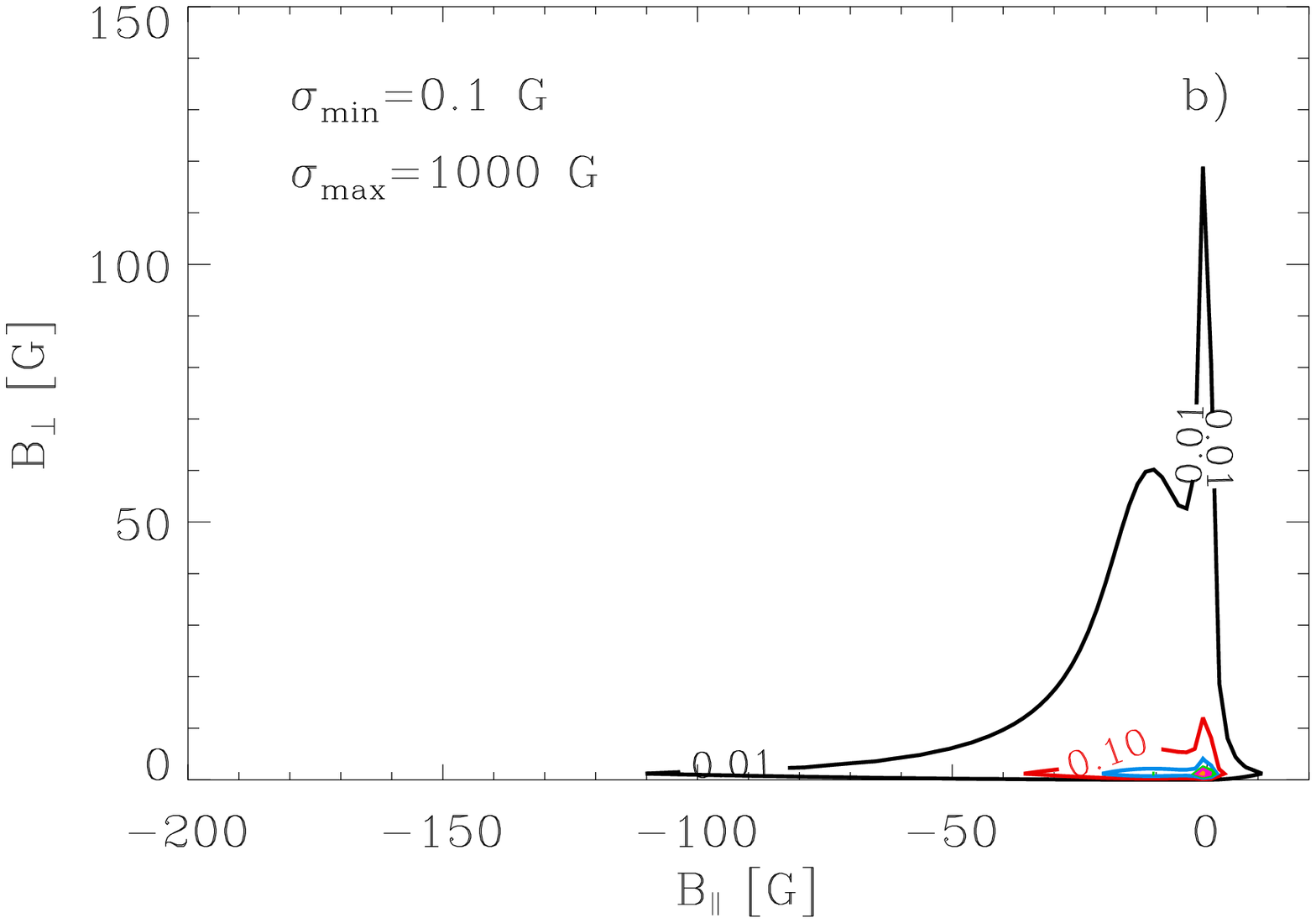}
\includegraphics*[viewport=12 10 547 372,width=0.31\textwidth]{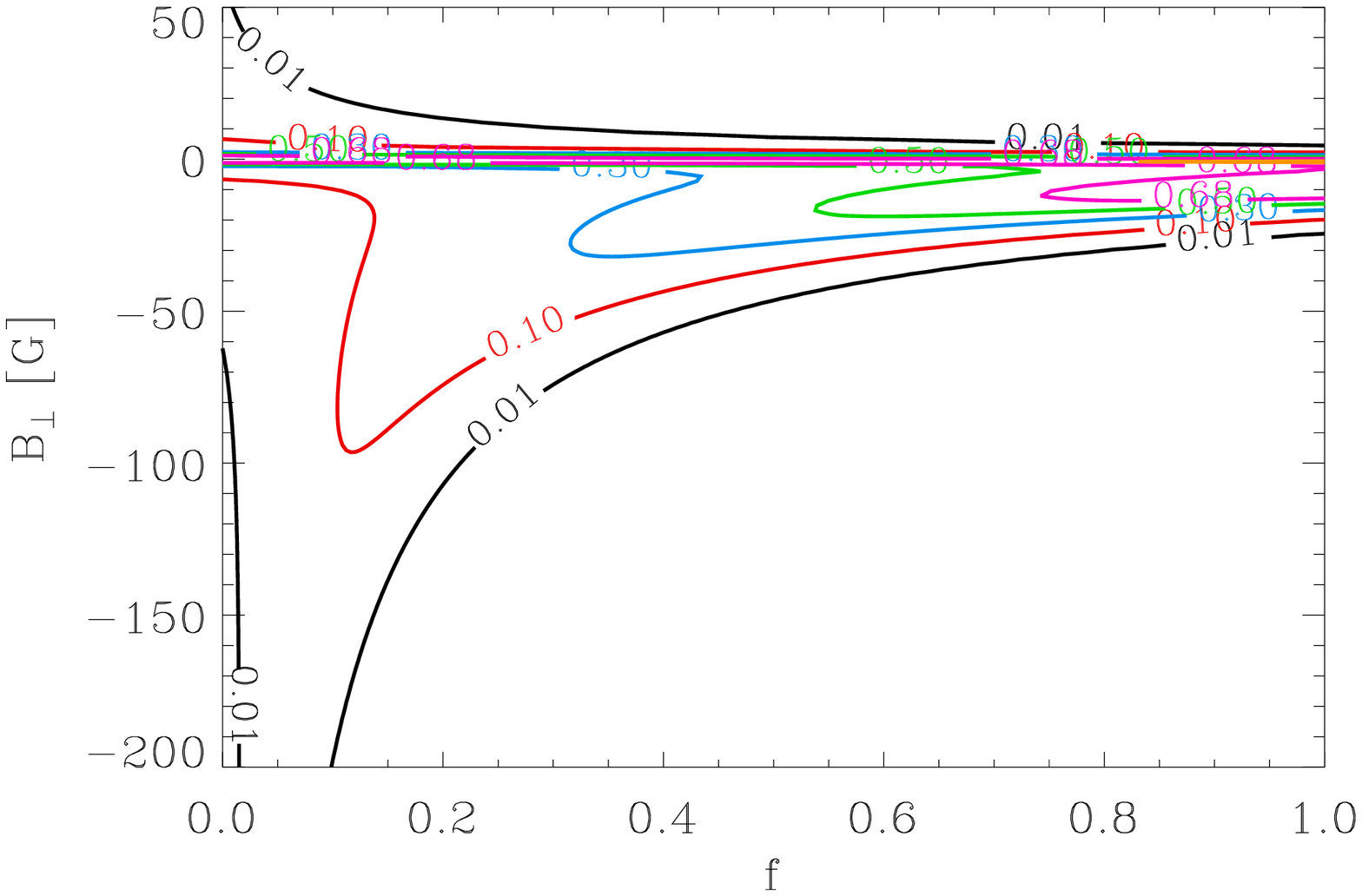}
\includegraphics*[viewport=12 10 547 372,width=0.31\textwidth]{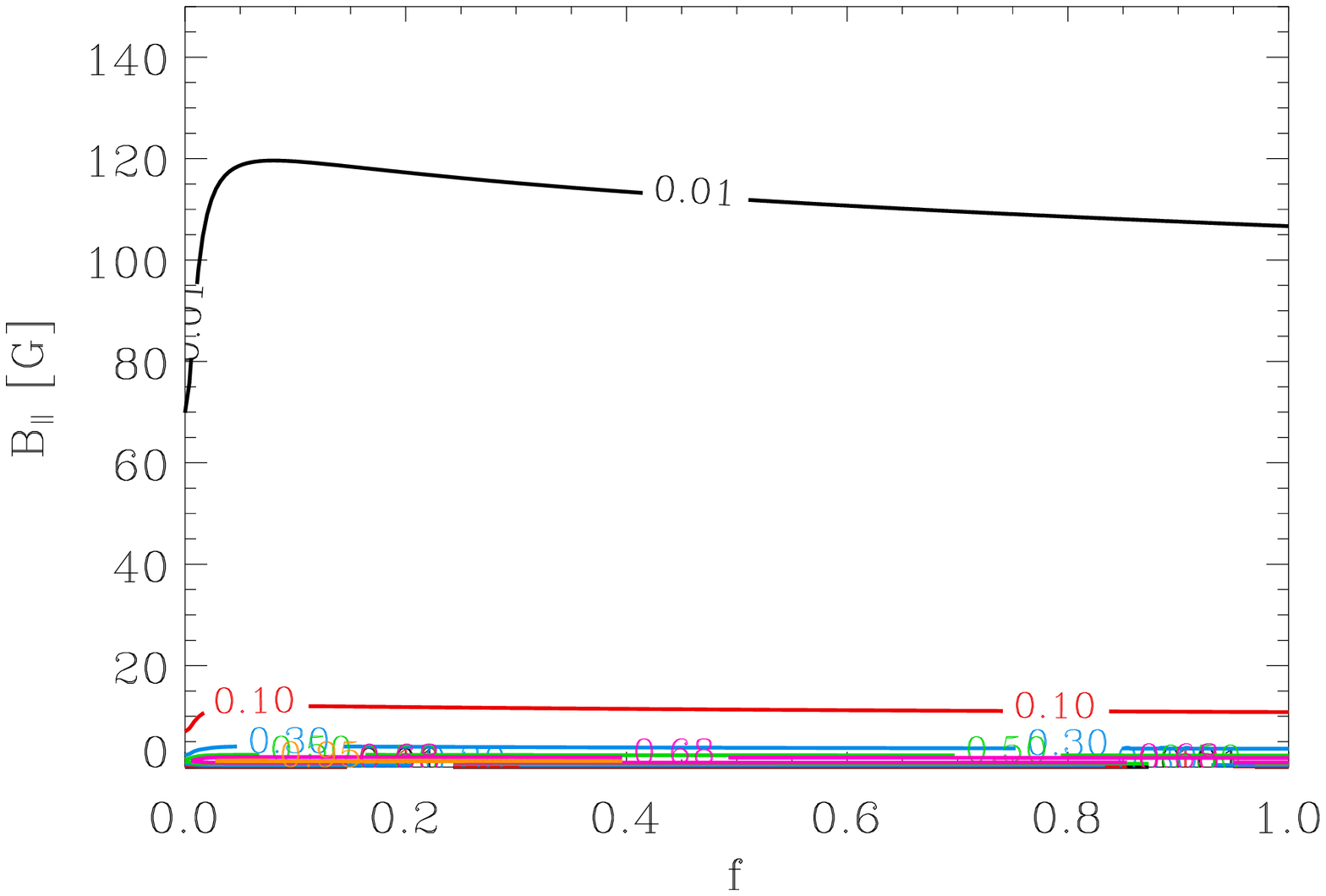}
\includegraphics*[viewport=12 10 547 372,width=0.31\textwidth]{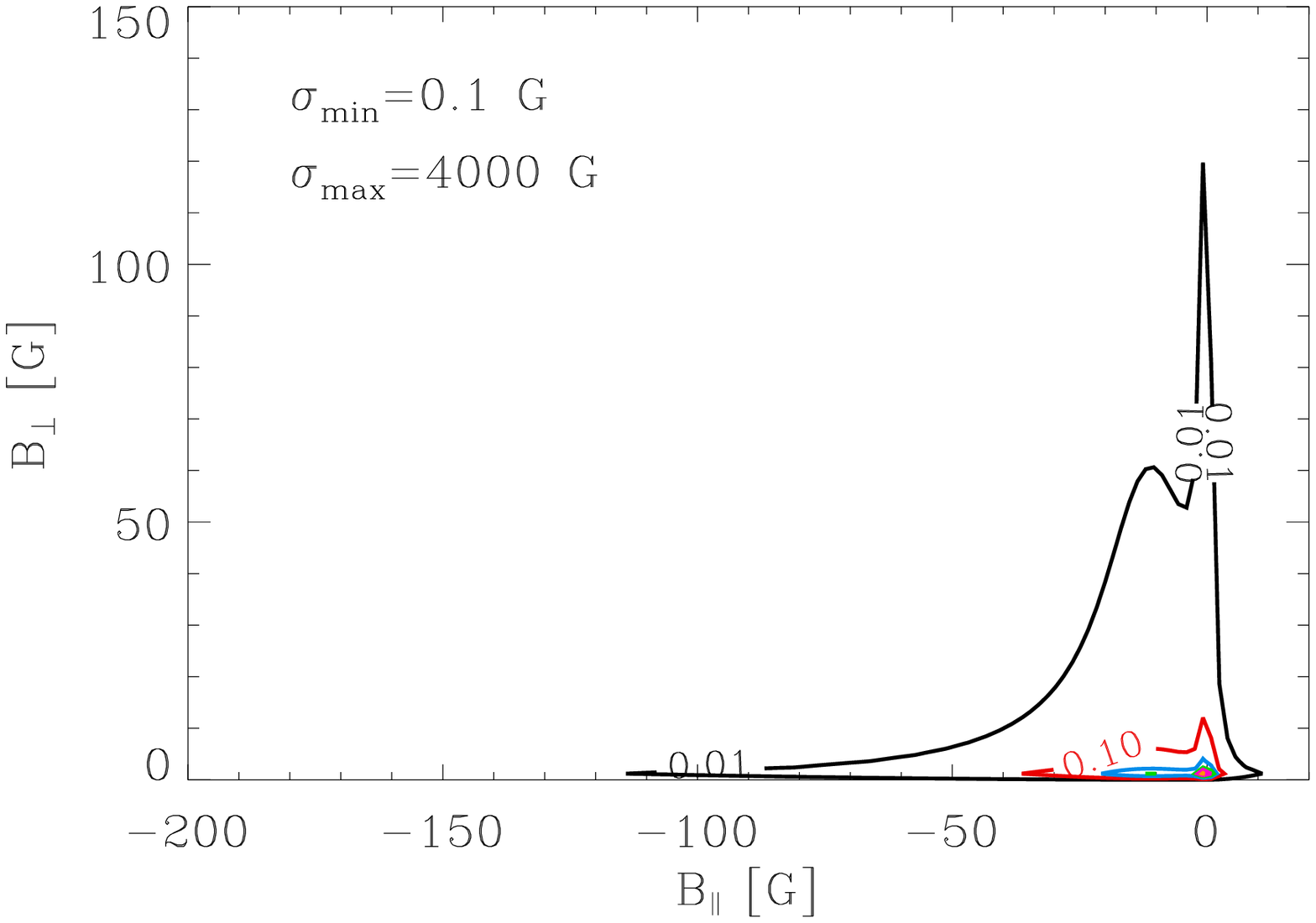}
\includegraphics*[viewport=12 10 547 372,width=0.31\textwidth]{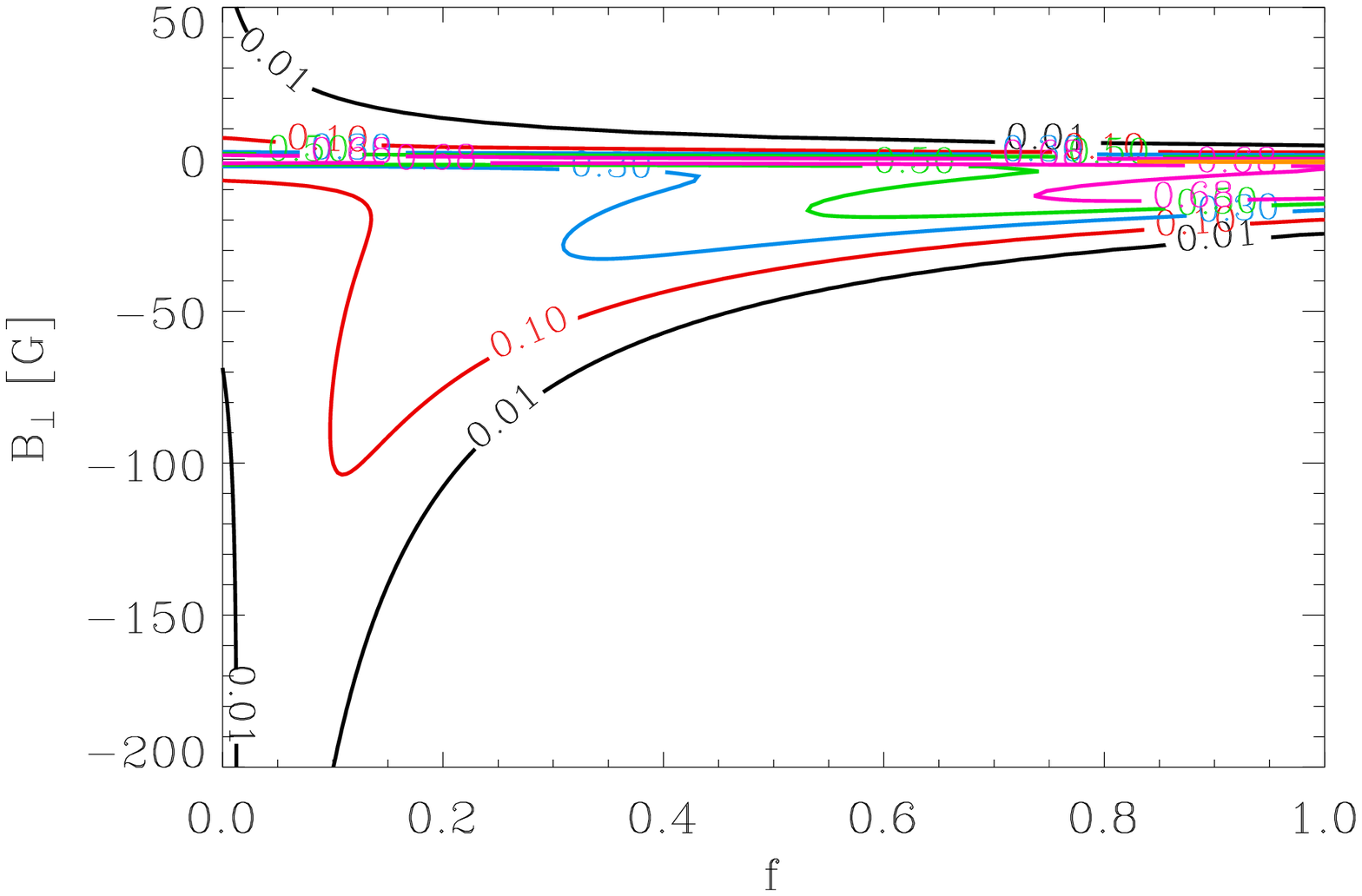}
\includegraphics*[viewport=12 10 547 372,width=0.31\textwidth]{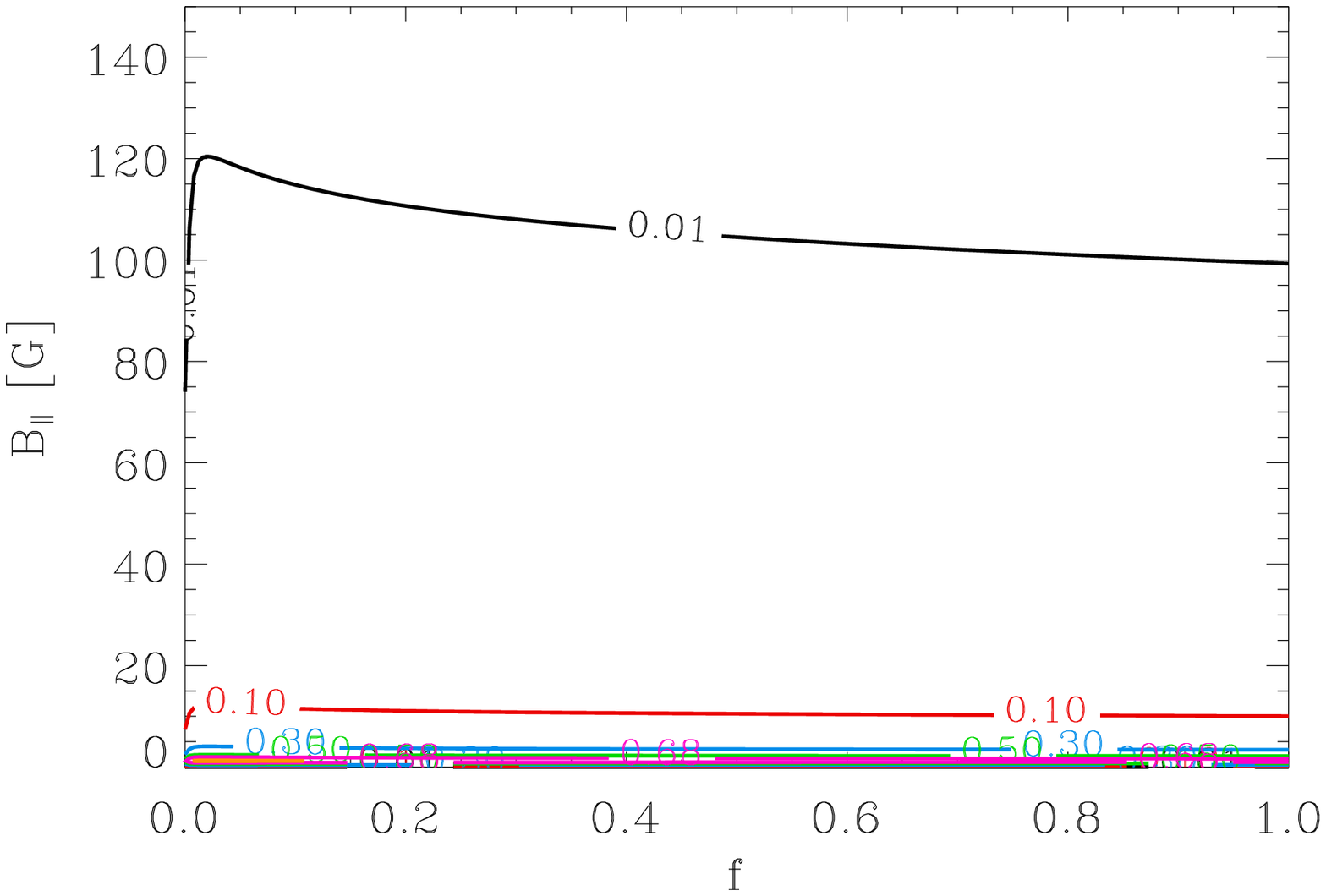}
\includegraphics*[viewport=12 10 547 372,width=0.31\textwidth]{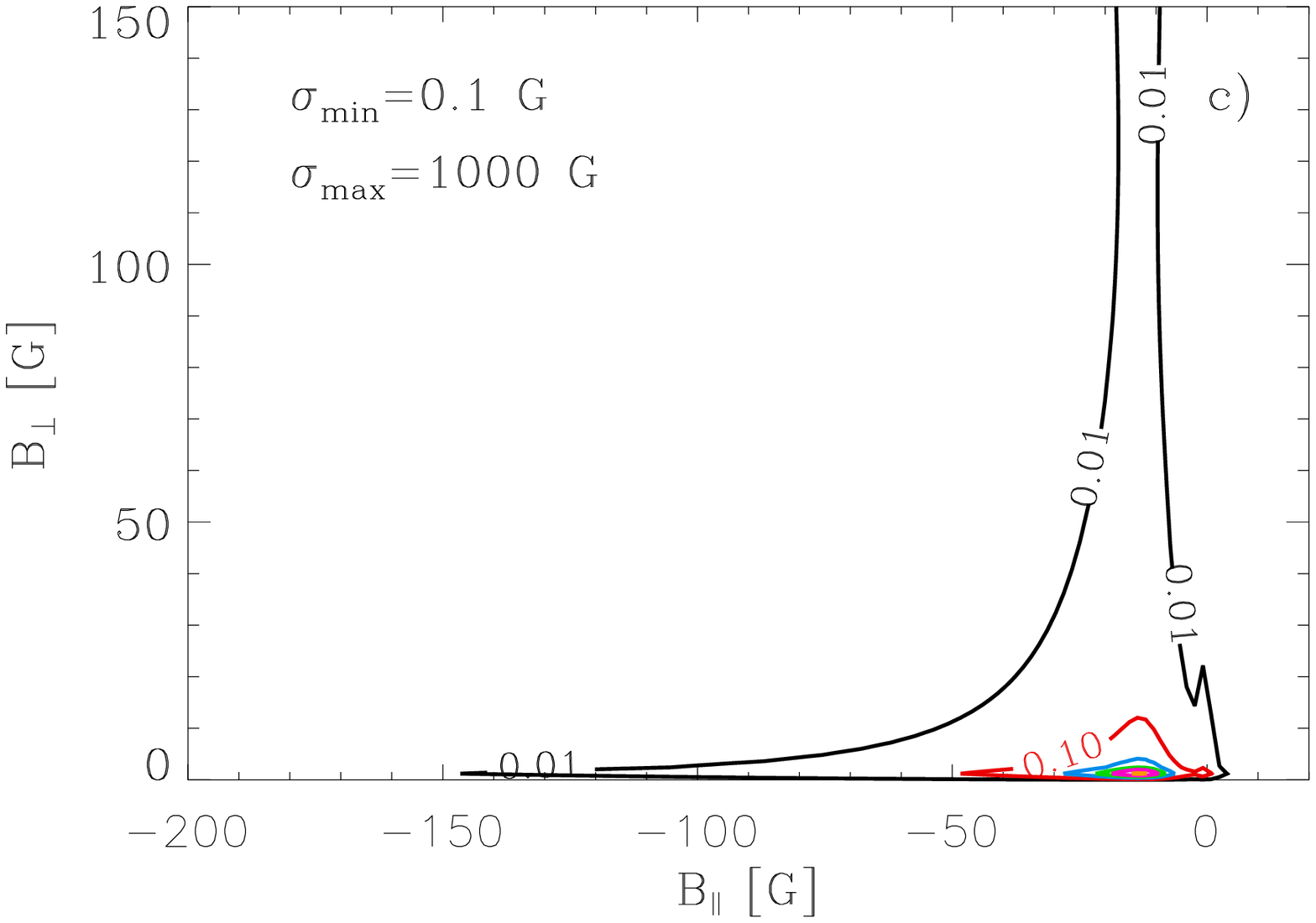}
\includegraphics*[viewport=12 10 547 372,width=0.31\textwidth]{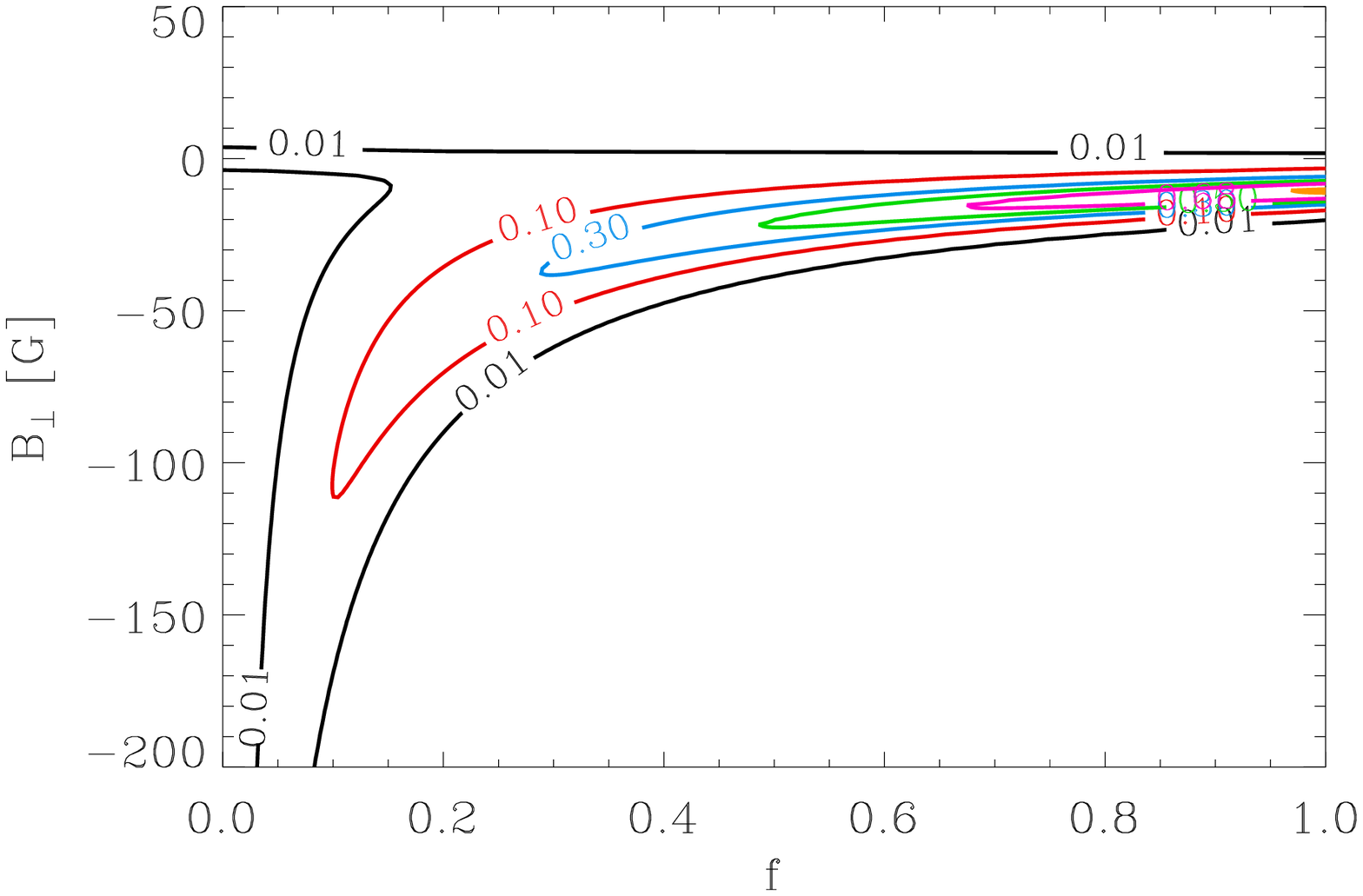}
\includegraphics*[viewport=12 10 547 372,width=0.31\textwidth]{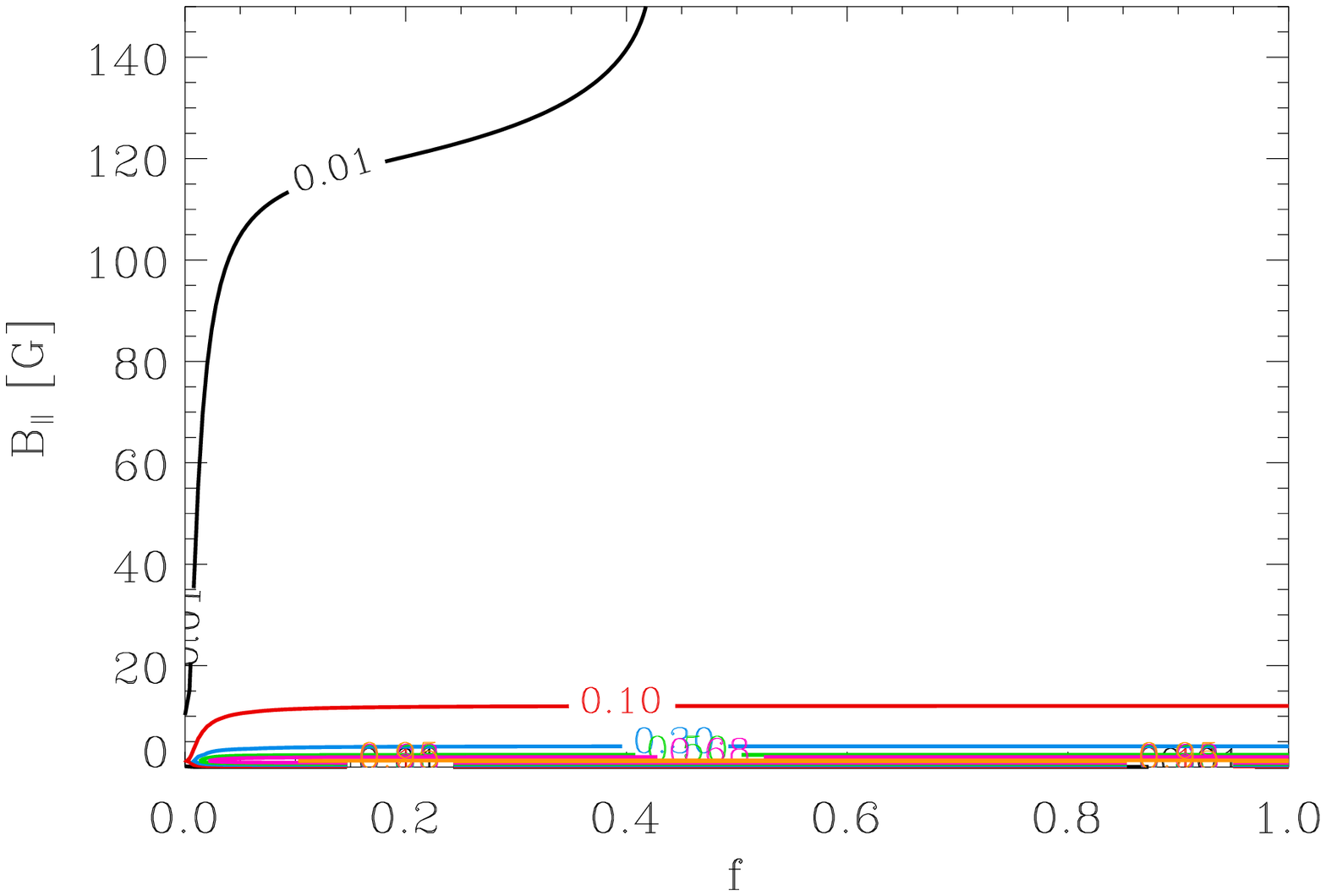}
\includegraphics*[viewport=12 10 547 372,width=0.31\textwidth]{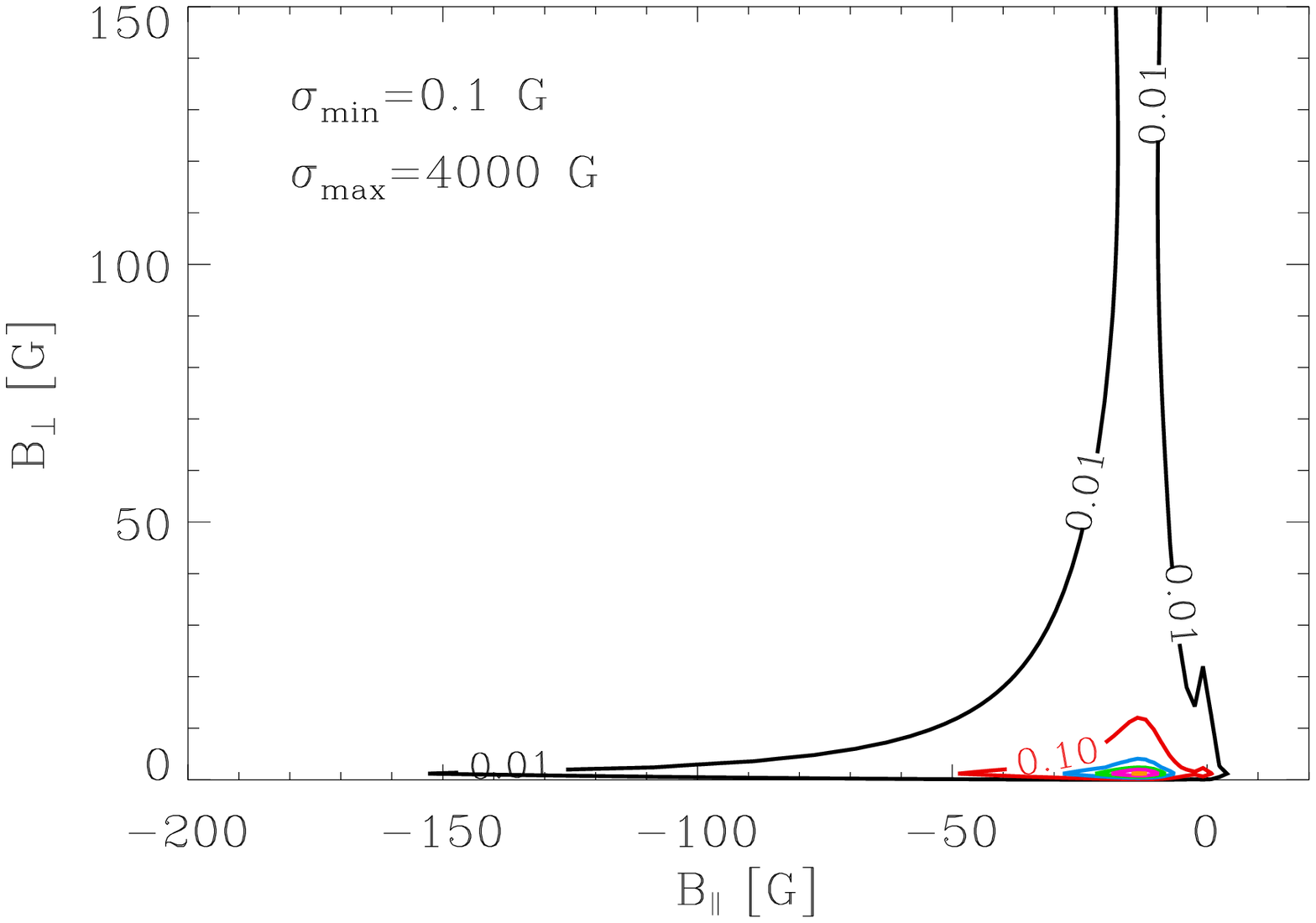}
\includegraphics*[viewport=12 10 547 372,width=0.31\textwidth]{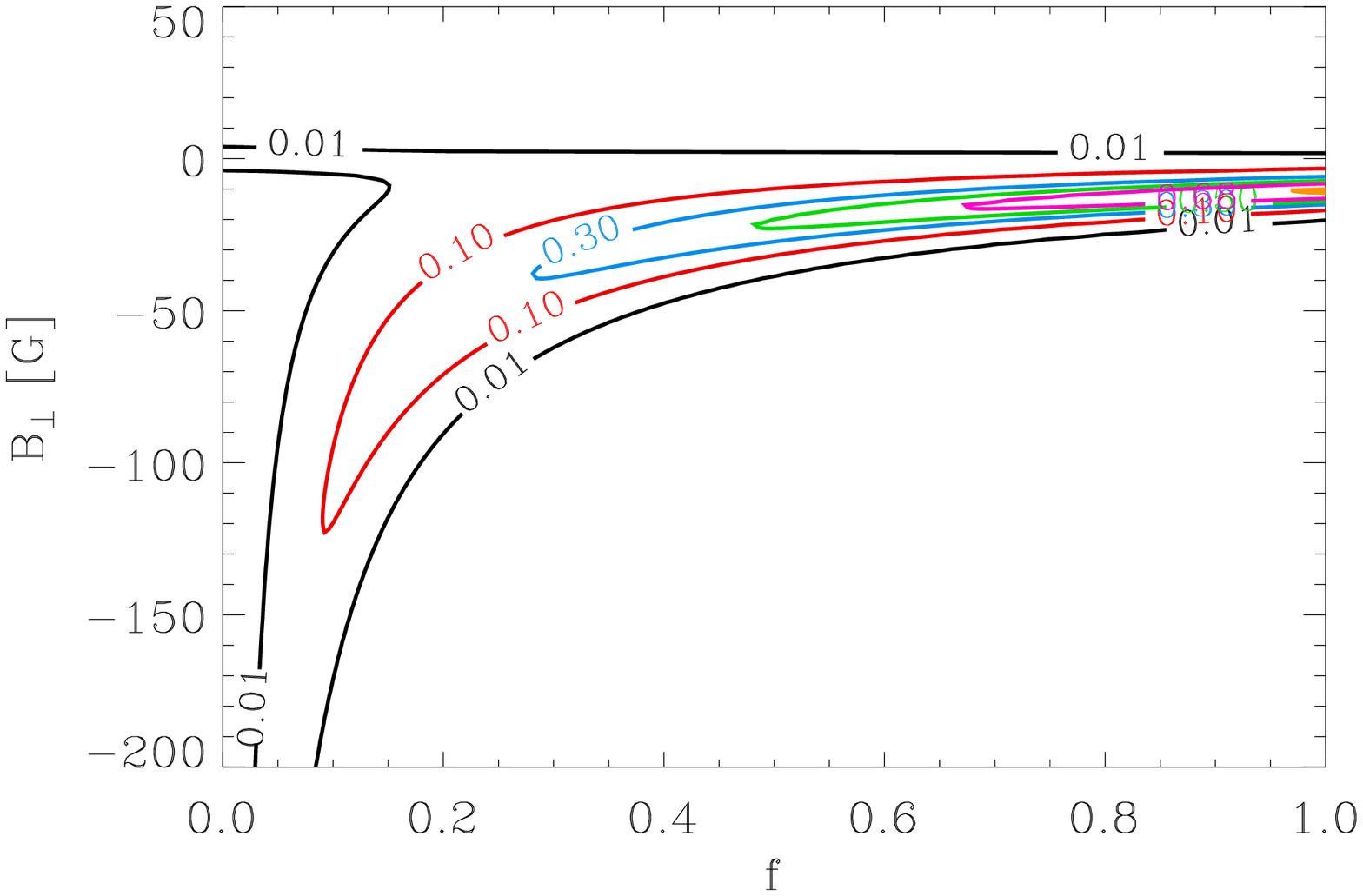}
\includegraphics*[viewport=12 10 547 372,width=0.31\textwidth]{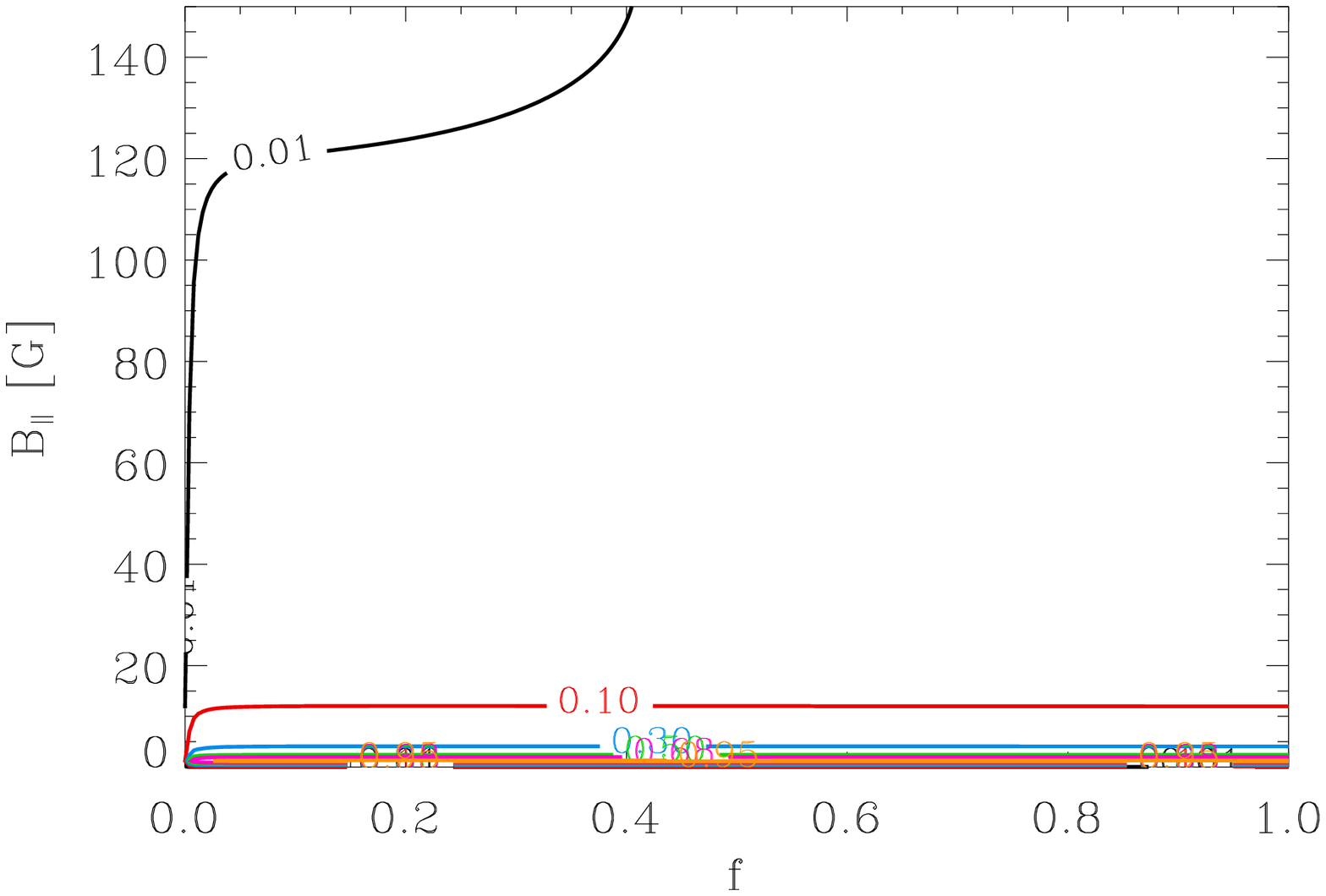}
\caption{Joint marginal posterior distributions for $f$, $\Bpar$ and $\Bperp$.
The top two rows, labeled with a), correspond to the non-hierarchical case
with two different values of $\sigma_\parallel$ and $\sigma_\perp$. The third and fourth rows,
labeled with b), show results with the hierarchical approach of Eq. (\ref{eq:posterior_hierarchical_known_noise})
for two values of $\sigma_\mathrm{max}$ while the last two rows, labeled with c), correspond to the
results of the hierarchical model of Eq. (\ref{eq:posterior_hierarchical_unknown_noise}) for fixed
values of $\sigma_{n_\mathrm{min}}=10^{-4}$ and $\sigma_{n_\mathrm{max}}=10^{-2}$ and the 
two same values for $\sigma_\mathrm{min}$ and $\sigma_\mathrm{max}$ considered in the previous case.
For clarity, the contours at normalized probability $0.01$, $0.1$, $0.3$, $0.5$, $0.68$, $0.95$
are marked in black, red, blue, green, magenta and orange, respectively.
\label{fig:joint_imax}}
\end{figure*}

Figure \ref{fig:joint_imax} shows the joint posterior marginal distributions $p(\Bpar,\Bperp|D)$ in the first column, 
$p(f,\Bpar|D)$ in the second one and $p(f,\Bperp|D)$ in the third. These marginal distributions
help us distinguish which parameters are degenerate and help us understand the behavior
of the full posterior. For comparison,
the first two rows (labeled \emph{a}) present cases obtained by marginalization of the non-hierarchical posterior of
Eq. (\ref{eq:posterior}). Specifically, these panels were obtained by numerical integration of 
Eq. (\ref{eq:marginal_3d_nonhier}).
The middle two rows (labeled \emph{b}) are obtained marginalizing the
hierarchical posterior of Eq. (\ref{eq:posterior_hierarchical_known_noise}) assuming Gaussian
noise with a standard deviation of $\sigma_n=2\times 10^{-3}$ in units of the continuum intensity. Specifically,
they are computed by numerical integration of Eq. (\ref{eq:posterior_hierarchical_known_noise_marginalchi}). Finally, the last
two rows (labeled \emph{c}) are obtained applying a numerical quadrature to the full hierarchical 
posterior of Eq. (\ref{eq:posterior_hierarchical_unknown_noise_final}) that marginalizes the noise.
Two typical values of the hyperparameters are considered in each one of the posteriors. The values of
the relevant parameters are indicated in each plot and in the caption. As a general rule, the non-hierarchical model has a 
stronger sensitivity to the prior hyperparameters than the hierarchical models \cite[e.g.,][]{gelman_bayesian03}.
In our case, the hierarchical models are almost insensitive to the exact values of the hyperparameters,
even though we have used two extreme values. The joint marginal posteriors clearly indicate 
the presence of strong degeneracies between the $f$ and $\Bpar$ and
$\Bperp$, as widely known. However, there is still some information available
in the observations about each parameter individually, at least sufficient to discard
regions of the space of parameters. Although part of this information is encoded in the priors,
it is transparently introduced and can be controlled at will. Furthermore, this information
is based on very general considerations about the behavior of the magnetic field (like typical
maximum values present in the solar atmosphere).

\begin{figure*}
\centering
\includegraphics*[viewport=32 10 555 372,width=0.32\textwidth]{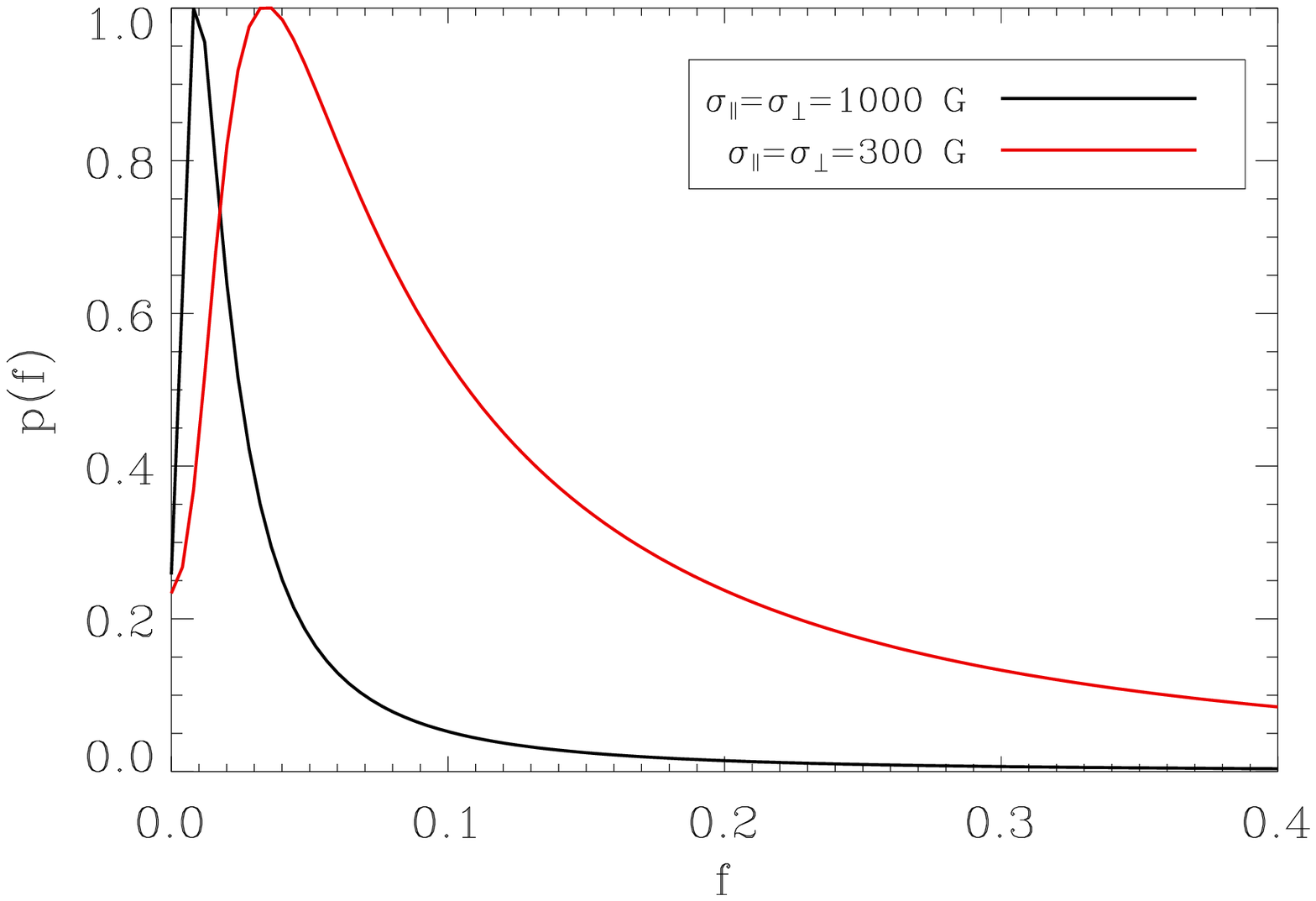}
\includegraphics*[viewport=32 10 555 372,width=0.32\textwidth]{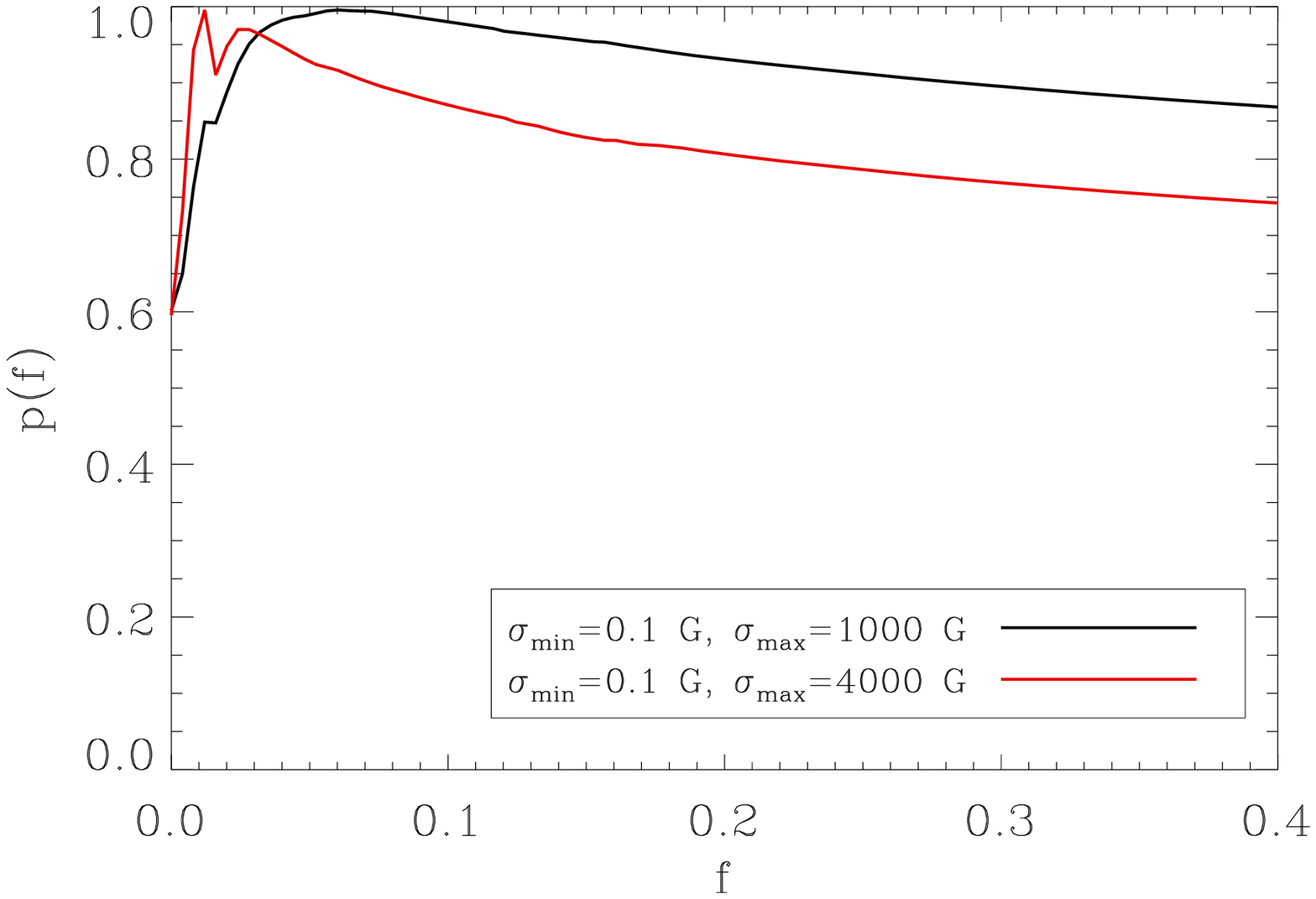}
\includegraphics*[viewport=32 10 555 372,width=0.32\textwidth]{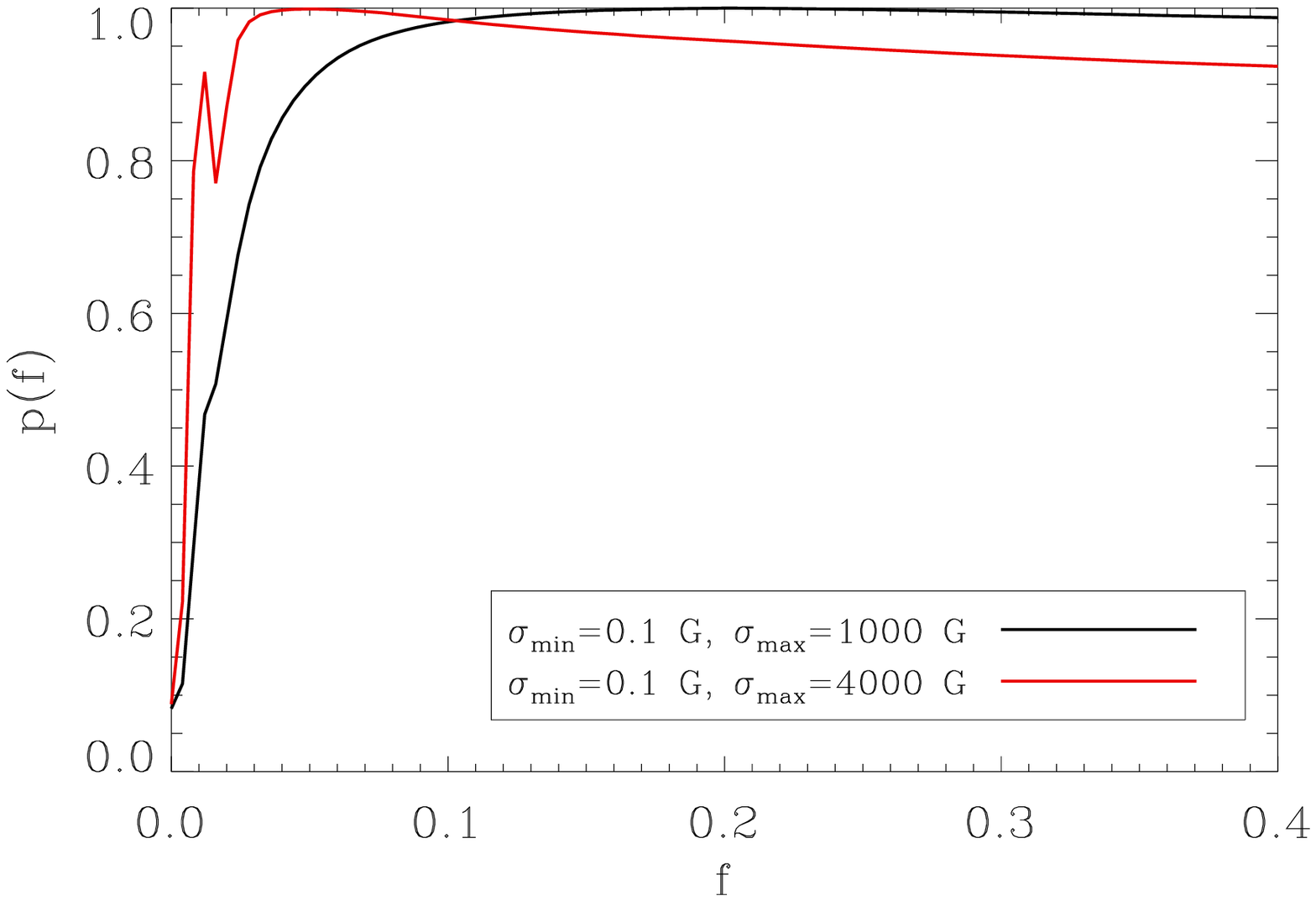}
\includegraphics*[viewport=32 10 555 372,width=0.32\textwidth]{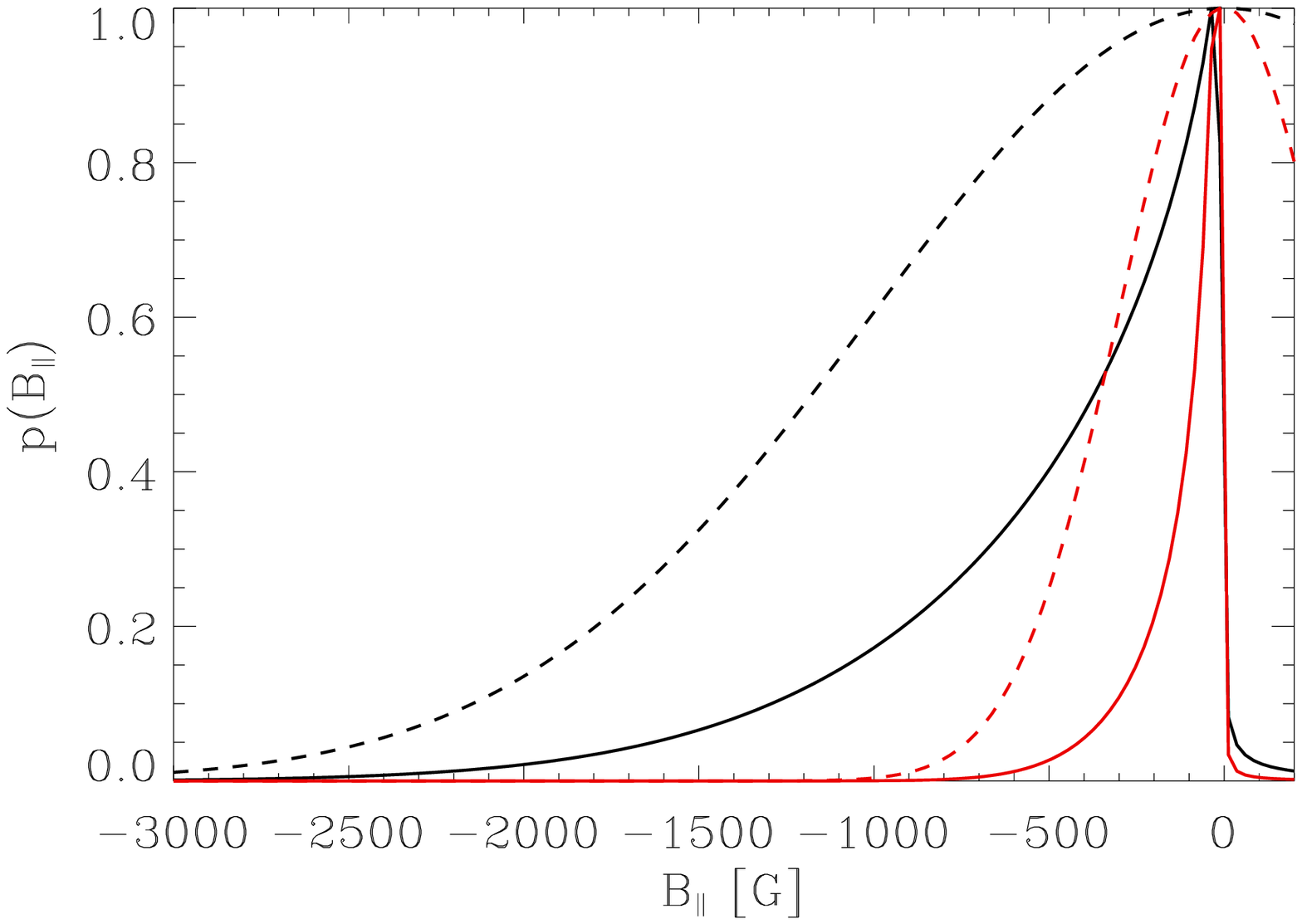}
\includegraphics*[viewport=32 10 555 372,width=0.32\textwidth]{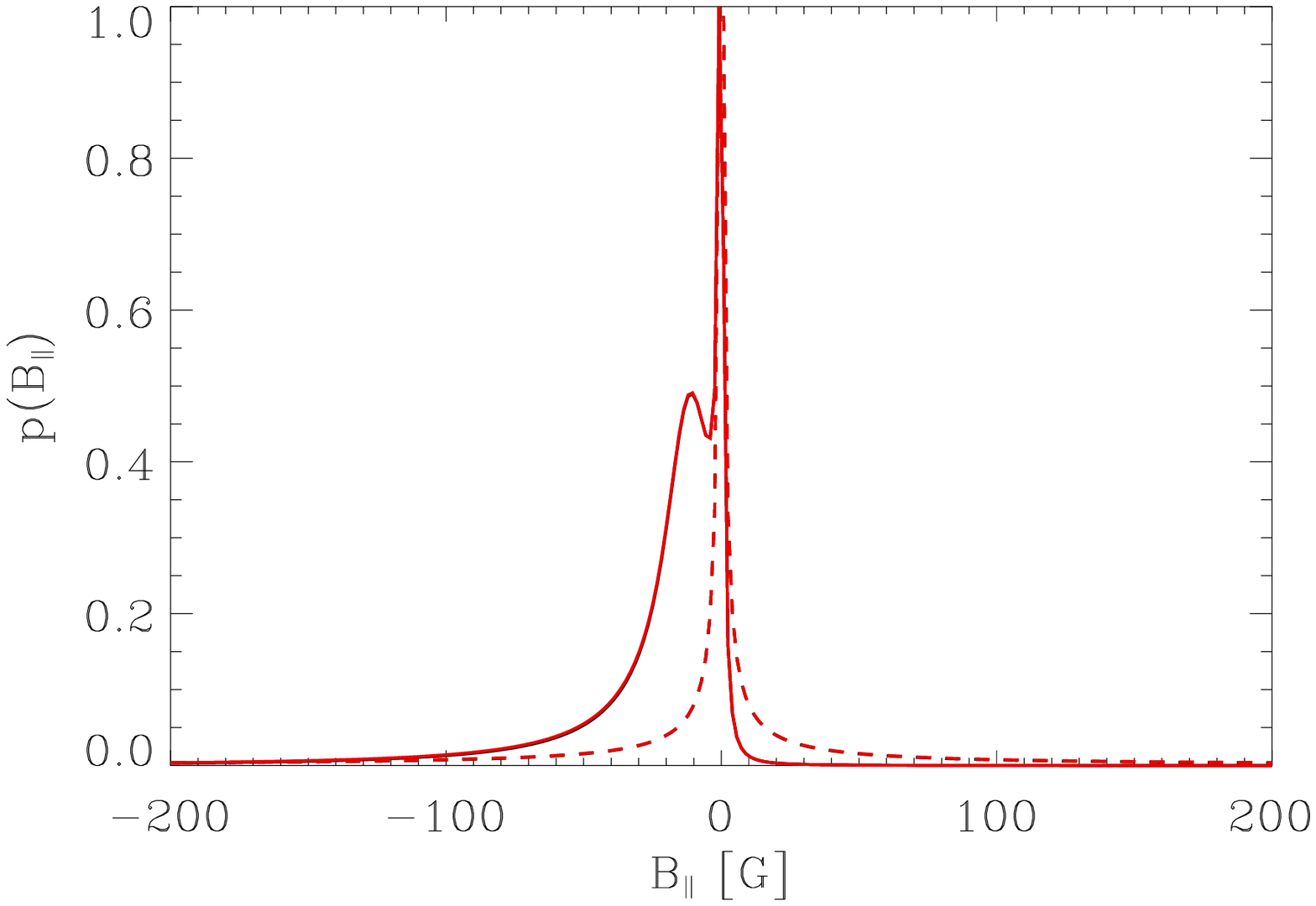}
\includegraphics*[viewport=32 10 555 372,width=0.32\textwidth]{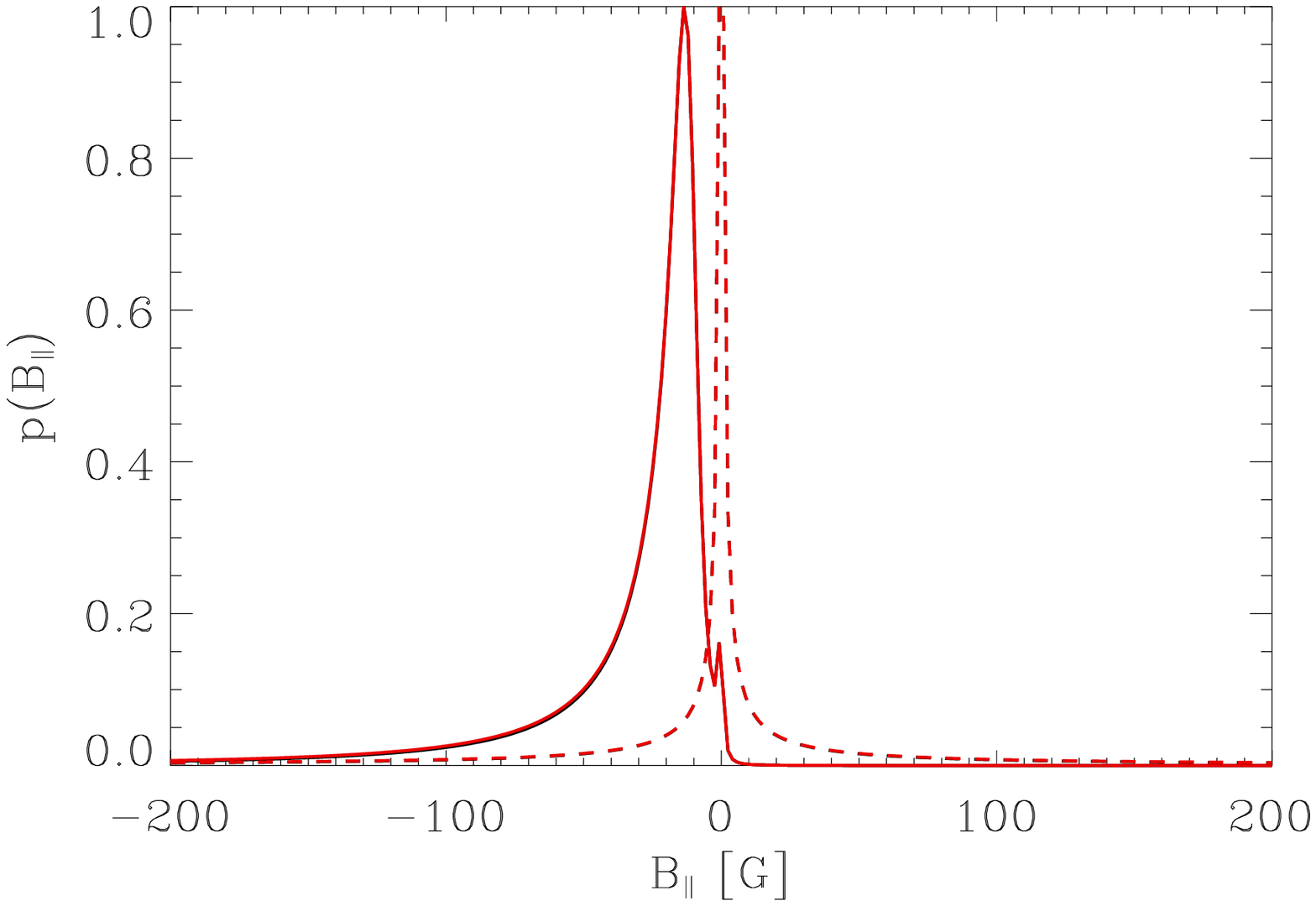}
\includegraphics*[viewport=32 10 555 372,width=0.32\textwidth]{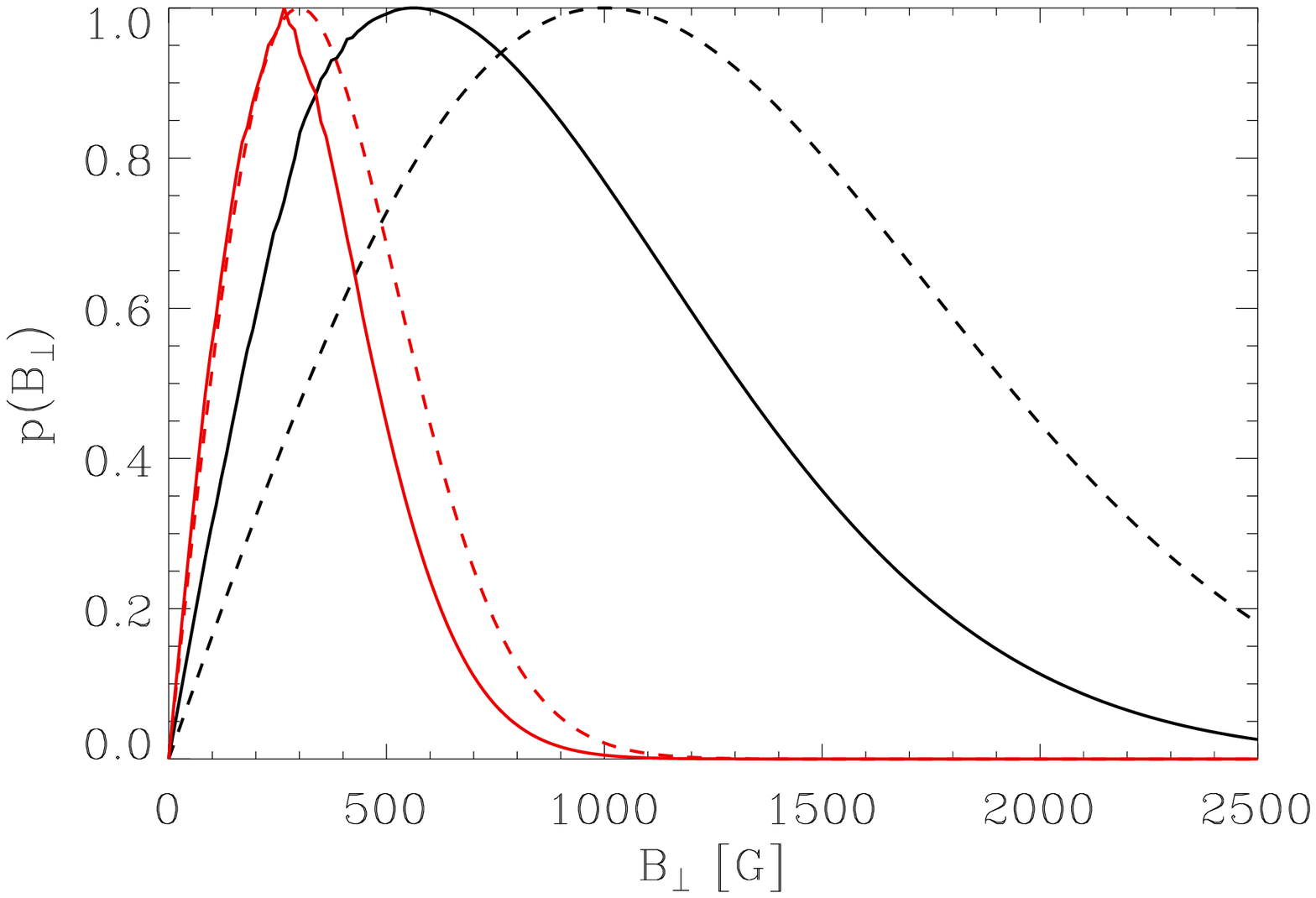}
\includegraphics*[viewport=32 10 555 372,width=0.32\textwidth]{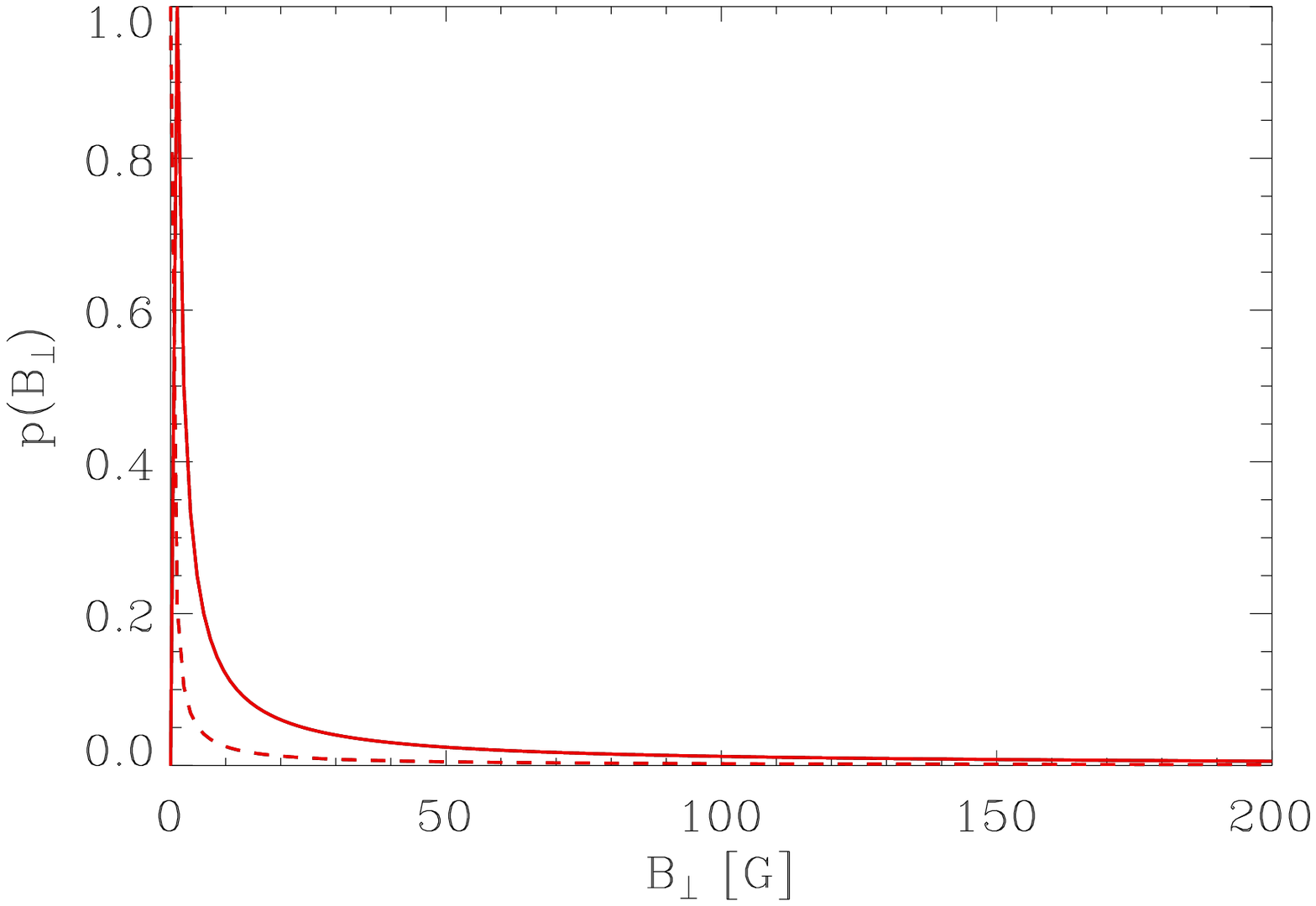}
\includegraphics*[viewport=32 10 555 372,width=0.32\textwidth]{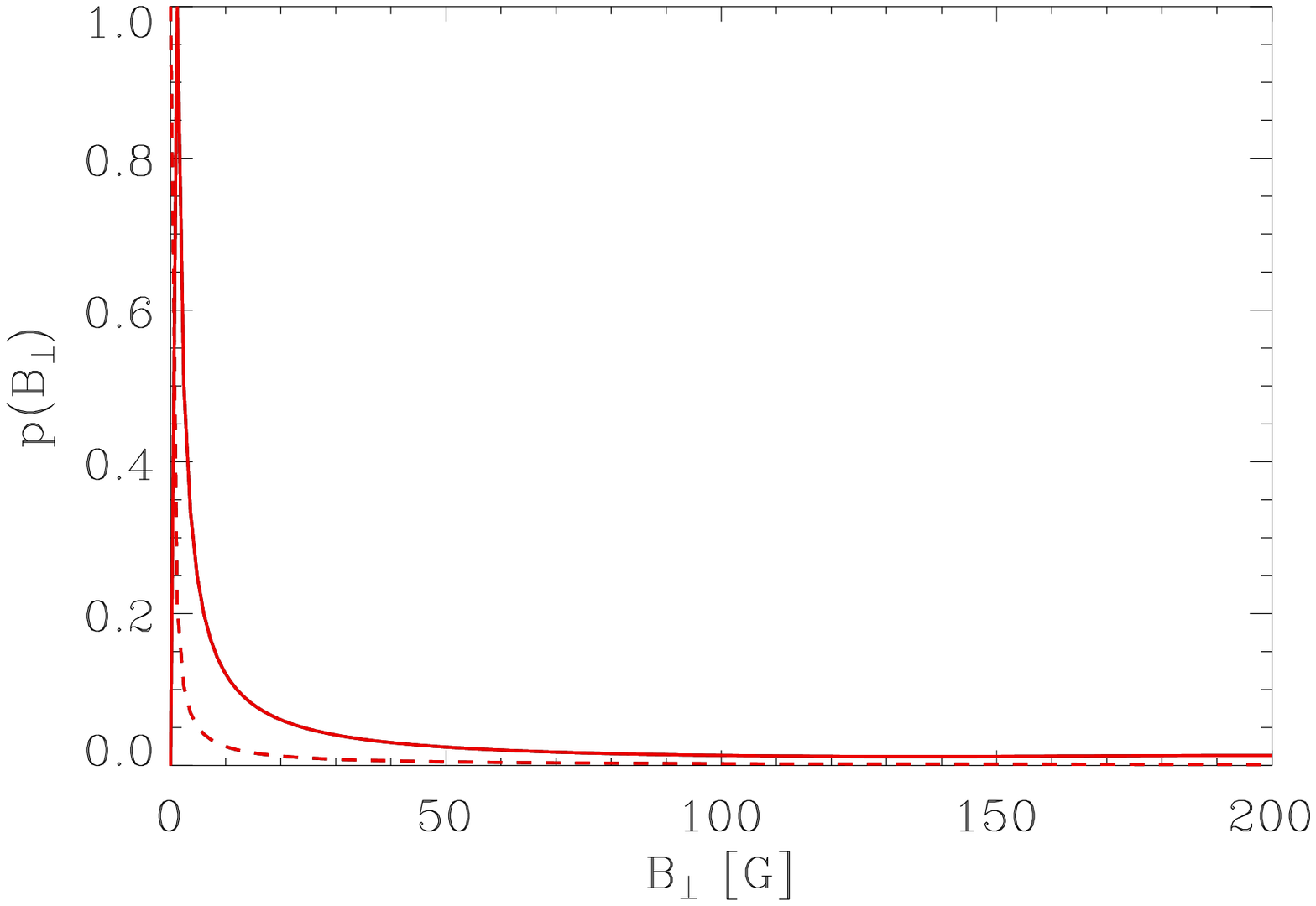}
\includegraphics*[viewport=32 10 555 372,width=0.32\textwidth]{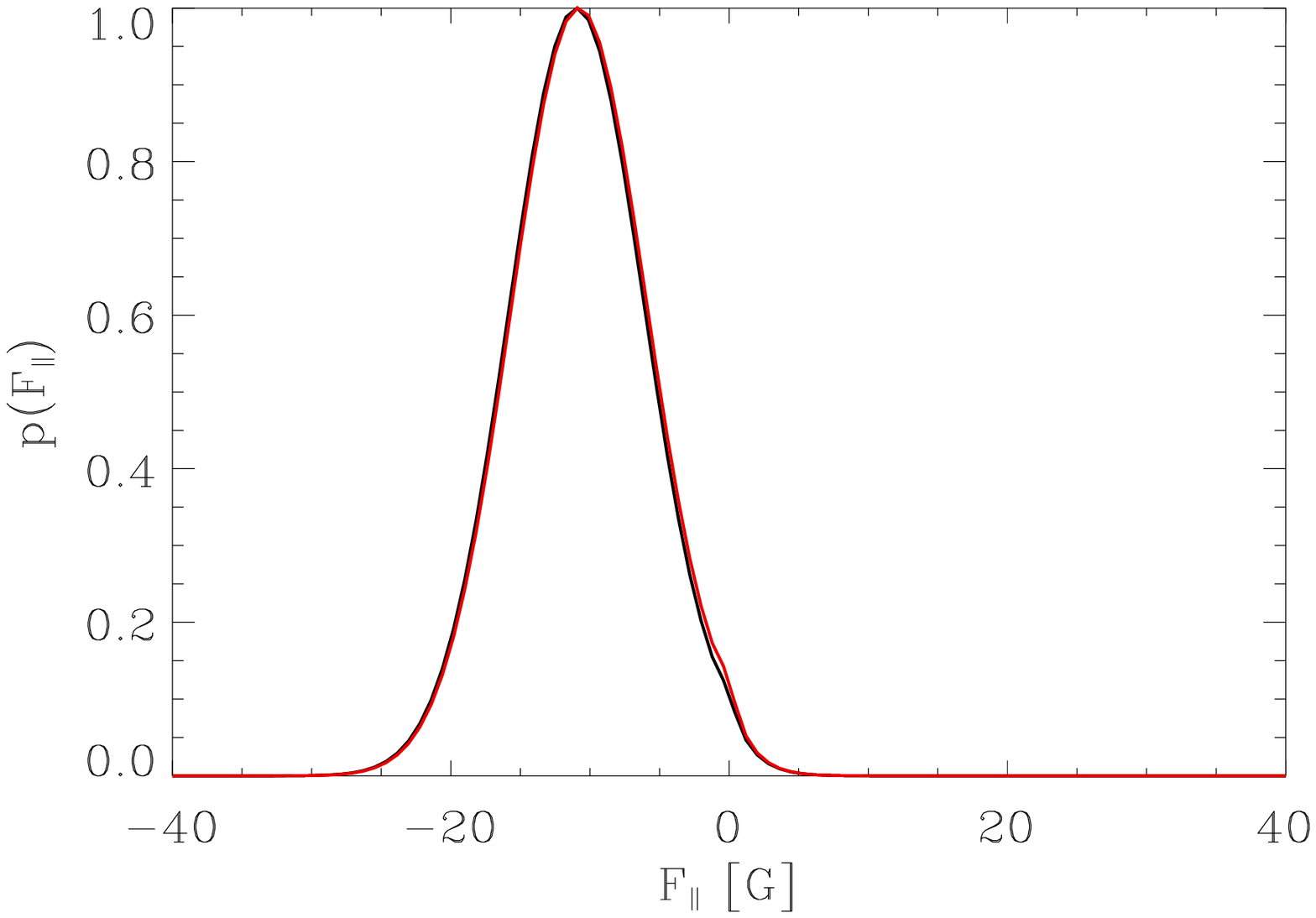}
\includegraphics*[viewport=32 10 555 372,width=0.32\textwidth]{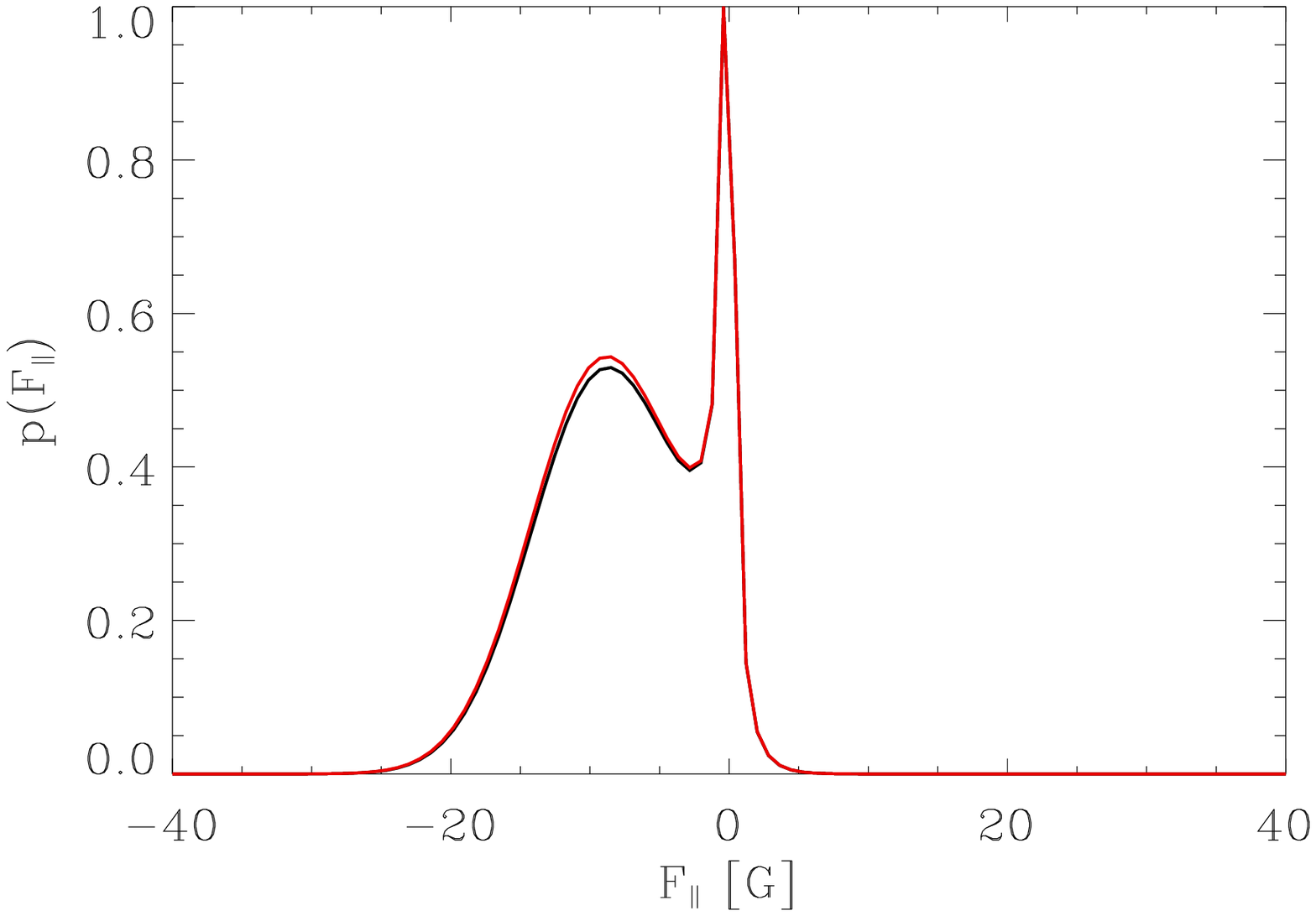}
\includegraphics*[viewport=32 10 555 372,width=0.32\textwidth]{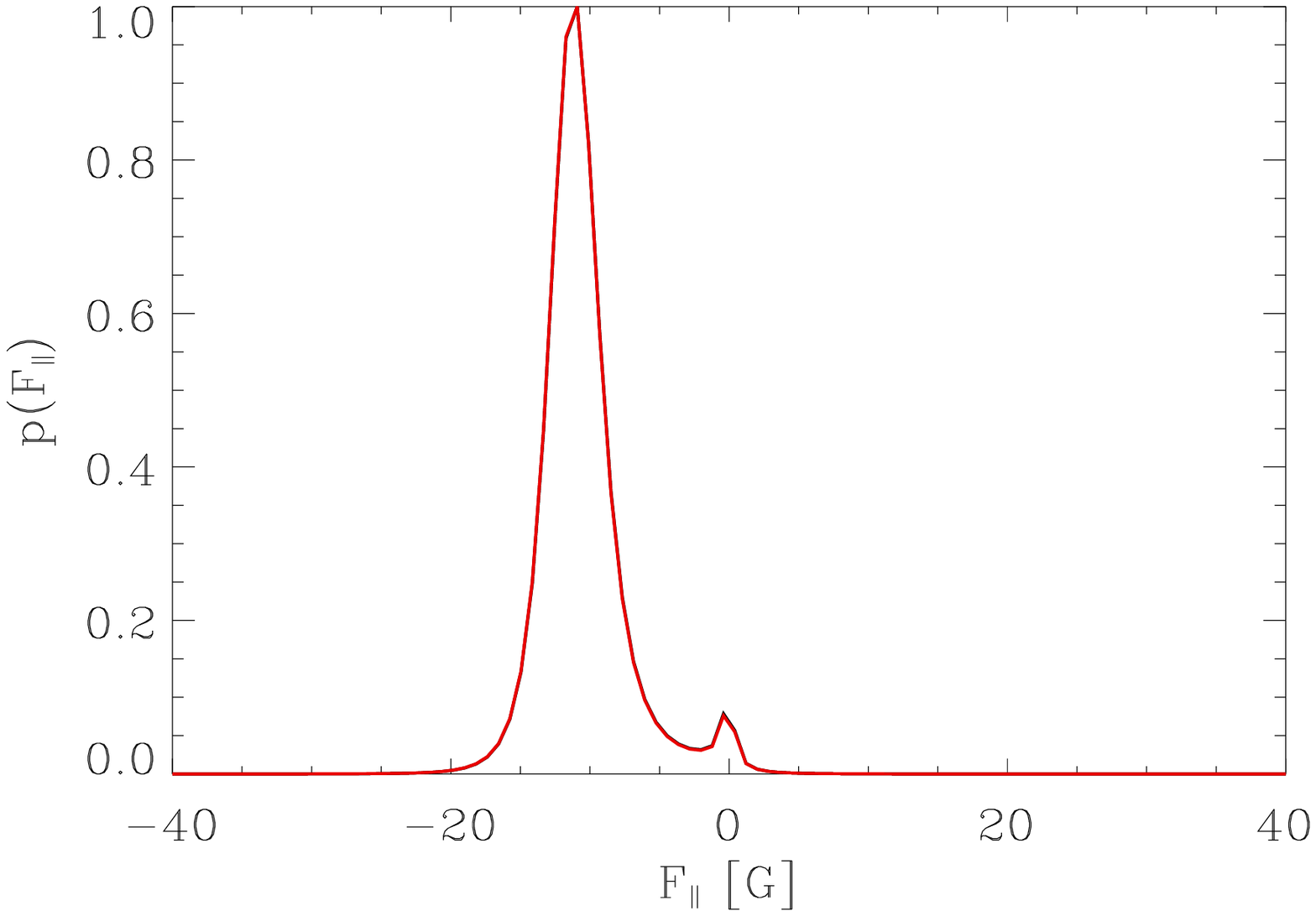}
\caption{One-dimensional marginal posteriors for $f$ (first row), $\Bpar$ (second
row), $\Bperp$ (third row) and $F_\parallel$ (fourth row). The left column
shows the non-hierarchical case, while the second and third columns present
results in the two considered hierarchical models. The colors are
associated with different values of the hyperparameters. Note the robustness
of the hierarchical models to the different values of the hyperparameters. The dashed
lines indicate the corresponding prior distribution. If the posterior of a given parameter is
clearly different from the prior, we can state that the data contain enough 
constraining information for this parameter.
\label{fig:marginal_imax}}
\end{figure*}

When extracting information for a given parameter, it is necessary to marginalize out the
rest of parameters. This marginalization carries out error propagation correctly and
degeneracies are evidenced as long tails in the marginal posteriors. 
Figure \ref{fig:marginal_imax} presents these posteriors. The first column
corresponds to the non-hierarchical model, while the second and third columns
are the results of the two hierarchical models. They were obtained numerically
by integrating the distributions shown in Fig. \ref{fig:joint_imax}. The first row is the marginal
posterior for $f$, the second for $\Bpar$, the third for $\Bperp$ and the
fourth for $F_\parallel$. The dashed lines are the priors considered for
each parameter.
In the non-hierarchical model, the MMAP value for $\Bpar$ seems to be relatively robust to 
the election of the prior, while it has a larger dependence on the prior for $\Bperp$ and $f$. 
According to the shapes of the marginal posterior distributions, it is possible to give the
most probable value together with asymmetric error bars which contain 68\% or
95\% of the total mass of the distribution. In this particular example, the lower
limit of $\Bpar$ will be much larger in magnitude than the upper limit. The reason
is that the sign of $\Bpar$ controls the sign of the Stokes $V$ profile, and 
positive fields are strongly discarded by data because the polarity is not
the correct one. The same applies to $f$ and $\Bperp$, which have a very asymmetric
confidence interval. Interestingly, since $F_\parallel$ is a quantity directly proportional
to the amplitude of Stokes $V$ in the weak-field regime, its marginal posterior distribution is very well determined and
its width is fundamentally characterized by the noise standard deviation.

Given the sensitivity to the election of the prior of the non-hierarchical model, we also
consider the two hierarchical models. The results are very robust to the 
election of the hyperparameters (the red and black curves overlap). The filling factor is 
not constrained at all by the observations, as expected in the weak-field regime.
In spite of that, the posterior for the component of the field along the line-of-sight is
constrained and one can define a MMAP value and a confidence interval. In any case, the
analysis discards positive values and very large negative values of $\Bpar$. Concerning
$\Bperp$, the posterior is very similar to the prior. Since the prior is based on 
physical considerations, we can safely give an upper limit to $\Bperp$, but always
taking into account that this comes essentially from a-priori knowledge.

The posterior for $F_\parallel$ is again very close to a Gaussian
with mean and variance given by Eq. (\ref{eq:mean_variance_fparallel}) in the 
case of the hierarchical model with known noise. Note also
that there is an additional peak at zero magnetic flux density. This peak is
produced by the marginalization over the priors, which opens the possibility
that the observed Stokes $V$ signal is compatible with the absence of
magnetic field. Concerning the hierarchical model with unknown noise standard deviation,
the posterior is slightly narrower, induced by the Jeffreys' prior.

\begin{figure*}
\centering
\includegraphics*[viewport=37 16 542 372,width=0.32\textwidth]{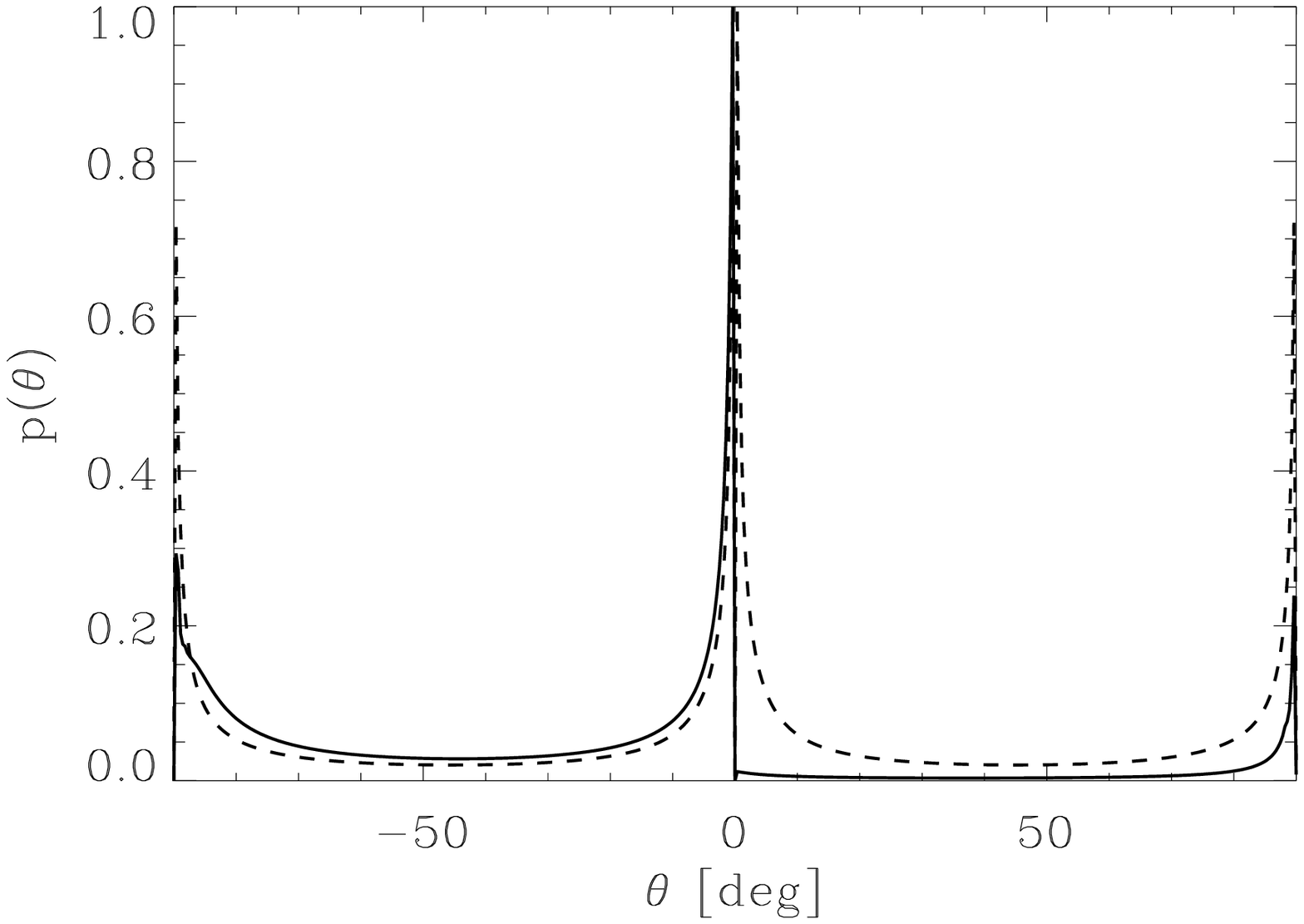}
\includegraphics*[viewport=37 16 542 372,width=0.32\textwidth]{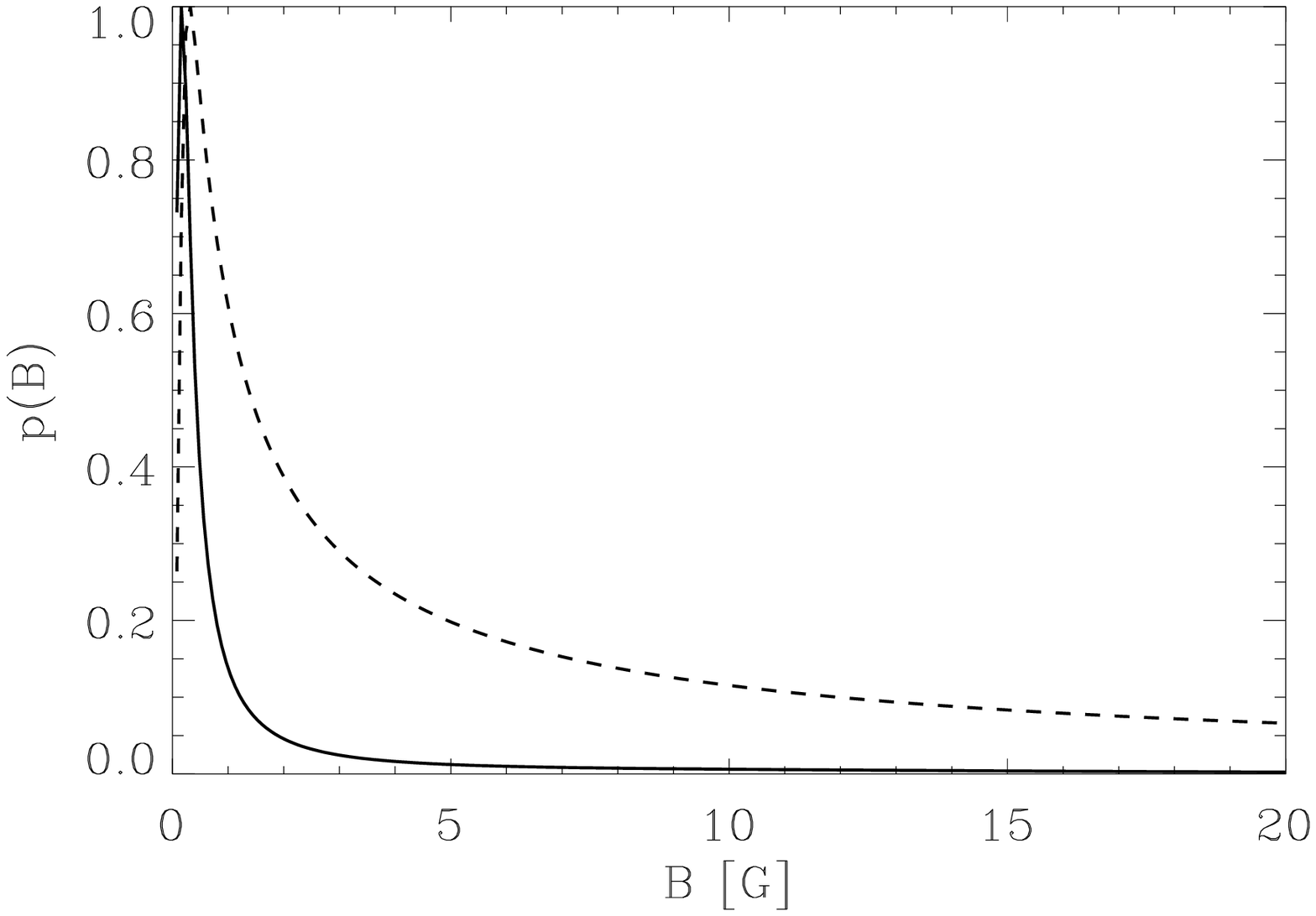}
\includegraphics*[viewport=37 16 542 372,width=0.32\textwidth]{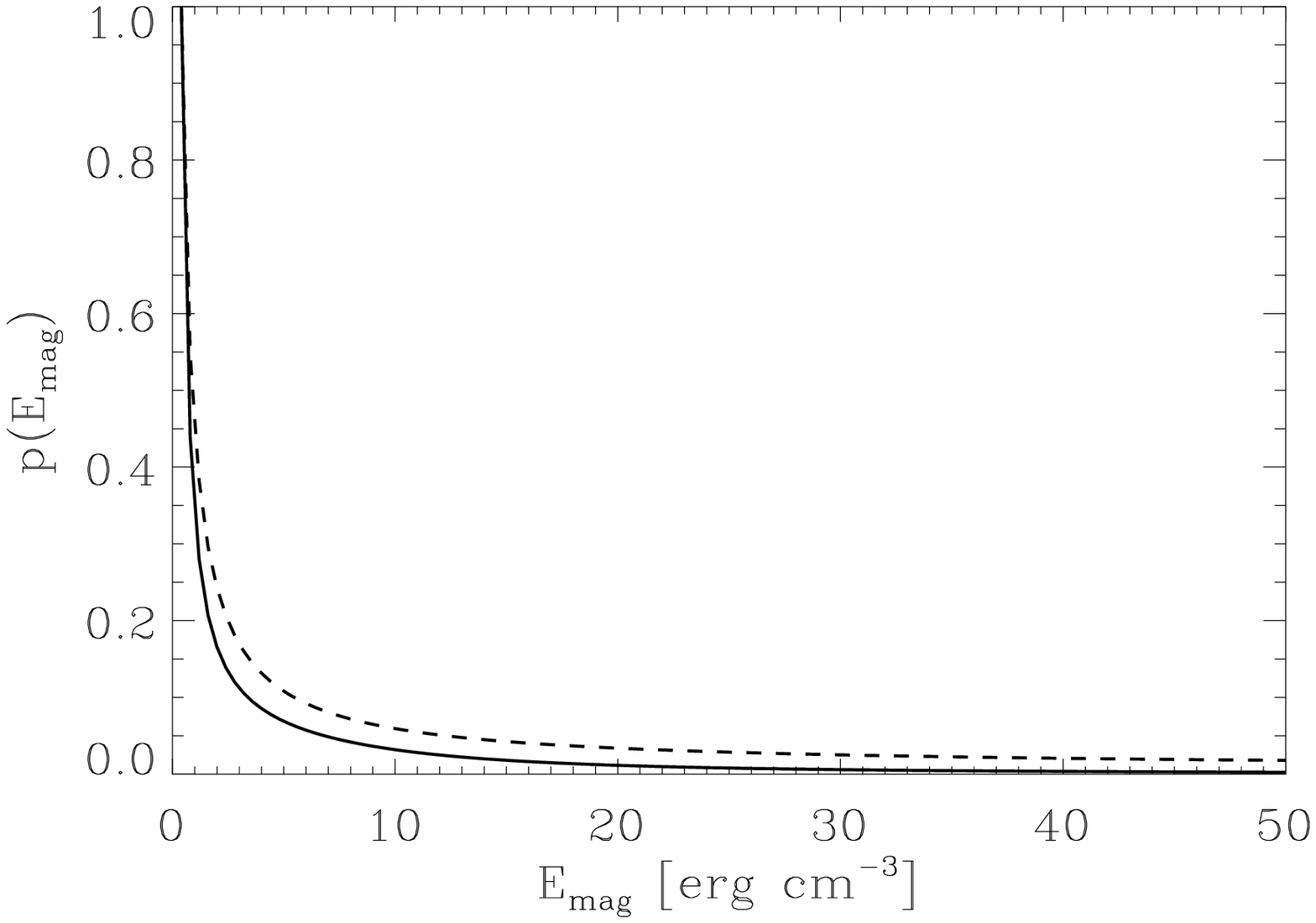}
\caption{Posterior for derived quantities for the profiles of Fig. \ref{fig:imax_profiles} using
the hierarchical model with $\sigma_n=2 \times 10^{-3}$, $\sigma_\mathrm{min}=0.1$ G and
$\sigma_\mathrm{max}=4000$ G. The
left panel shows the marginal posterior for the field inclination, $\theta$, in solid
line, while the prior is shown in dashed line. The middle panel presents
the marginal posterior for the field strength, while the right panel shows the same for the
magnetic energy.
\label{fig:derived_quantities}}
\end{figure*}

Finally, Fig. \ref{fig:derived_quantities} presents the marginal posteriors for the
derived quantities $\theta$, $B$ and $E_\mathrm{mag}$, that were obtained numerically
by integrating the distributions shown in Fig. \ref{fig:joint_imax} using the appropriate
change of variables. The results indicate that we
can put upper limits to the field strength and to the magnetic energy, and that
data contains some information on these two variables, with posterior clearly different
from the priors. Concerning
the field inclination, the marginal posterior looks like the prior distribution, 
so that we can only discard positive inclinations because they are not compatible with the
Stokes $V$ profile polarity.

\subsection{Degradation of information. The effect of noise}
\begin{figure*}
\centering
\includegraphics*[viewport=32 10 557 372,width=0.32\textwidth]{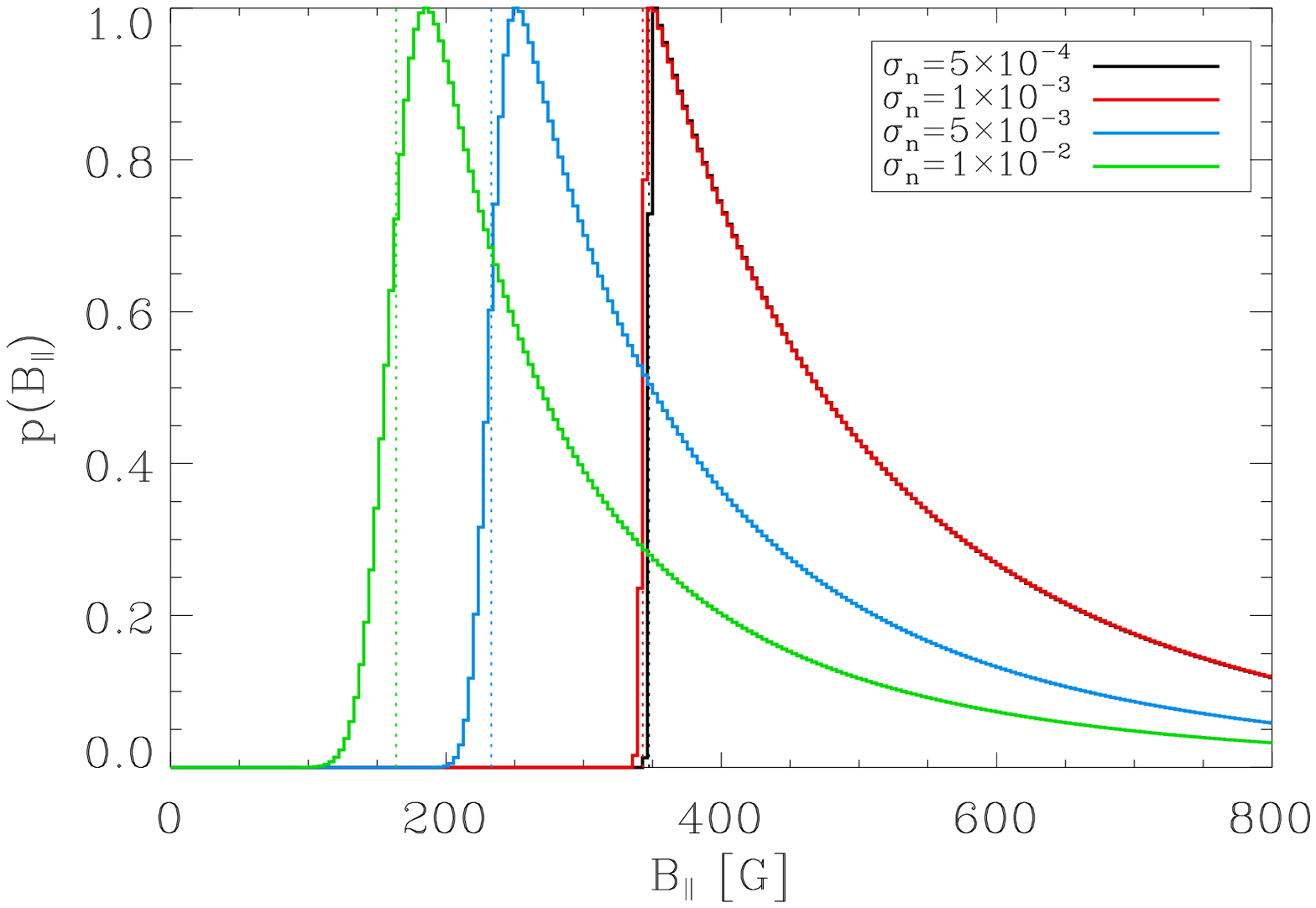}
\includegraphics*[viewport=32 10 557 372,width=0.32\textwidth]{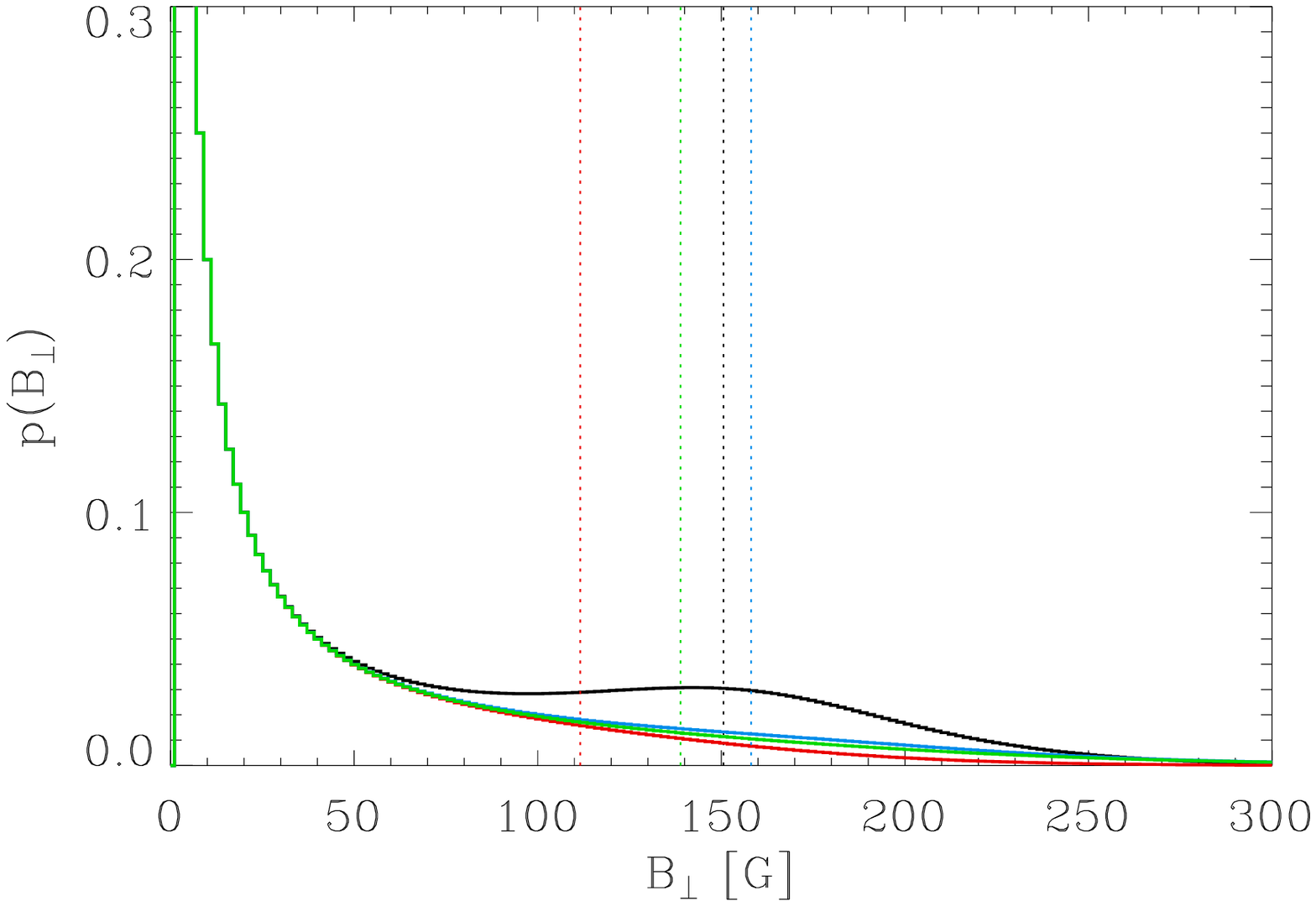}
\includegraphics*[viewport=32 10 557 372,width=0.32\textwidth]{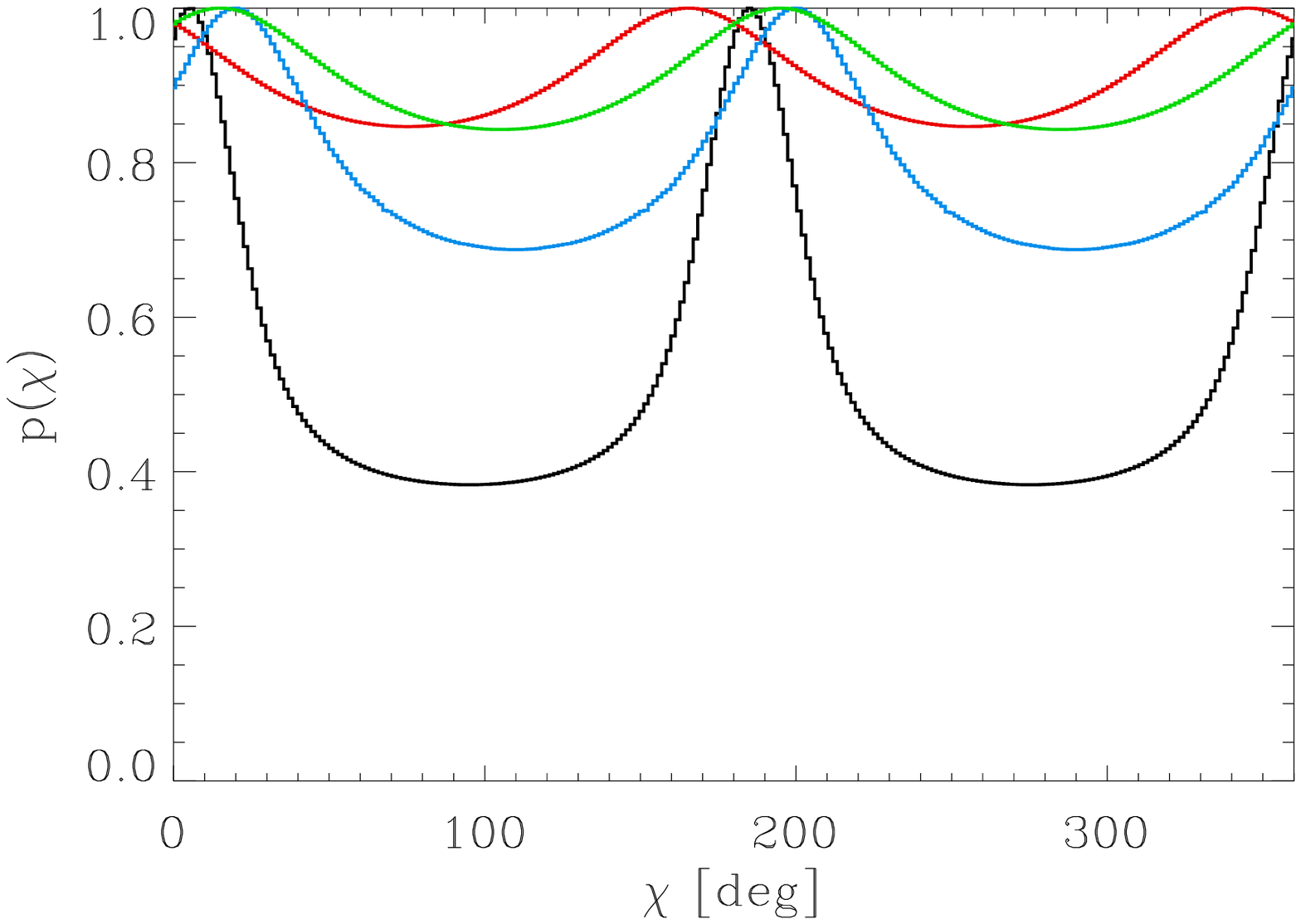}
\includegraphics*[viewport=32 10 557 372,width=0.32\textwidth]{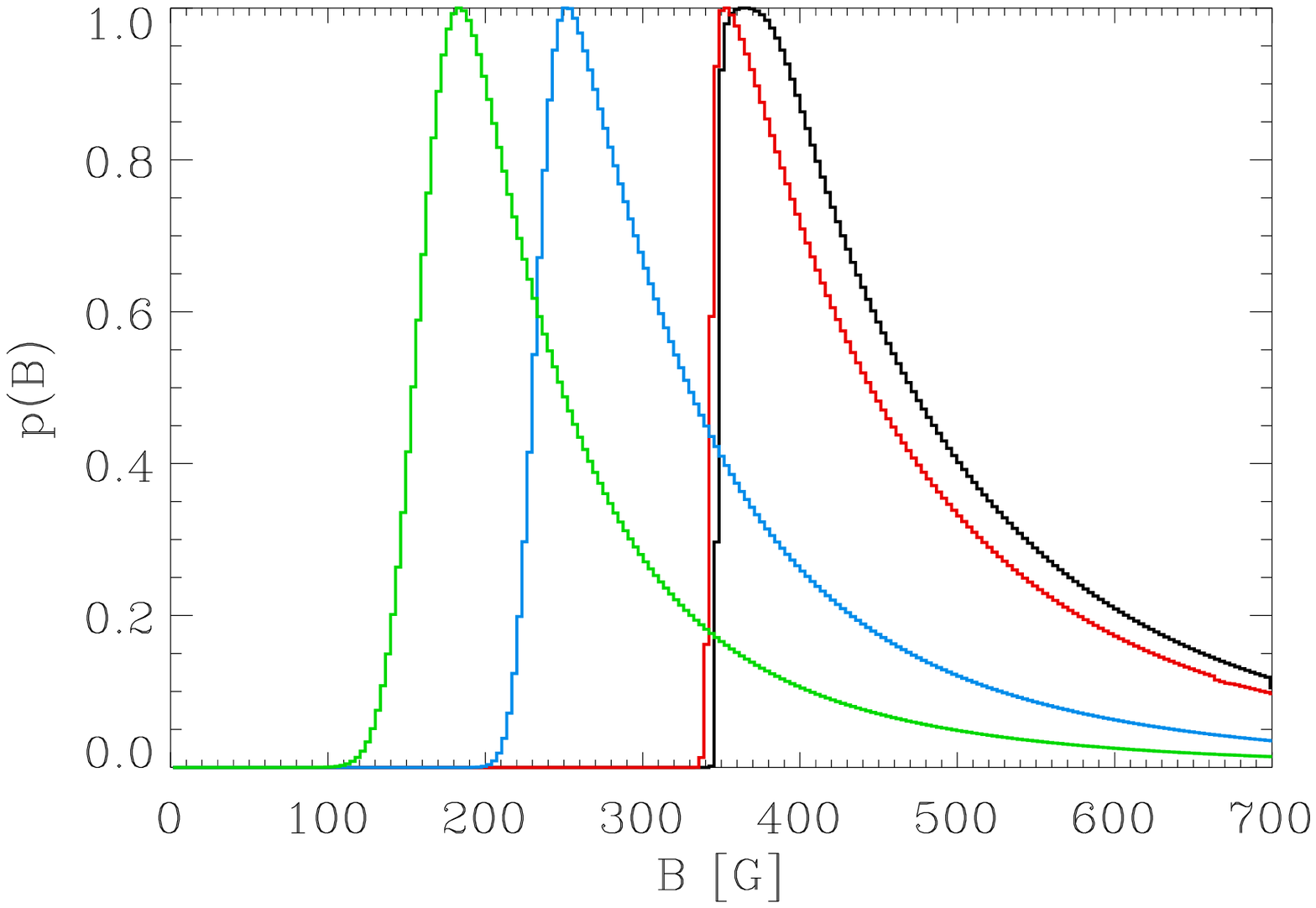}
\includegraphics*[viewport=32 10 557 372,width=0.32\textwidth]{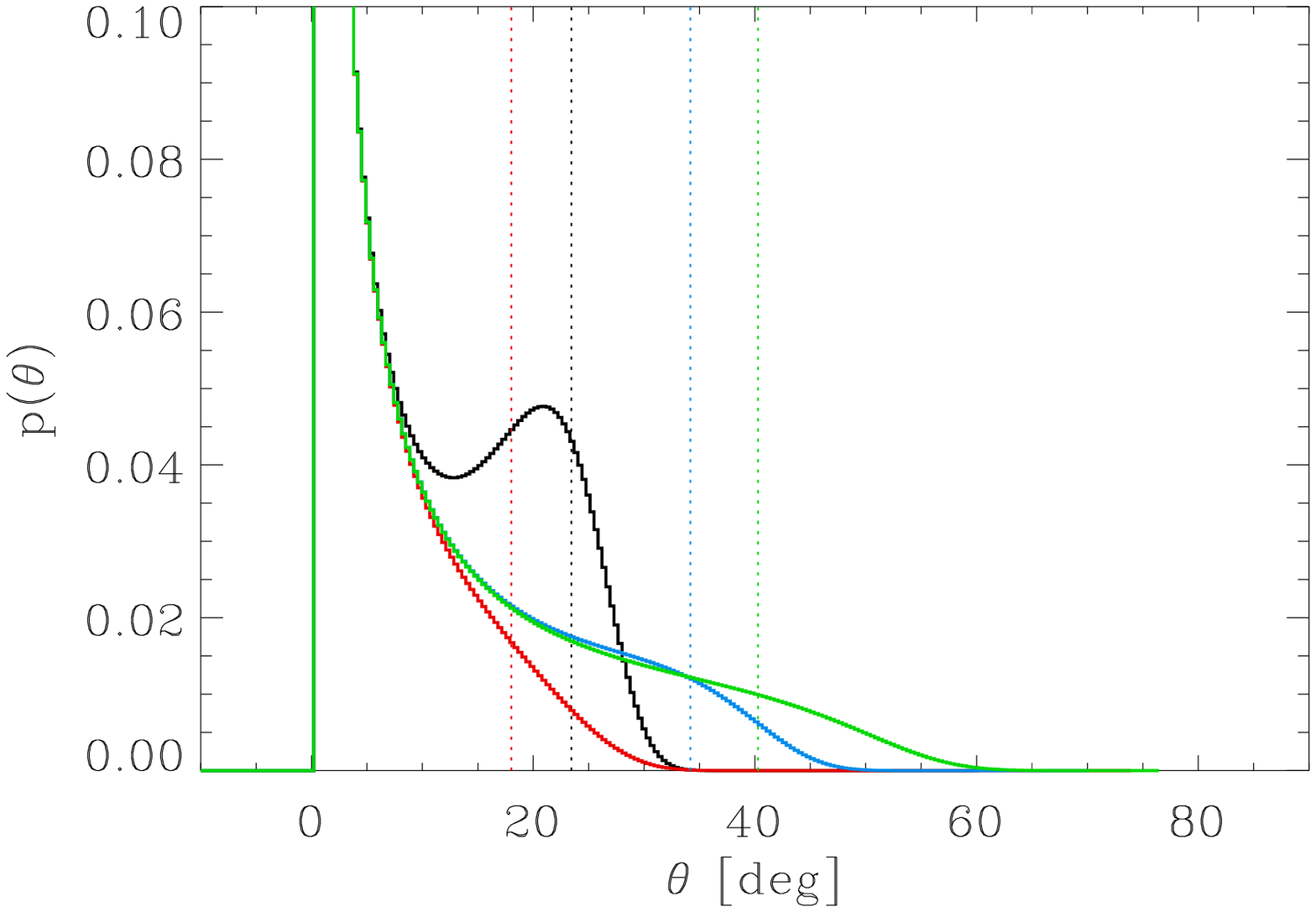}
\includegraphics*[viewport=32 10 557 372,width=0.32\textwidth]{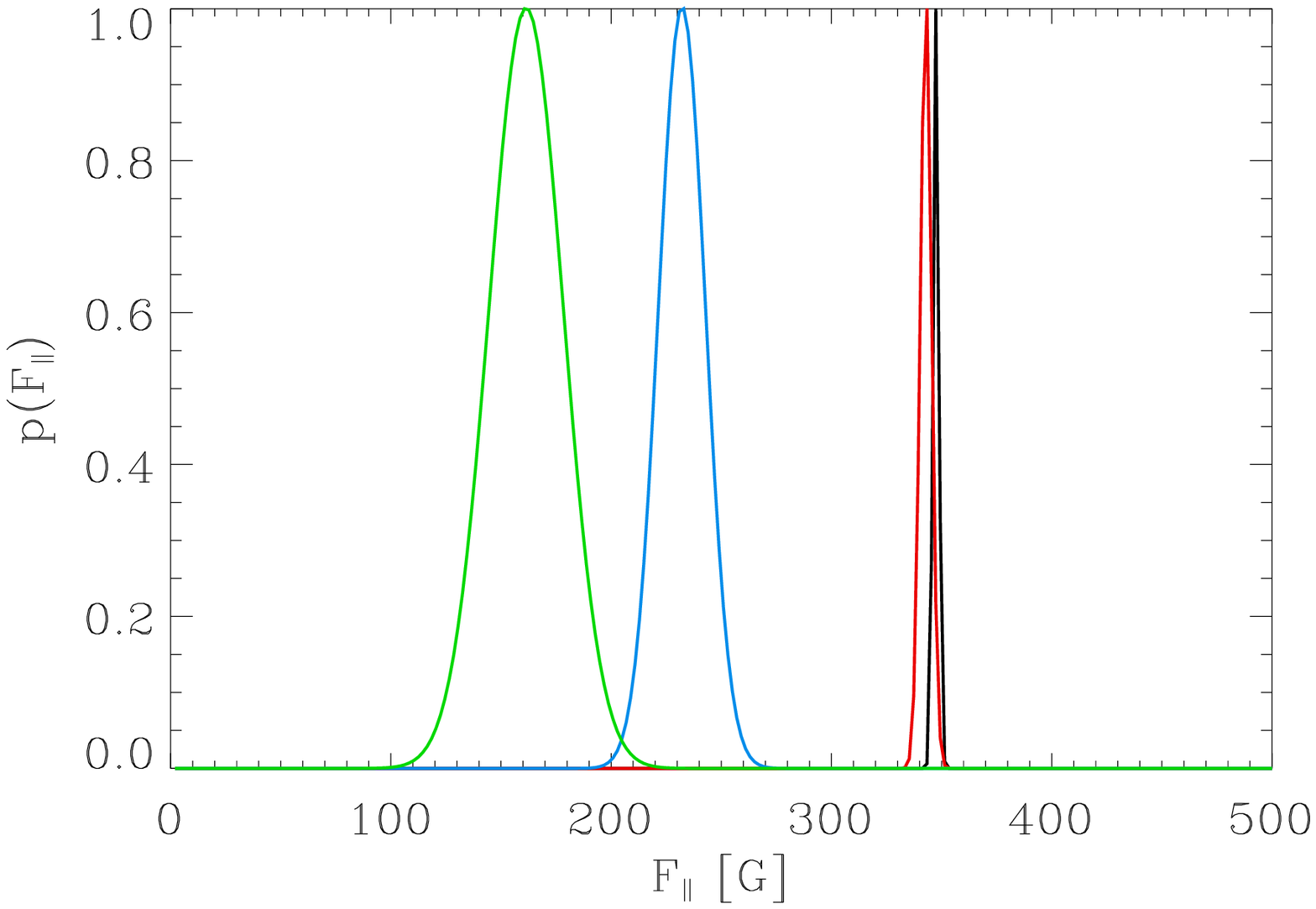}
\caption{Marginal posteriors for all physical parameters of relevance of a synthetic
Stokes profiles with increasingly higher noise levels. The different colors correspond
to different noise levels. The vertical dotted lines indicate the maximum-likelihood
solution.
\label{fig:degradation_noise}}
\end{figure*}

There are two important ingredients that differentiate a standard maximum-likelihood
(or least-squares) approach and a Bayesian approach. The first one is that
the solution to the inference problem is given in terms of marginal posterior
distributions that already encode error propagation. The second is that, thanks
to the effect of the priors, when noise is too large, the marginal posteriors
resemble the prior. This way, one can clearly state when a parameter is constrained
by data and avoid biased estimators. To show this effect, we have analyzed how the influence of noise
modifies the marginal posteriors. We consider the Stokes profiles of the Fe \textsc{i} line
at 6302.4904 \AA\ synthesized using the Milne-Eddington approximation with 
$B=400$ G, $\theta=30^\circ$ and $\chi=20^\circ$. This line has a large magnetic
sensitivity, with $\bar{g}=2.5$ and $\bar{G}=6.25$. Considering $f=1$, the amplitude
of Stokes $V$ is of the order of 2\% in units of the continuum intensity, $I_c$, while that of 
Stokes $Q$ and $V$ is of the order of 0.03\% in the same units. We corrupt these
profiles with Gaussian noise with zero mean and standard deviations $\sigma_n=\{5\times10^{-4}$,
$10^{-3}$, $5\times10^{-3}$, $10^{-2}\}$ in units $I_c$. The case with the smallest
$\sigma_n$ results in a noise amplitude of roughly the same amplitude of the linear polarization
signal. We use the hierarchical model with known noise variance of Eq. (\ref{eq:posterior_hierarchical_known_noise}) 
and numerically calculate
the marginal posteriors. The results are shown in Fig. \ref{fig:degradation_noise} for
all physical parameters of relevance. Each color corresponds to a different value
of $\sigma_n$.
The vertical dotted lines indicate the maximum-likelihood values
that maximize Eq. (\ref{eq:likelihood}). The curves are slightly dependent on the
actual noise realization because the $A_i$ coefficients change. We plot 
the results for only one noise realization.

As a consequence of the selected values for the magnetic field vector and $f$, the original
synthetic profile corresponds to $\Bpar=346.4$ G
and $\Bperp=200$ G. The marginal posterior for $\Bpar$ indicates that the
MMAP value is compatible with the correct value only when the noise is
not too large so that the Stokes $V$ signal is not too perturbed. As soon as
the noise increases, there is a shift towards smaller $\Bpar$. An interesting
property of the solution is that the confidence intervals are extremely asymmetric, with
small values of $\Bpar$ absolutely discarded and a very large upper limit. This is
produced, among other things, by the marginalization of the filling factor. Something
similar happens for the field strength. 

Concerning $\Bperp$, the marginal posterior is different
from the prior only for $\sigma_n=5\times10^{-4}$, with a bump roughly close to the correct
value although displaced towards smaller values. As soon as the noise increases,
there is no remaining information about this component of the field on the profiles and
one can only put an upper limit. A consequence of this is the fact that the 
marginal posterior for the azimuth is essentially flat. Some peaks of
larger probability are located close to the correct value but without statistical 
relevance. Since the field inclination depends on $\Bpar$ and $\Bperp$,
the marginal posterior for the inclination indicates that a reliable inclination
is only inferred when noise is sufficiently small. For large noises, only upper
limits can be correctly defined. Finally, the marginal posteriors for the magnetic flux 
density are Gaussian with a variance proportional to the noise standard deviation (in
G), as shown in Eq. (\ref{eq:mean_variance_fparallel}).

\section{Conclusions}
We have considered the complete Bayesian inference of the parameters of a model based on the 
weak-field approximation to explain observed Stokes profiles\footnote{Computer programs
that calculate all the quantities presented in this paper can be freely obtained from \texttt{http://www.iac.es/project/magnetism}.}. 
The simplicity of the
approximation has allowed us to obtain a closed analytical expression for the posterior
distribution function and for some of the ensuing marginal posteriors. Thanks to the
Bayesian approach, prior information is transparently introduced into the problem. We
have verified that results are sensitive to the hyperparameters of the prior and we have developed
a hierarchical approach based on physical arguments that introduce 
regularization into the problem. As a consequence, marginal posteriors are
almost insensitive to the exact value of the hyperparameters.
Using the Bayesian approach we are able to extract not only most probable values
but also confidence intervals for the model parameters. This Bayesian approach
can be of interest for filter-polarimetric data in which the line profiles are
only sampled in a reduced number of wavelengths. Likewise, signals very close
to the noise can be treated under this formalism avoiding biased 
estimations that plague least-squares solutions. It is left for the future
to carry out a profound analysis of the
biases introduced by least-squares solutions using the expressions introduced in this paper.

\begin{acknowledgements}
I thank R. Manso Sainz for fruitful discussions that led to improvements
on the original manuscript. I also thank C. Beck and M. J. Mart\'{\i}nez Gonz\'alez for useful suggestions
and V. Mart\'{\i}nez Pillet for providing the IMaX data used in Section \ref{sec:results}.
Financial support by the 
Spanish Ministry of Science and Innovation through projects AYA2010-18029 (Solar Magnetism and Astrophysical 
Spectropolarimetry) is gratefully acknowledged.
\end{acknowledgements}

\appendix

\section{Change of variables}
\label{sec:appendix}
Given two random variables $x$ and $y$ with joint distribution $p(x,y)$, the probability
distribution $q(w)$ for the derived quantity $w=f(x,y)$ with $f(x,y)$ invertible can
be obtained following the standard chain of variables rule. If we define
the auxiliar variable $z=g(x,y)$, we can write the following direct and inverse relations:
\begin{eqnarray}
w=f(x,y), \qquad x&=&\xi(z,w) \nonumber \\
z=g(x,y), \qquad y&=&\gamma(z,w).
\label{eq:transformation}
\end{eqnarray}
As a consequence, the joint probability distribution of $z$ and $w$ can be written as:
\begin{equation}
q(z,w) = p(\xi(z,w),\gamma(z,w)) |J|,
\label{eq:transformation_prob}
\end{equation}
where $J$ is the Jacobian of the transformation equation given by Eq. (\ref{eq:transformation}):
\begin{equation}
J = 
\left|
\begin{array}{cc}
\frac{\partial \xi(z,w)}{\partial z} & \frac{\partial \xi(z,w)}{\partial w} \\
\frac{\partial \gamma(z,w)}{\partial z} & \frac{\partial \gamma(z,w)}{\partial w}
\end{array}
\right| = 
\left|
\begin{array}{cc}
1 & 0 \\
\frac{\partial \gamma(z,w)}{\partial z} & \frac{\partial \gamma(z,w)}{\partial w}
\end{array}
\right| = \frac{\partial \gamma(z,w)}{\partial w},
\end{equation}
were the last two steps assumes $\xi(z,w)=z$ to simplify computations. The probability
distribution $q(w)$ is obtained by marginalizing $z$ from 
Eq. (\ref{eq:transformation_prob}):
\begin{equation}
q(w) = \int \mathrm{d}z \, p(z,\gamma(z,w)) \left| \frac{\partial \gamma(z,w)}{\partial w} \right|.
\end{equation}
For instance, if we want to carry out the transformation $w=f(x,y)=xy$, then $\gamma(z,w)=w/z$, so
that
\begin{equation}
q(w) = \int \mathrm{d}z \, p \left(z,\frac{w}{z} \right) \left| \frac{1}{z} \right|.
\end{equation}

\section{Marginal posteriors}
\label{sec:appendix_posteriors}
This appendix presents some of the analytical posteriors that can be obtained
from Eqs. (\ref{eq:posterior}) and (\ref{eq:posterior_hierarchical_known_noise}).
\subsection{Non-hierarchical model}
If we integrate the field azimuth from the posterior of the non-hierarchical model, we
obtain:
\begin{eqnarray}
p(f,\Bpar,\Bperp|D) &=& \Pi(f-1/2) \Pi\left(\frac{\chi-\pi}{2\pi} \right) (2\pi)^{-(3N+1)/2} \sigma_n^{-3N} \frac{1}{\sigma_\parallel \sigma_\perp^2} \Bperp 
I_0\left(2f\Bperp^2 \sqrt{A_5^2+A_6^2} \right) \nonumber \\
&\times& \exp \left\{ -\left[A_1 + \left(A_2 f^2+\frac{1}{2\sigma_\parallel^2} \right) 
\Bpar^2 + A_3 f^2 \Bperp^4 - 2 A_4 f \Bpar +
\frac{\Bperp^2}{2\sigma_\perp^2}  \right] \right\}
\label{eq:marginal_3d_nonhier}
\end{eqnarray}
where $I_0(x)$ is the modified Bessel function of the first kind \citep{abramowitz72}. It might be 
useful to apply the following series expansion of $I_0(x)$ to further integrate out the filling factor:
\begin{equation}
I_0(x) = \sum_{k=0}^\infty \left( \frac{1}{k!}\right)^2 \left( \frac{x}{2} \right)^{2k}=1 + \frac{x^2}{4} + \frac{x^4}{64}+\ldots.
\end{equation}
Likewise, marginalizing $\Bpar$ and $\Bperp$, we obtain:
\begin{eqnarray}
p(f,\chi|D ) &=& \frac{1}{8} \Pi(f-1/2) \Pi\left(\frac{\chi-\pi}{2\pi} \right) (2\pi)^{-(3N+1)/2} \sigma_n^{-3N} \frac{1}{\sigma_\parallel \sigma_\perp^2} 
\frac{1}{f \left[ A_3 \left(({2\sigma_\parallel^2})^{-1}+A_2f^2 \right) \right]^{1/2}} \nonumber \\
&\times& \exp \left[-A_1+\frac{\left(({2\sigma_\perp^2})^{-1}-2(A_5 \coschi + A_6 \sinchi) f \right)^2}{4A_3 f^2} + 
\frac{A_4^2 f^2}{({2\sigma_\parallel^2})^{-1} + A_2f^2} \right] \nonumber \\
&\times& \left[ 1+\mathrm{erf} \left( \frac{-({2\sigma_\perp^2})^{-1}+2(A_5 \coschi + A_6 \sinchi)f}{2\sqrt{A_3}f} \right) \right].
\end{eqnarray}

Concerning the marginal posterior for the field components, we have been unable
to find a close analytical expression for the probability distributions
$p(\Bpar|D)$ and $p(\Bperp|D)$ because $\Bpar$ and $\Bperp$ appear very 
intricately. However, the marginal posteriors can be obtained using adequate
numerical quadrature methods on any of the following joint posteriors: 
\begin{eqnarray}
p(\Bpar,\Bperp,\chi|D) &=& \frac{1}{2\sqrt{2}} \Pi\left(\frac{\chi-\pi}{2\pi} \right) (2\pi)^{-(3N+2)/2} \sigma_n^{-3N} 
\frac{1}{\sigma_\parallel \sigma_\perp^2} \Bperp
(A_2 \Bpar^2 + A_3 \Bperp^4)^{-1/2} \nonumber \\
&\times& \exp\left(-A_1 + \frac{\Bpar^2}{2\sigma_\parallel^2}  + 
\frac{\Bperp^2}{2\sigma_\perp^2} +\frac{(A_4 \Bpar + (A_5 \coschi + A_6 \sinchi) \Bperp^2)^2}{A_2 \Bpar^2 + A_3 \Bperp^4}\right) \nonumber \\
&\times& \left[ \mathrm{erf} \left( \frac{A_4 \Bpar + (A_5 \coschi + A_6 \sinchi) \Bperp^2}{\sqrt{A_2 \Bpar^2 + A_3 \Bperp^4}} \right) \right. \nonumber \\
&-& \left. \mathrm{erf} \left( \frac{A_4 \Bpar + (A_5 \coschi + A_6 \sinchi) \Bperp^2-A_2 \Bpar^2 - A_3 \Bperp^4}{\sqrt{A_2 \Bpar^2 + A_3 \Bperp^4}} 
\right)\right] \label{eq:joint_Bpar_Bperp} \\
p(f,\Bperp,\chi|D) &=& \frac{1}{\sqrt{2}} \Pi(f-1/2) \Pi\left(\frac{\chi-\pi}{2\pi} \right) (2\pi)^{-(3N+2)/2} \sigma_n^{-3N} 
\frac{1}{\sigma_\parallel \sigma_\perp^2} \frac{\Bperp}{\sqrt{({2\sigma_\parallel^2})^{-1}+A_2f^2}} \nonumber \\
&\times& \exp \left\{ \frac{A_4^2 f^2}{({2\sigma_\parallel^2})^{-1}+A_2f^2} - A_1 + 
\left[2(A_5 \coschi + A_6 \sinchi)f-\frac{1}{2\sigma_\perp^2} \right]\Bperp^2
-A_3f^2 \Bperp^4 \right\} \\
p(f,\Bpar,\chi|D) &=& \frac{1}{4\sqrt{2}} \Pi(f-1/2) \Pi\left(\frac{\chi-\pi}{2\pi} \right) (2\pi)^{-(3N+2)/2} \sigma_n^{-3N} 
\frac{1}{\sigma_\parallel \sigma_\perp^2} \frac{1}{\sqrt{A_3}f} \nonumber \\
&\times&\exp\left[ \frac{({4\sigma_\perp^2})^{-2}-(A_5 \coschi + A_6 \sinchi)({2\sigma_\perp^2})^{-1} - 
(A_1 A_3+A_5^2-\frac{A_3 \Bpar^2}{8\sigma_\parallel^2}) f^2
+2A_4 f^3- A_2 \Bpar f^4}{A_3 f^2}\right] \nonumber \\
&\times& \left[ 1+\mathrm{erf} \left( \frac{-({2\sigma_\perp^2})^{-1}+2(A_5 \coschi + A_6 \sinchi)f}{2\sqrt{A_3}f} \right) \right].
\label{eq:p_f_Bpar}
\end{eqnarray}
The previous expressions can be particularized to the specific case of a longitudinal magnetograph,
in which only circular polarization is observed. We obtain such a case making $A_3=A_5=A_6=0$, dropping
$Q_i$ and $U_i$ from $A_1$ and making $\sigma_\perp \to \infty$.

\subsection{Hierarchical model}
Starting from Eq. (\ref{eq:posterior_hierarchical_known_noise}), we have integrated out
the field azimuth, yielding:
\begin{eqnarray}
p(f,\Bpar,\Bperp|D) &=& \frac{1}{2} (2\pi)^{-(3N-1)/2} \sigma_n^{-3N} \Pi(f-1/2) \nonumber \\
&\times& \exp\left[-\left(A_1 + A_2 f^2 \Bpar^2 + A_3 f^2 \Bperp^4 - 2 A_4 f \Bpar\right)\right] I_0\left(2f\Bperp^2\sqrt{A_5^2+A_6^2}\right)\nonumber \\
&\times& \frac{1}{\Bpar \Bperp} \left[ 
\mathrm{erf} \left( \frac{\Bpar}{\sqrt{2}\sigma_\mathrm{min}} \right) -
\mathrm{erf} \left( \frac{\Bpar}{\sqrt{2}\sigma_\mathrm{max}} \right)\right] 
\left[ 
\exp \left( -\frac{\Bperp^2}{2\sigma_\mathrm{max}^2} \right) -
\exp \left( -\frac{\Bperp^2}{2\sigma_\mathrm{min}^2} \right)\right]
\label{eq:posterior_hierarchical_known_noise_marginalchi}
\end{eqnarray}
Finally, it is possible to analytically integrate out $f$:
\begin{eqnarray}
p(\Bpar,\Bperp,\chi|D) &=& \frac{1}{2} (2\pi)^{-3N/2} \sigma_n^{-3N} \Pi\left(\frac{\chi-\pi}{2\pi} \right) (A_2 \Bpar^2 + A_3 \Bperp^4)^{-1/2} \nonumber \\
&\times& \exp\left(-A_1+\frac{(A_4 \Bpar + (A_5 \coschi + A_6 \sinchi) \Bperp^2)^2}{A_2 \Bpar^2 + A_3 \Bperp^4}\right) \nonumber \\
&\times& \left[ \mathrm{erf} \left( \frac{A_4 \Bpar + (A_5 \coschi + A_6 \sinchi) \Bperp^2}{\sqrt{A_2 \Bpar^2 + A_3 \Bperp^4}} \right) \right.\nonumber \\
&-& \left. \mathrm{erf} \left( \frac{A_4 \Bpar + (A_5 \coschi + A_6 \sinchi) \Bperp^2-(A_2 \Bpar^2 + A_3 \Bperp^4)}{\sqrt{A_2 \Bpar^2 + A_3 \Bperp^4}} 
\right)\right] \nonumber \\
&\times&
\frac{1}{\Bpar \Bperp} \left[ 
\mathrm{erf} \left( \frac{\Bpar}{\sqrt{2}\sigma_\mathrm{min}} \right) -
\mathrm{erf} \left( \frac{\Bpar}{\sqrt{2}\sigma_\mathrm{max}} \right)\right] 
\left[ 
\exp \left( -\frac{\Bperp^2}{2\sigma_\mathrm{max}^2} \right) -
\exp \left( -\frac{\Bperp^2}{2\sigma_\mathrm{min}^2} \right)\right].
\label{eq:marginal_bpar_bperp_knownnoise}
\end{eqnarray}

\begin{thebibliography}{38}
\expandafter\ifx\csname natexlab\endcsname\relax\def\natexlab#1{#1}\fi

\bibitem[{{Abramowitz} \& {Stegun}(1972)}]{abramowitz72}
{Abramowitz}, M., \& {Stegun}, I.~A. 1972, Handbook of Mathematical Functions
  (New York: Dover)

\bibitem[{{Asensio Ramos}(2009)}]{asensio_hinode09}
{Asensio Ramos}, A. 2009, \apj, 701, 1032

\bibitem[{{Asensio Ramos}(2010)}]{asensio_spw6_10}
{Asensio Ramos}, A. 2010, in Solar Polarization 6, ASP Conf. Ser.

\bibitem[{{Asensio Ramos} {et~al.}(2007){Asensio Ramos}, {Mart\'{\i}nez
  Gonz\'alez}, \& {Rubi\~no Mart\'{\i}n}}]{asensio_martinez_rubino07}
{Asensio Ramos}, A., {Mart\'{\i}nez Gonz\'alez}, M.~J., \& {Rubi\~no
  Mart\'{\i}n}, J.~A. 2007, A\&A, 476, 959

\bibitem[{{Asensio Ramos} {et~al.}(2008){Asensio Ramos}, {Trujillo Bueno}, \&
  {Landi Degl'Innocenti}}]{asensio_trujillo_hazel08}
{Asensio Ramos}, A., {Trujillo Bueno}, J., \& {Landi Degl'Innocenti}, E. 2008,
  ApJ, 683, 542

\bibitem[{{Auer} {et~al.}(1977){Auer}, {House}, \&
  {Heasley}}]{auer_heasly_house77}
{Auer}, L.~H., {House}, L.~L., \& {Heasley}, J.~N. 1977, Sol. Phys., 55, 47

\bibitem[{{Aznar Cuadrado} {et~al.}(2004){Aznar Cuadrado}, {Jordan},
  {Napiwotzki}, {Schmid}, {Solanki}, \& {Mathys}}]{aznar_cuadrado04}
{Aznar Cuadrado}, R., {Jordan}, S., {Napiwotzki}, R., {Schmid}, H.~M.,
  {Solanki}, S.~K., \& {Mathys}, G. 2004, \aap, 423, 1081

\bibitem[{{Bagnulo} {et~al.}(2002){Bagnulo}, {Szeifert}, {Wade}, {Landstreet},
  \& {Mathys}}]{bagnulo02}
{Bagnulo}, S., {Szeifert}, T., {Wade}, G.~A., {Landstreet}, J.~D., \& {Mathys},
  G. 2002, \aap, 389, 191

\bibitem[{{Bellot Rubio}(2006)}]{bellotrubio_spw4_06}
{Bellot Rubio}, L.~R. 2006, in ASP Conf. Ser., Vol. 358, Solar Polarization
  Workshop 4, ed. R.~{Casini} \& B.~W. {Lites}, 107

\bibitem[{{Borrero} {et~al.}(2010){Borrero}, {Tomczyk}, {Kubo},
  {Socas-Navarro}, {Schou}, {Couvidat}, \& {Bogart}}]{borrero_vfisv10}
{Borrero}, J.~M., {Tomczyk}, S., {Kubo}, M., {Socas-Navarro}, H., {Schou}, J.,
  {Couvidat}, S., \& {Bogart}, R. 2010, \solphys, 35

\bibitem[{{Donati} {et~al.}(1997){Donati}, {Semel}, {Carter}, {Rees}, \&
  {Collier Cameron}}]{donati97}
{Donati}, J.-F., {Semel}, M., {Carter}, B.~D., {Rees}, D.~E., \& {Collier
  Cameron}, A. 1997, MNRAS, 291, 658

\bibitem[{{Frutiger} {et~al.}(2000){Frutiger}, {Solanki}, {Fligge}, \&
  {Bruls}}]{frutiger00}
{Frutiger}, C., {Solanki}, S.~K., {Fligge}, M., \& {Bruls}, J.~H.~M.~J. 2000,
  A\&A, 358, 1109

\bibitem[{Gelman {et~al.}(2003)Gelman, Carlin, Stern, \&
  Rubin}]{gelman_bayesian03}
Gelman, A., Carlin, J.~B., Stern, H.~S., \& Rubin, D.~B. 2003, Bayesian Data
  Analysis, Second Edition (Chapman \& Hall/CRC Texts in Statistical Science)
  (Chapman \& Hall)

\bibitem[{{Gregory}(2005)}]{gregory05}
{Gregory}, P.~C. 2005, Bayesian Logical Data Analysis for the Physical Sciences
  (Cambridge: Cambridge University Press)

\bibitem[{{Jeffreys}(1961)}]{jeffreys61}
{Jeffreys}, H. 1961, Theory of Probability (Oxford: Oxford University Press)

\bibitem[{{Jordan} {et~al.}(2005){Jordan}, {Werner}, \& {O'Toole}}]{jordan05}
{Jordan}, S., {Werner}, K., \& {O'Toole}, S.~J. 2005, \aap, 432, 273

\bibitem[{{Keller} {et~al.}(1990){Keller}, {Steiner}, {Stenflo}, \&
  {Solanki}}]{keller90}
{Keller}, C.~U., {Steiner}, O., {Stenflo}, J.~O., \& {Solanki}, S.~K. 1990,
  A\&A, 233, 583

\bibitem[{{Lagg} {et~al.}(2004){Lagg}, {Woch}, {Krupp}, \& {Solanki}}]{Lagg04}
{Lagg}, A., {Woch}, J., {Krupp}, N., \& {Solanki}, S.~K. 2004, A\&A, 414, 1109

\bibitem[{{Landi Degl'Innocenti} \& {Landi Degl'Innocenti}(1973)}]{landi73}
{Landi Degl'Innocenti}, E., \& {Landi Degl'Innocenti}, M. 1973, Sol. Phys., 31,
  319

\bibitem[{{Landi Degl'Innocenti} \& {Landi
  Degl'Innocenti}(1977)}]{landi_response77}
---. 1977, A\&A, 56, 111

\bibitem[{{Landi Degl'Innocenti} \& {Landolfi}(2004)}]{landi_landolfi04}
{Landi Degl'Innocenti}, E., \& {Landolfi}, M. 2004, Polarization in Spectral
  Lines (Kluwer Academic Publishers)

\bibitem[{{Lites} \& {Skumanich}(1990)}]{lites_skumanich90}
{Lites}, B.~W., \& {Skumanich}, A. 1990, ApJ, 348, 747

\bibitem[{{MacKay}(2003)}]{mackay03}
{MacKay}, D. J.~C. 2003, {Information Theory, Inference, and Learning
  Algorithms} (Cambridge University Press)

\bibitem[{{Mart{\'{\i}}nez Gonz{\'a}lez} {et~al.}(2008){Mart{\'{\i}}nez
  Gonz{\'a}lez}, {Asensio Ramos}, {L{\'o}pez Ariste}, \& {Manso
  Sainz}}]{marian_clv08}
{Mart{\'{\i}}nez Gonz{\'a}lez}, M.~J., {Asensio Ramos}, A., {L{\'o}pez Ariste},
  A., \& {Manso Sainz}, R. 2008, A\&A, 479, 229

\bibitem[{{Mart{\'{\i}}nez Gonz{\'a}lez} \& {Bellot
  Rubio}(2009)}]{marian_luis09}
{Mart{\'{\i}}nez Gonz{\'a}lez}, M.~J., \& {Bellot Rubio}, L.~R. 2009, \apj,
  700, 1391

\bibitem[{{Mart\'{\i}nez Pillet} {et~al.}(2004){Mart\'{\i}nez Pillet}, {Bonet},
  {Collados}, {Jochum}, {Mathew}, {Medina Trujillo}, {Ruiz Cobo}, {del Toro
  Iniesta}, {Lopez Jimenez}, {Castillo Lorenzo}, {Herranz}, {Jeronimo},
  {Mellado}, {Morales}, {Rodriguez}, {Alvarez-Herrero}, {Belenguer},
  {Heredero}, {Menendez}, {Ramos}, {Reina}, {Pastor}, {Sanchez}, {Villanueva},
  {Domingo}, {Gasent}, \& {Rodriguez}}]{imax04}
{Mart\'{\i}nez Pillet}, V., {Bonet}, J.~A., {Collados}, M.~V., {Jochum}, L.,
  {Mathew}, S., {Medina Trujillo}, J.~L., {Ruiz Cobo}, B., {del Toro Iniesta},
  J.~C., {Lopez Jimenez}, A.~C., {Castillo Lorenzo}, J., {Herranz}, M.,
  {Jeronimo}, J.~M., {Mellado}, P., {Morales}, R., {Rodriguez}, J.,
  {Alvarez-Herrero}, A., {Belenguer}, T., {Heredero}, R.~L., {Menendez}, M.,
  {Ramos}, G., {Reina}, M., {Pastor}, C., {Sanchez}, A., {Villanueva}, J.,
  {Domingo}, V., {Gasent}, J.~L., \& {Rodriguez}, P. 2004, in Society of
  Photo-Optical Instrumentation Engineers (SPIE) Conference Series, Vol. 5487,
  Society of Photo-Optical Instrumentation Engineers (SPIE) Conference Series,
  ed. {J.~C.~Mather}, 1152--1164

\bibitem[{{Mart{\'{\i}}nez Pillet} {et~al.}(2011){Mart{\'{\i}}nez Pillet}, {Del
  Toro Iniesta}, {{\'A}lvarez-Herrero}, {Domingo}, {Bonet}, {Gonz{\'a}lez
  Fern{\'a}ndez}, {L{\'o}pez Jim{\'e}nez}, {Pastor}, {Gasent Blesa}, {Mellado},
  {Piqueras}, {Aparicio}, {Balaguer}, {Ballesteros}, {Belenguer}, {Bellot
  Rubio}, {Berkefeld}, {Collados}, {Deutsch}, {Feller}, {Girela}, {Grauf},
  {Heredero}, {Herranz}, {Jer{\'o}nimo}, {Laguna}, {Meller}, {Men{\'e}ndez},
  {Morales}, {Orozco Su{\'a}rez}, {Ramos}, {Reina}, {Ramos},
  {Rodr{\'{\i}}guez}, {S{\'a}nchez}, {Uribe-Patarroyo}, {Barthol}, {Gandorfer},
  {Knoelker}, {Schmidt}, {Solanki}, \& {Vargas Dom{\'{\i}}nguez}}]{imax11}
{Mart{\'{\i}}nez Pillet}, V., {Del Toro Iniesta}, J.~C., {{\'A}lvarez-Herrero},
  A., {Domingo}, V., {Bonet}, J.~A., {Gonz{\'a}lez Fern{\'a}ndez}, L.,
  {L{\'o}pez Jim{\'e}nez}, A., {Pastor}, C., {Gasent Blesa}, J.~L., {Mellado},
  P., {Piqueras}, J., {Aparicio}, B., {Balaguer}, M., {Ballesteros}, E.,
  {Belenguer}, T., {Bellot Rubio}, L.~R., {Berkefeld}, T., {Collados}, M.,
  {Deutsch}, W., {Feller}, A., {Girela}, F., {Grauf}, B., {Heredero}, R.~L.,
  {Herranz}, M., {Jer{\'o}nimo}, J.~M., {Laguna}, H., {Meller}, R.,
  {Men{\'e}ndez}, M., {Morales}, R., {Orozco Su{\'a}rez}, D., {Ramos}, G.,
  {Reina}, M., {Ramos}, J.~L., {Rodr{\'{\i}}guez}, P., {S{\'a}nchez}, A.,
  {Uribe-Patarroyo}, N., {Barthol}, P., {Gandorfer}, A., {Knoelker}, M.,
  {Schmidt}, W., {Solanki}, S.~K., \& {Vargas Dom{\'{\i}}nguez}, S. 2011,
  \solphys, 268, 57

\bibitem[{{Merenda} {et~al.}(2006){Merenda}, {Trujillo Bueno}, {Landi
  Degl'Innocenti}, \& {Collados}}]{merenda06}
{Merenda}, L., {Trujillo Bueno}, J., {Landi Degl'Innocenti}, E., \& {Collados},
  M. 2006, ApJ, 642, 554

\bibitem[{{Orozco Su{\'a}rez} {et~al.}(2007){Orozco Su{\'a}rez}, {Bellot
  Rubio}, {del Toro Iniesta}, {Tsuneta}, {Lites}, {Ichimoto}, {Katsukawa},
  {Nagata}, {Shimizu}, {Shine}, {Suematsu}, {Tarbell}, \&
  {Title}}]{orozco_hinode07}
{Orozco Su{\'a}rez}, D., {Bellot Rubio}, L.~R., {del Toro Iniesta}, J.~C.,
  {Tsuneta}, S., {Lites}, B.~W., {Ichimoto}, K., {Katsukawa}, Y., {Nagata}, S.,
  {Shimizu}, T., {Shine}, R.~A., {Suematsu}, Y., {Tarbell}, T.~D., \& {Title},
  A.~M. 2007, \apjl, 670, L61

\bibitem[{{O'Toole} {et~al.}(2005){O'Toole}, {Jordan}, {Friedrich}, \&
  {Heber}}]{otoole05}
{O'Toole}, S.~J., {Jordan}, S., {Friedrich}, S., \& {Heber}, U. 2005, \aap,
  437, 227

\bibitem[{{Ruiz Cobo} \& {del Toro Iniesta}(1992)}]{sir92}
{Ruiz Cobo}, B., \& {del Toro Iniesta}, J.~C. 1992, ApJ, 398, 375

\bibitem[{{Silvester} {et~al.}(2009){Silvester}, {Neiner}, {Henrichs}, {Wade},
  {Petit}, {Alecian}, {Huat}, {Martayan}, {Power}, \& {Thizy}}]{silvester09}
{Silvester}, J., {Neiner}, C., {Henrichs}, H.~F., {Wade}, G.~A., {Petit}, V.,
  {Alecian}, E., {Huat}, A., {Martayan}, C., {Power}, J., \& {Thizy}, O. 2009,
  \mnras, 398, 1505

\bibitem[{{Skumanich} \& {Lites}(1987)}]{skumanich_lites87}
{Skumanich}, A., \& {Lites}, B.~W. 1987, ApJ, 322, 473

\bibitem[{{Socas-Navarro} {et~al.}(2000){Socas-Navarro}, {Trujillo Bueno}, \&
  {Ruiz Cobo}}]{socas_trujillo_ruiz00}
{Socas-Navarro}, H., {Trujillo Bueno}, J., \& {Ruiz Cobo}, B. 2000, ApJ, 530,
  977

\bibitem[{{Solanki} {et~al.}(2010){Solanki}, {Barthol}, {Danilovic}, {Feller},
  {Gandorfer}, {Hirzberger}, {Riethm{\"u}ller}, {Sch{\"u}ssler}, {Bonet},
  {Mart{\'{\i}}nez Pillet}, {del Toro Iniesta}, {Domingo}, {Palacios},
  {Kn{\"o}lker}, {Bello Gonz{\'a}lez}, {Berkefeld}, {Franz}, {Schmidt}, \&
  {Title}}]{sunrise10}
{Solanki}, S.~K., {Barthol}, P., {Danilovic}, S., {Feller}, A., {Gandorfer},
  A., {Hirzberger}, J., {Riethm{\"u}ller}, T.~L., {Sch{\"u}ssler}, M., {Bonet},
  J.~A., {Mart{\'{\i}}nez Pillet}, V., {del Toro Iniesta}, J.~C., {Domingo},
  V., {Palacios}, J., {Kn{\"o}lker}, M., {Bello Gonz{\'a}lez}, N., {Berkefeld},
  T., {Franz}, M., {Schmidt}, W., \& {Title}, A.~M. 2010, \apjl, 723, L127

\bibitem[{{Spirock} {et~al.}(2001){Spirock}, {Denker}, {Varsik}, {Shumko},
  {Qiu}, {Gallagher}, {Chae}, {Goode}, \& {Wang}}]{spirock01}
{Spirock}, T.~J., {Denker}, C., {Varsik}, J., {Shumko}, S., {Qiu}, J.,
  {Gallagher}, P., {Chae}, J., {Goode}, P., \& {Wang}, H. 2001, AGU Spring
  Meeting Abstracts, 51

\bibitem[{{Varsik}(1995)}]{varsik95}
{Varsik}, J.~R. 1995, \solphys, 161, 207

\bibitem[{{Wade} {et~al.}(2000){Wade}, {Donati}, {Landstreet}, \&
  {Shorlin}}]{wade00}
{Wade}, G.~A., {Donati}, J., {Landstreet}, J.~D., \& {Shorlin}, S.~L.~S. 2000,
  \mnras, 313, 851

\end{thebibliography}

\end{document}